\documentclass[11pt]{article}
\usepackage{graphicx} % Required for inserting images
\usepackage[inline,shortlabels]{enumitem}
\usepackage{float}
\usepackage{mathtools}
\usepackage{subdepth}
\usepackage{graphicx}
\usepackage{amsmath}
\usepackage{array}
\usepackage{tikz}
\usetikzlibrary{shapes.geometric, arrows.meta, positioning, calc}
\usepackage{adjustbox}
\usepackage[mathscr]{euscript}
\usepackage{algorithm,algpseudocode}
\counterwithin{figure}{section}
\counterwithin{table}{section}
\usepackage[square,numbers,sort&compress]{natbib}
\usepackage{bbm}
\usepackage[margin=1in]{geometry}
\usepackage{tikz}
\usetikzlibrary{arrows.meta}
\usetikzlibrary{decorations.markings}
\tikzset{+ /.tip = {Bar[sep=-3pt 2,width=3pt 4]_[sep=0]}}
\usepackage[english]{babel}
\usepackage{longtable}
\usepackage{color}
\usepackage{mathtools}
\usepackage{amssymb,amsmath,amsthm}
\usepackage{multirow}
\usepackage[titletoc,title]{appendix}
\usepackage{authblk}
\usepackage{bm}
\usepackage{setspace}
\usepackage{dsfont}
\usepackage[OT1]{fontenc}
\usepackage{subcaption}
\usepackage{refcount}
\usepackage{booktabs}
\usepackage{indentfirst}
\usepackage{hyperref}
\usepackage{caption}
\usepackage{float}

\newtheorem{thm}{Theorem}[section]
\newtheorem{lem}{Lemma}[section]
\newtheorem{assumption}{Assumption}[section]

\newcommand{\norm}[1]{\left\lVert#1\right\rVert}
\newcommand{\ep}{\varepsilon}

\title{Regularized High-Dimensional Additive Tensor Autoregressive Model}
\author[1]{Debika Ghosh}
\affil[1]{\small Indian Institute of Management Udaipur}
\author[2]{Nilanjana Chakraborty}
\affil[2]{\small Indian Institute of Management Udaipur}
\author[3]{Samrat Roy$^{*}$}
\affil[3]{\small Indian Institute of Management Ahmedabad}

\begin{document}
\maketitle
\begin{abstract}
    High-dimensional time series has diverse applications in econometrics and finance. Recent models for capturing temporal dependence have employed a bilinear representation for matrix time series, or the Tucker-decomposition based representation in case of tensor time series. A Tucker-decomposition based temporal effect is difficult to interpret on many occasions, along with its computational complexity due to the non-convex nature of the underlying optimization problem. Moreover, the existing tensor models have not sufficiently explored the possibilities of imposing any lower-dimensional pattern on the transition matrices. In this work, we propose a regularized additive tensor autoregressive model with additive interaction of row-wise, column-wise and tube-wise temporal dependence, that offers more interpretability, less computational burden due to its convex nature and estimation of the underlying low rank plus sparse pattern of its transition matrices. We address the issue of identifiability of the various components in our model and subsequently develop a scalable alternating block minimization algorithm for estimating the parameters. We provide a finite sample error bound under high-dimensional scaling for the model parameters. Finally, the efficacy of the proposed model is demonstrated on synthetic and real data.
\end{abstract}
\maketitle

\section{Introduction}
\label{intro}

 The study of high-dimensional time series models has emerged as a significant area of research in recent years, driven by advances in high-dimensional theoretical inference \citep{basu2015regularized, zhang2017gaussian, wang2022high, adamek2023lasso}, and the growing availability of high-dimensional temporal data. Such models have found widespread applications in diverse areas, including finance and macroeconomics \citep{de2008forecasting, bernanke2005measuring, blanchard2002empirical}, demography \citep{gao2019high}, functional genomics \citep{michailidis2013autoregressive}, transportation networks \citep{chen2019modeling} and neuroscience \citep{seth2015granger}.

Most existing approaches model temporal dependence in high-dimensional vector-valued time series through regularized vector autoregressive (VAR) models \citep{banbura2010large,basu2015regularized,kock2015oracle,ghosh2018high}. More recently, analogous models have been developed for matrix and tensor-valued time series, where each observation is represented in the form of a matrix \citep{chen2021autoregressive} or a tensor \citep{li2021multi} respectively. Treating such observations as vectors ignores their inherent multi-way structure and discards valuable information encoded in the interactions among different modes of the data, potentially resulting in a loss of statistical efficiency and interpretability. \cite{chen2021autoregressive} proposed a matrix autoregressive (MAR) model in which they used a bilinear multiplicative form $AY_{t-1}B^\prime$ to represent the temporal dependence between the data matrices $\{Y_t\}_{t=1}^T$, and the transition matrices $A$ and $B$ captured the row-wise and column-wise temporal dependence. Along the same line, \cite{li2021multi} considered a similar autoregressive model for tensor-variate time series (TAR), where they used a Tucker decomposed structure \citep{kolda2009tensor} to capture the underlying temporal dependence in the data. In particular, they employed Tucker-based multiplicative form $\mathcal{Y}_{t-1} \times_1 A_1 \times_2 A_2 \times_3 A_3$ to model the temporal dependence in the three-dimensional tensor data $\{\mathcal{Y}_t\}_{t=1}^T$, where the matrices $A_1$, $A_2$ and $A_3$ captured temporal dependence along three modes of the tensor. To facilitate dimension reduction in the above-mentioned multiplicative MAR or multiplicative TAR models, both reduced rank structure and sparsity structure of the transition matrices have been explored \citep{xiao2022reduced, hsu2021matrix, wang2022high, chen2025dynamic, cai2025efficient, boonen2026low}. While these approaches help in reducing high-dimensionality, they may suffer from the following problems:

\begin{figure}[H]
    \centering
    \includegraphics[scale=0.5]{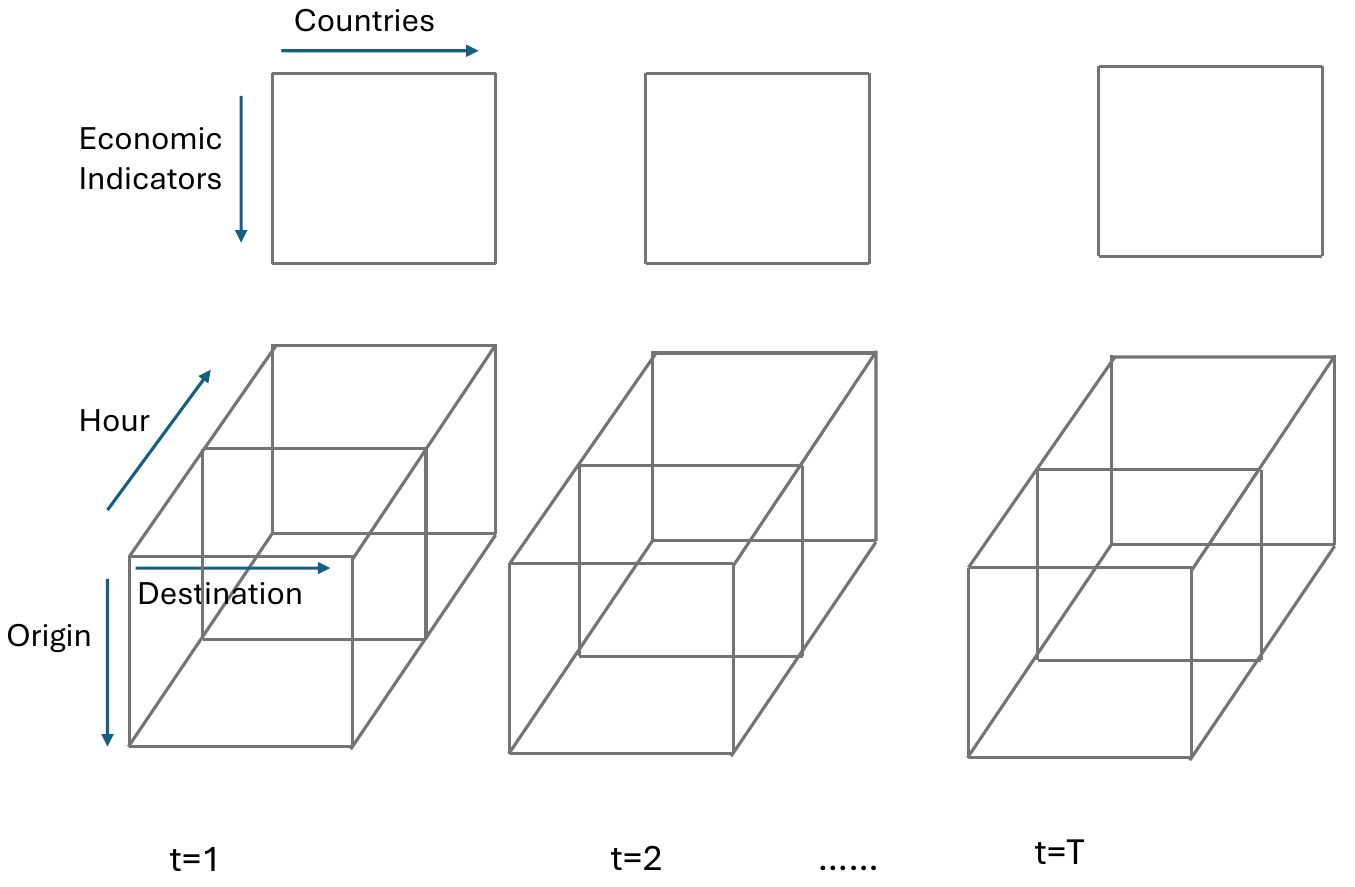}
    \caption{Examples of matrix and tensor-valued time series data: the upper panel displays a matrix-valued macroeconomic data with values of different economic indicators across the rows for different countries across the columns. The lower panel illustrates a three-dimensional tensor-valued time series, where the first and second modes correspond to the origin and destination boroughs in New York City, respectively, while the third mode represents the operational hour of the day. Each tensor entry records the number of trips between the corresponding origin–destination pair during the specified hour.}  
    \label{fig:examplepic}
\end{figure}

\begin{enumerate}
    \item [(a)] In case of Tucker-based multiplicative representation $\mathcal{Y}_{t-1} \times_1 A_1 \times_2 A_2 \times_3 A_3$, the temporal effects along the three modes, that is row-wise, column-wise and tube-wise temporal effects, are convoluted in multiplicative interaction form, and it becomes difficult to disjoin and interpret the three effects separately. As illustrated in Section \ref{model}, while modeling the temporal dependence of tensor-valued origin-destination demand for yellow taxis in NYC with different trip-origin boroughs across the rows, trip-destination boroughs across the columns and different operational hours along the tubes, one may be interested in coherently estimating the three sources of temporal dependence $\textendash$ along different origin boroughs, different destination boroughs and different operational hours. However, a Tucker-based convoluted structure will not serve that purpose as it entangles the three effects through multiplicative interactions. 
    
    \item[(b)] Though a reduced-rank or sparse structure imposed on the transition matrices $A_1$, $A_2$ and $A_3$ of the Tucker-based multiplicative form $\mathcal{Y}_{t-1} \times_1 A_1 \times_2 A_2 \times_3 A_3$ alleviates the high-dimensionality of the parameters, it can be inadequate to represent the desired low-dimensional pattern on many occasions. In particular, under the reduced-rank assumption, the aforementioned Tucker-based structure is represented as $\sum_{r=1}^R\mathcal{Y}_{t-1} \times_1 A_1^{(r)} \times_2 A_2^{(r)} \times_3 A_3^{(r)}$ \citep{li2021multi}, and additionally $A_1^{(r)}$, $A_2^{(r)}$ and $A_3^{(r)}$ are assumed to be sparse which helps further dimension reduction. However, this low-dimensional representation may not always be suitable to capture the underlying pattern. For instance, in the context of aforementioned taxi-demand data (see Figure \ref{fig:examplepic}) with trip-origin boroughs along the rows, trip-destination boroughs along the columns and operational hours along the tubes, it is reasonable to assume that some boroughs will have similar temporal dependence in terms of trip origination. For example, core commercial boroughs like Manhattan and Brooklyn may share some similar temporal patterns as trip-origin points, while residential boroughs like Bronx and Staten Island share some similarities. In other words, there will be a baseline component of origin-wise temporal dependence which will be `shared' or `similar' across the boroughs. Likewise, there will be a baseline `shared' component of temporal dependence along the dimension of trip-destination and operational hours too. In addition to these baseline components, there can be some additional idiosyncratic parts of temporal dependence as well. For example, high demand of taxis in Staten Island in case of disruption in ferry service can attribute to such idiosyncratic component of temporal dependence. However, the Tucker-based reduced-rank structure $\sum_{r=1}^R\mathcal{Y}_{t-1} \times_1 A_1^{(r)} \times_2 A_2^{(r)} \times_3 A_3^{(r)}$ with sparse $A_1^{(r)}$, $A_2^{(r)}$ and $A_3^{(r)}$ does not explicitly accommodate the aforementioned low-dimensional pattern that is decomposed into baseline and idiosyncratic components.           
    
    \item [(c)]Finally, with Tucker-based representation of the temporal dependence, the estimation process becomes computationally involved, and often the underlying optimization turns out to be a non-convex one.     
\end{enumerate}

In this paper, we propose a high-dimensional regularized additive tensor autoregressive model that overcomes the above-mentioned drawbacks. Our model captures the temporal dependence among the tensor-valued time series by employing an additive interaction form, wherein the overall temporal connection is represented as the sum of row-wise, column-wise and tube-wise temporal dependence in the data. To accommodate high-dimensionality of the parameters, we then impose low-rank plus sparse decomposed structures on row-wise, column-wise and tube-wsie transition matrices. As discussed above, this additive interaction form, as opposed to convoluted bilinear representation, offers more comprehensible interpretation of the row-wise, column-wise and tube-wise temporal dependence \citep{zhang2024additive}. Also, with additive form, the penalized transition matrices help in extracting meaningful low-dimensional pattern in the data, whereas, the same with bilinear form provides only dimension reduction. We develop a scalable alternating minimization algorithm to estimate the model parameters under high-dimensional setting that solves a convex optimization problem. We also address the issue of identifiability by employing a novel incoherence condition on the low-rank and sparse components of our model parameters. Finally, in terms of theoretical developments, we provide a detailed derivation and interpretation of the non-asymptotic upper bound of the estimation error under high dimensional scaling of the model parameters. To the best of our knowledge, the proposed methodology and the subsequent theoretical developments are novel contributions to the field of high-dimensional tensor time series analysis.

 The remainder of the paper is organized as follows. Section \ref{model} provides a detailed description of our proposed model, illustrating all the steps involved in it, and also describes our algorithm to estimate the model parameters. Section \ref{theo} provides theoretical results related to the upper bound of the estimation error under high-dimensional scaling of the parameters. We then illustrate the performance of our posited method based on both synthetic and real data in Sections \ref{simu} and \ref{real_data} respectively, which is then followed by a concluding discussion in Section \ref{disc}. 

\section{Regularized Additive Tensor Autoregressive Model}
\label{model}
Suppose $\{\mathcal Y_t \in \mathbb{R}^{d_1 \times d_2 \times d_3}\}_{t=1}^T$ denotes a three-dimensional tensor valued time series and the objective is to model the underlying temporal dependence in the data. As mentioned in Section \ref{intro}, an example of such tensor-valued time series can be three-dimensional tensor with trip-origin boroughs along the rows, trip-destination boroughs across the columns and operational hours along the tubes, and the $(i,j,k)^{th}$ element of $\mathcal{Y}_t$ represents the number of trips originated from the $i^{th}$ borough, ending at the $j^{th}$ borough, during the $k^{th}$ operational hour at the time point $t$. As explained earlier in Section \ref{intro}, existing literature use a Tucker-based multiplicative form $\mathcal{Y}_{t-1} \times_1 A_1 \times_2 A_2 \times_3 A_3$ to capture the temporal dependence in the data. However, in this representation, row-wise, column-wise and tube-wise temporal effects are convoluted with each other in multiplicative interaction form, and it becomes difficult to disjoin and interpret the three effects separately. 

\begin{figure}[H]
    \centering
    \includegraphics[scale=0.5]{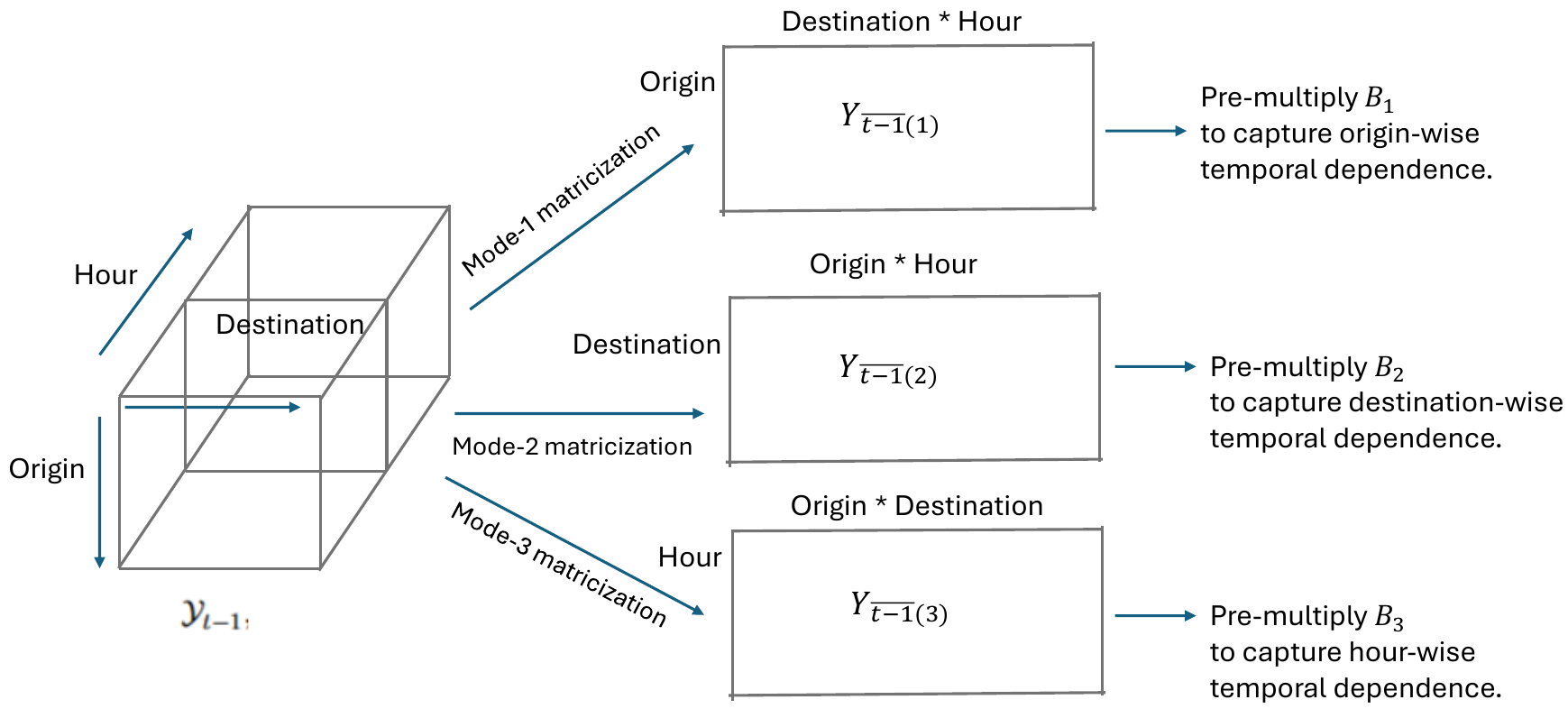}
    \caption{Mode-wise additive temporal dependence structure: past tensor data is first `matricized' \citep{kolda2009tensor} along the three modes. The transition matrices are then multiplied to those matricized versions in order to capture temporal dependence along the three modes. Finally those matricized temporal depndence along each mode is converted back to the tensor form by `fold' operation.}  
    \label{fig:matricization}
\end{figure}

To alleviate this issue, we propose a mode-wise additive tensor autoregressive model as discussed next. For the past tensor data $\mathcal{Y}_{t-1}$, we first construct its `matricized' versions \citep{kolda2009tensor} along its three different modes, namely, row, column and tube. As defined in \cite{kolda2009tensor}, a mode-1 or row-wise matricization of a three-dimensional tensor in $\mathbb{R}^{d_1 \times d_2 \times d_3}$ converts it into a matrix with $d_1$ rows and $d_2\times d_3$ columns. Thus, for our example, row-wise matricized version of past demand tensor $\mathcal{Y}_{t-1}$ will be a matrix $Y_{\overline{t-1}(1)}$ with $d_1$ origin boroughs along the rows and $d_2d_3$ combinations of destination boroughs and operational hours along the columns. Likewise, mode-2 or column-wise matricized version of $\mathcal{Y}_{t-1}$ will arrange the tensor data as a matrix $Y_{\overline{t-1}(2)}$ with $d_2$ destination boroughs along the rows and $d_1d_3$ combinations of origin boroughs and operational hours along the columns. Finally, mode-3 or tube-wise matricization $Y_{\overline{t-1}(3)}$ arranges the data as a matrix with $d_3$ hours along the rows and $d_1d_2$ origin and destination boroughs across the columns. We now use the matrices $B_1\in\mathbb{R}^{d_1\times d_1}$, $B_2\in\mathbb{R}^{d_2\times d_2}$ and $B_3\in\mathbb{R}^{d_3\times d_3}$ to capture temporal dependence along these three modes, and propose an additive tensor autoregressive model with additive interaction of row-wise, column-wise and tube-wise temporal dependence as follows:               
\begin{flalign}
    \mathcal{Y}_{t} &= \operatorname{fold}_1 \Bigg[ B_1 Y_{\overline{t-1} (1)}\Bigg] + \operatorname{fold}_2 \Bigg[ B_2 Y_{\overline{t-1} (2)}\Bigg] + \operatorname{fold}_3 \Bigg[ B_3 Y_{\overline{t-1} (3)}\Bigg] + \mathcal{E}_{t} \hspace{0.3cm} 
    \label{eq1}
\end{flalign}

\noindent for $t=1,2,\cdots,T$, where the operation $\text{fold}_1(\cdot)$ arranges the matrix $B_1Y_{\overline{t-1}(1)} \in \mathbb{R}^{d_1\times d_2d_3}$ into a three-dimensional tensor in $\mathbb{R}^{d_1 \times d_2 \times d_3}$. The operators $\text{fold}_2(\cdot)$ and $\text{fold}_3(\cdot)$ are defined in a similar fashion and the tensor $\mathcal{E}_t\in \mathbb{R}^{d_1 \times d_2 \times d_3}$ denotes the error tensor.  

To alleviate high-dimensionality of the model parameters $B_1$, $B_2$ and $B_3$, we assume low-dimensional pattern on the same in the following fashion. As discussed earlier in Section \ref{intro}, in case of taxi-demand data, it is reasonable to assume that some boroughs will have similar temporal dependence patterns in terms of trip origination. For example, core commercial boroughs like Manhattan and Brooklyn may share some similar pattern as trip-origin points, while residential boroughs like Bronx and Staten Island share some similarities. As discussed in Section \ref{intro}, this `similar' or `shared' component of temporal dependence is the baseline component. So, there will be baseline temporal dependence along all three modes $\textendash$ origin-wise, destination-wise and operational hour-wise. In addition to these baseline components, there can be some additional idiosyncratic parts of temporal dependence as well. As mentioned in Section \ref{intro}, a high demand for taxis in Staten Island in case of disruption in ferry service can attribute to one such potential idiosyncratic component of temporal dependence. To capture such baseline plus idiosyncratic temporal dependence along each mode, we decompose each transition matrix $B_k$ into a low-rank matrix $L_k$ and a sparse matrix $S_k$ for $k=1,2,3$. The low-rank matrix $L_k$, due to its inherent linear dependence structure, is aimed at capturing the similar baseline temporal dependence along the $k^{th}$ mode. On the other hand, the sparse matrix $S_k$, with only a few non-zero elements in it, is used to capture the additional idiosyncratic temporal dependence along the $k^{th}$ mode. Thus \eqref{eq1} translates to  
\begin{flalign}
    \mathcal{Y}_{t} &= \operatorname{fold}_1 \Bigg[(L_1 + S_1) Y_{\overline{t-1} (1)}\Bigg] + \operatorname{fold}_2 \Bigg[ (L_2 + S_2) Y_{\overline{t-1} (2)}\Bigg] + \operatorname{fold}_3 \Bigg[ (L_3 + S_3)Y_{\overline{t-1} (3)}\Bigg] + \mathcal{E}_{t} \hspace{0.3cm}
    \label{eq2}
\end{flalign}
for $t=1,2,\cdots,T$. Using the nuclear norm $\norm{\cdot}_{*}$ and $\ell_1$ norm $\norm{\cdot}_{1}$ as suitable convex surrogates for low-rank and sparsity constraints respectively, our aim is to minimize the following jointly convex objective function. 
\begin{equation}
\label{obj_func}
\begin{aligned}
    \frac{1}{2T} \sum_{t=1}^{T} \left \lVert \mathcal{Y}_{t} - \operatorname{fold}_1 \Bigg[ (L_1 + S_1) Y_{\overline{t-1} (1)}\Bigg]  - \operatorname{fold}_2 \Bigg[ (L_2 + S_2) Y_{\overline{t-1} (2)}\Bigg] - \operatorname{fold}_3 \Bigg[ (L_3 + S_3) Y_{\overline{t-1} (3)}\Bigg]\right \rVert_{F}^{2} \\
    + \lambda_{S_1} \| S_1 \|_{1} + \lambda_{S_2} \| S_2 \|_{1}+ \lambda_{S_3} \| S_3 \|_{1} +\lambda_{L_1} \| L_1 \|_{*}
     + \lambda_{L_2} \| L_2 \|_{*} + \lambda_{L_3} \| L_3 \|_{*}
     \end{aligned}
\end{equation}
\noindent where $\lambda_{L_1}$, $\lambda_{L_2}$, $\lambda_{L_3}$ and $\lambda_{S_1}$, $\lambda_{S_2}$, $\lambda_{S_3}$ are non-negative regularization parameters for the low-rank and sparse components respectively. Later in Section \ref{theo}, we discuss the ideas to ensure identifiability of these low-rank and sparse components. 

\subsection{Estimation of the parameters}
We use the notation $f(L_1,S_1,L_2,S_2,L_3,S_3)$ to denote the objective function \eqref{obj_func}. It is easy to verify that `$f$' is jointly convex in its arguments and hence the following alternating block  minimization procedure summarized in Algorithm \ref{Algo}, will obtain the desired minimizer.
\begin{algorithm}[H]
\caption{Alternating Block Minimization for minimizing
$f(L_1,S_1,L_2,S_2,L_3,S_3)$}
\label{Algo}
\begin{algorithmic}
\State \textbf{Input:} data $\{\mathcal{Y}_t\}_{t=1}^T$,
$\lambda_{L_1},\lambda_{L_2},\lambda_{L_3},
\lambda_{S_1},\lambda_{S_2},\lambda_{S_3}$

\State \textbf{Initialize:}
$L_1^{(0)},S_1^{(0)},L_2^{(0)},S_2^{(0)},L_3^{(0)},S_3^{(0)}$

\State
\textbf{Repeat}

\State Step 1: Update
\Statex \hspace{\algorithmicindent}
$L_1^{(t+1)}
=\arg\min_{L_1}
f(L_1,S_1^{(t)},L_2^{(t)},S_2^{(t)},L_3^{(t)},S_3^{(t)})$.

\vspace{0.1in}

\State Step 2: Update
\Statex \hspace{\algorithmicindent}
$S_1^{(t+1)}
=\arg\min_{S_1}
f(L_1^{(t+1)},S_1,L_2^{(t)},S_2^{(t)},L_3^{(t)},S_3^{(t)})$.

\vspace{0.1in}

\State Step 3: Update
\Statex \hspace{\algorithmicindent}
$L_2^{(t+1)}
=\arg\min_{L_2}
f(L_1^{(t+1)},S_1^{(t+1)},L_2,S_2^{(t)},L_3^{(t)},S_3^{(t)})$.

\vspace{0.1in}

\State Step 4: Update
\Statex \hspace{\algorithmicindent}
$S_2^{(t+1)}
=\arg\min_{S_2}
f(L_1^{(t+1)},S_1^{(t+1)},L_2^{(t+1)},S_2,L_3^{(t)},S_3^{(t)})$.

\vspace{0.1in}

\State Step 5: Update
\Statex \hspace{\algorithmicindent}
$L_3^{(t+1)}
=\arg\min_{L_3}
f(L_1^{(t+1)},S_1^{(t+1)},L_2^{(t+1)},S_2^{(t+1)},L_3,S_3^{(t)})$.

\vspace{0.1in}

\State Step 6: Update
\Statex \hspace{\algorithmicindent}
$S_3^{(t+1)}
=\arg\min_{S_3}
f(L_1^{(t+1)},S_1^{(t+1)},L_2^{(t+1)},S_2^{(t+1)},L_3^{(t+1)},S_3)$.

\vspace{0.1in}

\State
\textbf{Until}
$f(L_1^{(t+1)},S_1^{(t+1)},L_2^{(t+1)},S_2^{(t+1)},L_3^{(t+1)},S_3^{(t+1)})$
converges.

\end{algorithmic}
\end{algorithm}

In steps 1, 3 and 5 of the above algorithm, we update the low-rank component $L_1$, $L_2$ and $L_3$ with nuclear norm penalization. This minimization problem shows up in various applications of machine learning, such as matrix classification, multi-task learning and matrix completion (see \cite{argyriou2008convex, tomioka2007classifying, roy2022regularized}). \cite{ji2009accelerated} considered a general class of optimization problems that includes the above formulation and proposed an Extended Gradient Algorithm and Accelerated Gradient Algorithm to obtain the minimizer. A direct application of the aforementioned algorithms provides the optimal solution in our case. On the other hand, in steps 2, 4 and 6, when we update $S_1$, $S_2$ and $S_3$, we use the algorithm for penalized multivariate regression used in \cite{lin2016penalized}.

\section{Theoretical Results}
\label{theo}
Denoting the minimizer of the objective function in \eqref{obj_func} as $\hat{L}_1$, $\hat{S}_1$, $\hat{L}_2$, $\hat{S}_2$, $\hat{L}_3$ and $\hat{S}_3$, we define the estimation error as follows:
\begin{equation}
\begin{aligned}
\label{def}
 e^2(\hat{L}_1, \hat{L}_2, \hat{L}_3, \hat{S}_1, \hat{S}_2, \hat{S}_3) &= \left \lVert \hat{L}_{1} - L_1 \right \rVert_F^{2} + \left \lVert \hat{L}_{2} - L_2 \right \rVert_F^{2} + \left \lVert \hat{L}_{3} - L_3 \right \rVert_F^{2} \\
 &+\left \lVert \hat{S}_{1} - S_1 \right \rVert_F^{2} + \left \lVert \hat{S}_{2} - S_2 \right \rVert_F^{2} + \left \lVert \hat{S}_{3} - S_3 \right \rVert_F^{2}
 \end{aligned}
\end{equation}

The key theoretical contribution of our work is to derive a non-asymptotic upper bound to the estimation error of the proposed additive tensor autoregressive model parameters under high-dimensional scaling. We derive that estimation error bound under both Gaussian and sub-Exponential distributional assumptions. As discussed in Section~\ref{intro}, the only significant prior work that explicitly models temporal dependence in tensor-variate time series is \cite{li2021multi}, which adopts a Tucker-based multiplicative form. That formulation, as discussed earlier, presents challenges in terms of interpretability and incurs significant computational costs. A further point of distinction lies in the theoretical treatment. While \cite{li2021multi} estimate the mode-wise transition matrices $A_1^{(r)}$, $A_2^{(r)}$ and $A_3^{(r)}$ (see Section \ref{intro}) through an alternating optimization procedure, their high-dimensional consistency result is established only for the Kronecker-product representation $\sum_{r=1}^R A_3^{(r)}\otimes A_2^{(r)} \otimes A_1^{(r)}$, which limits its interpretation (Theorem 3 in \cite{li2021multi}). Consequently, their theoretical analysis characterizes the estimation error in terms of only the overall dimensionality of the tensor relative to the sample size. In contrast, our theoretical results explicitly consider estimation of all three mode-wise low-rank and sparse components and the corresponding error bound characterizes the rates at which the dimensions of each of these low-rank and sparse components can grow relative to the sample size while ensuring consistency and identifiability. As is standard in proving high-dimensional consistency results (see \cite{slsbook}), our derivation proceeds in two steps. In the first step, we derive an upper bound to the estimation error under non-random errors $\mathcal E_t$, that is, when the sequence $\{\mathcal E_t\}_{t=1}^T$ is treated as deterministic. Ignoring constants, this upper bound involves the regularization parameters $\lambda_{L_1}$, $\lambda_{L_2}$, $\lambda_{L_3}$, $\lambda_{S_1}$, $\lambda_{S_2}$ and $\lambda_{S_3}$, the ranks of the low-rank components $L_1$, $L_2$ and $L_3$ and the numbers of nonzero elements in the sparse components $S_1$, $S_2$ and $S_3$ (see Lemma~\ref{lem31}). However, this step requires assuming suitable lower bounds on the regularization parameters, and those bounds depend on $\{\mathcal E_t\}_{t=1}^T$. Such assumptions are fairly standard in the high-dimensional literature \citep{slsbook,basu2015regularized,roy2022regularized}. In the second step, we introduce the distributional assumptions on $\{\mathcal E_t\}_{t=1}^T$ and, consequently, it behooves us to find suitable choices of $\lambda_{L_1}$, $\lambda_{L_2}$, $\lambda_{L_3}$, $\lambda_{S_1}$, $\lambda_{S_2}$ and $\lambda_{S_3}$ such that the assumed lower bounds from the first step hold with high probability. Finally, the estimation error bound is obtained by substituting these probabilistically valid choices of the regularization parameters into the deterministic bound derived in the first step (see Theorems~\ref{thm:ebound}). To the best of our knowledge, these estimation error bound derivations and their subsequent interpretations are fairly new in the high-dimensional tensor autoregressive literature.
\begin{figure}
\centering
\begin{adjustbox}{width=\textwidth}
\begin{tikzpicture}[
    node/.style={
        rectangle,
        draw,
        rounded corners,
        fill=gray!10,
        thick,
        text width=4.3cm,
        align=center,
        font=\small,
        minimum height=2.4cm
    },
    arrow/.style={-Stealth, thick}
]

% Nodes
\node[node] (assump) {
\textbf{Assumptions}\\[1mm]
\begin{itemize}
    \item Restricted strong convexity\\
    \item Incoherence condition for identifiability\\
     \item Lower bound on regularization parameters
\end{itemize}
};

\node[node, right=2cm of assump] (lemma1) {
\textbf{Lemma 3.1}\\
Estimation error bound under deterministic (non-random) error.
};

\node[node, above right=1.4cm and 2cm of lemma1] (theorem1) {
\textbf{Theorem 3.1}\\
Estimation error bound under Gaussian errors.
};

\node[node, below right=1.4cm and 2cm of lemma1] (lemmas) {
\textbf{Lemmas A3--A6}\\
Tools to establish high-dimensional results under sub-exponential errors.
};

\node[node, right=2cm of lemmas] (theorem2) {
\textbf{Theorem 3.2}\\
Estimation error bound under sub-exponential errors.
};

% Arrows
\draw[arrow] (assump) -- (lemma1);
\draw[arrow] (lemma1) -- (theorem1);
\draw[arrow] (lemma1) -- (lemmas);
\draw[arrow] (lemmas) -- (theorem2);

\end{tikzpicture}
\end{adjustbox}
\caption{A schematic of the theoretical developments in this paper, highlighting the key contributions and outlining the overall roadmap.}
\label{fig:flow_diagram}
\end{figure}
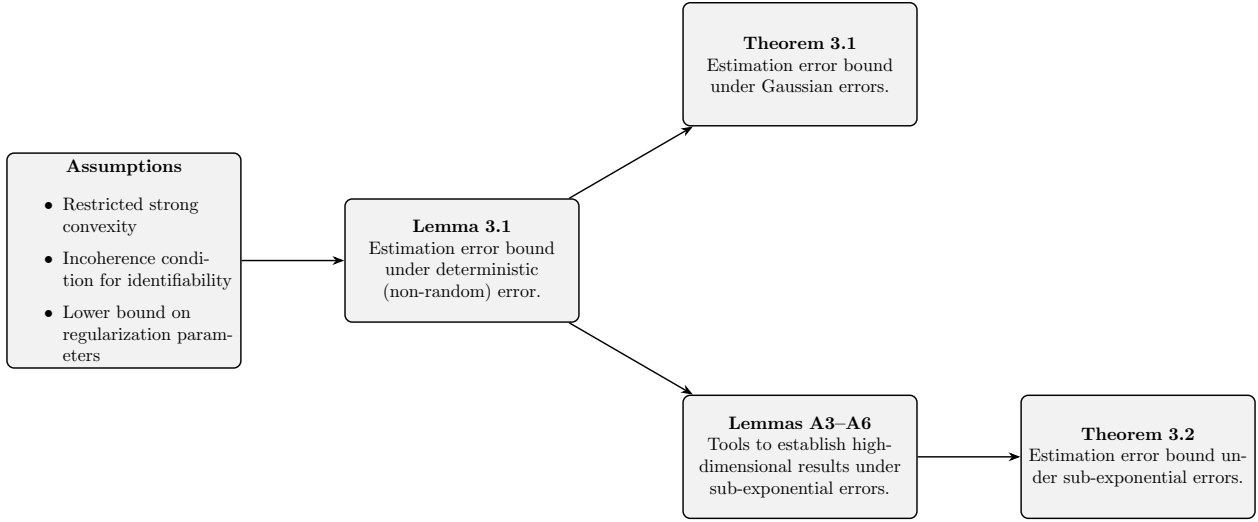

Figure \ref{fig:flow_diagram} illustrates the theoretical developments in this paper, highlighting the novel contributions. As depicted there, the roadmap for theoretical developments in this section is as follows: we start with three key assumptions, namely, restricted strong convexity of the loss function (Assumption \ref{ass 1}), incoherence condition for identifiability of the low-rank and sparse components (Assumption \ref{ass 2}) and lower bound on the regularization parameters (Assumption \ref{ass 3}). For restricted strong convexity of the loss function, it is enough to assume that our loss function has strong convexity over some `restricted set' of interest \citep{negahban2012unified} -- a set where the errors ($\hat{L_1}-L_1$,$\hat{S_1}-S_1$,$\hat{L_2}-L_2$,$\hat{S_2}-S_2$,$\hat{L_3}-L_3$,$\hat{S_3}-S_3$) belong. Lemmas \ref{lem-1} and \ref{lem-2} in the Appendix characterize that restricted set of interest in our case. With these standard assumptions, Lemma \ref{lem31} establishes the bound on the estimation error $e^2(\hat{L}_1, \hat{L}_2, \hat{L}_3, \hat{S}_1, \hat{S}_2, \hat{S}_3)$ under  deterministic or non-random realizations of the errors. Theorem \ref{thm:ebound} extends the previous result and provides the error-bound under the Gaussian error distribution. We then develop technical tools in Lemmas \ref{lemma5}–\ref{lemma8} to generalize the deviation bounds from the Gaussian to the sub-Exponential setting, which are then used to derive the estimation error bound under sub-Exponential distribution in Theorem \ref{theo_x_rand}.  

\vspace{0.1in}
\noindent \textbf{Additional notation:} We begin by introducing some additional notation that will be used throughout the remainder of this section. Let $R_1 \ll d_1$, $R_2 \ll d_2$ and $R_3 \ll d_3$ denote the ranks of $L_1$, $L_2$ and $L_3$ respectively. We assume that $S_1$, $S_2$ and $S_3$ have $s_1 \ll d_1^2$, $s_2 \ll d_2^2$ and $s_3 \ll d_3^2$ non-zero elements respectively. More specifically, suppose that $S_1$ is supported on a subset $H \subseteq \{1,2,\cdots,d_1^2\}$, with $|H|=s_1$. We define a pair of subspaces $(\mathbb{M}(H),\mathbb{M}^{\perp}(H))$, such that, $\mathbb{M}(H)=\{M\in \mathbb{R}^{d_1 \times d_1} \mid k^{th} \text{ element of } M =0, \forall k \notin H\}$ and $\mathbb{M}^{\perp}(H)= (\mathbb{M}(H))^{\perp}$. As shown in \cite{r1} and \cite{negahban2012unified}, one can easily verify that for any $M_1 \in \mathbb{M}(H)$ and $M_2 \in \mathbb{M}^{\perp}(H)$, $\norm{M_1+M_2}_{1}=\norm{M_1}_{1}+\norm{M_2}_{1}$. This ensures that the regularizer $\norm{\cdot}_{1}$ is decomposable (see \cite{negahban2012unified}) with respect to the subspace pair $(\mathbb{M}(H),\mathbb{M}^{\perp}(H))$. Simplifying the notation from $(\mathbb{M}(H),\mathbb{M}^{\perp}(H))$ to $(\mathbb{M},\mathbb{M}^{\perp})$, it is evident that, $S_1 \in \mathbb{M}$. Let $\pi$ be the orthogonal projection operator defined with respect to the standard inner product on the matrix space. Therefore, any $A \in \mathbb{M}^\perp(H)$ will have zeros on all entries in $H$. Thus $\pi_{\mathbb{M}^{\perp}} (S_1)=0$ and $\pi_{\mathbb{M}} (S_1)= S_1$. We define $\hat{\Delta}_{L_1}=\hat{L}_1-L_1$, $\hat{\Delta}_{S_1}=\hat{S}_1-S_1$, $\hat{\Delta}_{L_2}=\hat{L}_2-L_2$, $\hat{\Delta}_{S_2}=\hat{S}_2-S_2$, $\hat{\Delta}_{L_3}=\hat{L}_3-L_3$ and $\hat{\Delta}_{S_3}=\hat{S}_3-S_3$. Also, $\hat{\Delta}_{S_1}^{\mathbb{M}} = \pi_{\mathbb{M}}( \hat{\Delta}_{S_1})$ and $\hat{\Delta}_{S_1}^{\mathbb{M}^{\perp}} = \pi_{\mathbb{M}^{\perp}}( \hat{\Delta}_{S_1})$. Similarly, for a pair of subspaces $(\mathbb{N},\mathbb{N}^{\perp})$, we define $\hat{\Delta}_{S_2}^{\mathbb{N}} = \pi_{\mathbb{N}}( \hat{\Delta}_{S_2})$ and  $\hat{\Delta}_{S_2}^{\mathbb{N}^{\perp}} = \pi_{\mathbb{N}^{\perp}}( \hat{\Delta}_{S_2})$, and for a pair of subspaces $(\mathbb{G},\mathbb{G}^{\perp})$, we define $\hat{\Delta}_{S_3}^{\mathbb{G}} = \pi_{\mathbb{G}}( \hat{\Delta}_{S_3})$ and  $\hat{\Delta}_{S_3}^{\mathbb{G}^{\perp}} = \pi_{\mathbb{G}^{\perp}}( \hat{\Delta}_{S_3})$  The $\ell_1$ and $\ell_{\infty}$ norm of a matrix $A$ are defined by $\norm{A}_1=\underset{i}{\sum} \underset{j}{\sum}\lvert{a_{ij}}\rvert$ and $\norm{A}_{\infty}=\underset{i,j}{max}\lvert{a_{ij}}\rvert$ respectively. Denoting the singular values of $A\in \mathbb{R}^{m_1 \times m_2}$ by $\sigma_1(A), \sigma_2(A),\cdots, \sigma_m(A)$, where $m = min\{m_1, m_2\}$, we define the Nuclear Norm of $A$ by $\norm{A}_{*} = \sum_{j=1}^m \sigma_j(A)$ and the Spectral Norm of $A$ by $\norm{A}_{sp} = \underset{1\leq j \leq m}{max} \{\sigma_j(A)\}$. 
\\

\vspace{0.2in}

%===============================Assumptions===============

\noindent \textbf{Restricted strong convexity assumption:}\\
\vspace{0.03in}\\
This is a fairly standard assumption in the high-dimensional literature \cite{r1, roy2022regularized}, which ensures that the loss function exhibits strong convexity over some `restricted set' of interest. In other words, this implies that the loss function should have a sharp curvature around the optimal solution, ensuring that a small difference in loss implies a small error. Otherwise, if there is not sufficient curvature of the loss function around the optimal solution, then the error can be large even if the difference in loss is small, which is undesirable. As explained in \cite{negahban2012unified}, this sharp curvature or strong convexity of our loss function $\frac{1}{2T} \sum_{t=1}^{T} \left \lVert \mathcal{Y}_{t} - \operatorname{fold}_1 [ (L_1 + S_1) Y_{\overline{t-1} (1)}]  - \operatorname{fold}_2 [ (L_2 + S_2) Y_{\overline{t-1} (2)}] - \operatorname{fold}_3 [ (L_3 + S_3) Y_{\overline{t-1} (3)}]\right \rVert_{F}^{2}$ is ensured by imposing a suitable lower bound to the first-order Taylor series expansion of the loss function (see \eqref{eqa2}). However, note that it is impossible to ensure global strong convexity under high-dimensional setup and thus, a common practice is to ensure strong convexity on some `restricted set' of interest \citep{negahban2012unified}. In our case, that set is the one where the errors ($\hat{L_1}-L_1$,$\hat{S_1}-S_1$,$\hat{L_2}-L_2$,$\hat{S_2}-S_2$,$\hat{L_3}-L_3$,$\hat{S_3}-S_3$) belong and as derived in Lemmas \ref{lem-1} and \ref{lem-2}, that set is essentially characterized by equations \eqref{eq3}, \eqref{eq4}, \eqref{eq5}, \eqref{rsset}, \eqref{rsset1} and \eqref{rsset2}. Thus, Assumption \ref{ass 1} ensures strong convexity of the loss function over that restricted set.

\vspace{0.1in}
\begin{assumption}
 \label{ass 1} 
 The loss function\\
 $\frac{1}{2T} \sum_{t=1}^{T} \left \lVert \mathcal{Y}_{t} - \operatorname{fold}_1 [ (L_1 + S_1) Y_{\overline{t-1} (1)}]  - \operatorname{fold}_2 [ (L_2 + S_2) Y_{\overline{t-1} (2)}] - \operatorname{fold}_3 [ (L_3 + S_3) Y_{\overline{t-1} (3)}]\right \rVert_{F}^{2}$ \\
satisfies the \textit{Restricted Strong Convexity} condition with with curvature $\gamma > 0$. In other words,  there exists a positive constant $\gamma>0$ such that

\begin{flalign}
    &\frac{1}{2T} \sum_{t=1}^{T} \left\lVert \operatorname{fold}_1 [ ({\Delta}_{L_{1}} + {\Delta}_{S_{1}})  Y_{\overline{t-1} (1)}] + \operatorname{fold}_2 [ ({\Delta}_{L_{2}} + {\Delta}_{S_{2}})  Y_{\overline{t-1} (2)}]+ \operatorname{fold}_3 [ ({\Delta}_{L_{3}} + {\Delta}_{S_{3}})  Y_{\overline{t-1} (3)} ] \right\rVert_{F}^{2} \nonumber \\
    & \geq
    \frac{\gamma}{2} [ \left \lVert \Delta_{L_{1}}+ \Delta_{{S_{1}}}  \right \rVert ^{2}_{F} + \left \lVert \Delta_{{L_{2}}} + \Delta_{{S_{2}}}  \right \rVert ^{2}_{F} + \left \lVert \Delta_{{L_{3}}} + \Delta_{{S_{3}}}  \right \rVert ^{2}_{F}]
    \label{eqa2}
\end{flalign}
for $\Delta_{{L_{1}}}$, $\Delta_{{L_{2}}}$, $\Delta_{{L_{3}}}$, $\Delta_{{S_{1}}}$. $\Delta_{{S_{2}}}$ and $\Delta_{{S_{3}}}$ satisfying \eqref{eq3}, \eqref{eq4}, \eqref{eq5}, \eqref{rsset}, \eqref{rsset1} and \eqref{rsset2}.
\end{assumption}

%================= 2nd Assumption==================
\vspace{0.2in}

\noindent \textbf{Incoherence assumption for identifiability:}\\
\vspace{0.03in}\\
This assumption aims to ensure that the low-rank components $L_1$, $L_2$, $L_3$ and the sparse components $S_1$, $S_2$ and $S_3$ are identifiable. In other words, the low-rank components are incoherent with the sparse components. This assumption is a straightforward application of the `spikiness' restriction on the the low-rank matrix, as introduced in \cite{r1}. As described in \cite{r1}, when all the `mass' of $L_1$ (or, $L_2$ and $L_3$) is distributed equally among its $d_1^2$ (or, $d_2^2$ and $d_3^2$) elements, then $L_1$ (or, $L_2$ and $L_3$) will have `minimal spikiness'. This corresponds to the case when the parameter $\alpha_1$ (similarly, $\alpha_2$ and $\alpha_3$) $\approx$ 1 in \eqref{as2eq} . On the other extreme, when the parameter $\alpha_1 \approx \sqrt{d_1d_1}$ (or, $\alpha_2 \approx \sqrt{d_2d_2}$ and $\alpha_3 \approx \sqrt{d_3d_3}$), then all the mass of $L_1$ (or, of $L_2$ and $L_3$) will be concentrated only on one element and the other elements will be zeros. In this latter case, $L_1$, $L_2$ and $L_3$ will have `maximal spikiness', implying that they will essentially become sparse matrices, which is undesirable. In practice, the values of $\alpha_1$, $\alpha_2$ and $\alpha_3$ are set between the above two extremes. Thus, by controlling the spikiness of the low-rank matrices, the parameters $\alpha_1$, $\alpha_2$ and $\alpha_3$ ensure sufficient `coherence' between the low-rank and sparse components. This incoherence assumption is milder than other incoherence conditions in the existing literature, including those in \cite{liu2019low,zhang2016exact}, which involve the components of singular value decomposition.

\begin{assumption}
    \label{ass 2}
    \begin{equation}
    \label{as2eq}
        \left\lVert L_1 \right\rVert_{\infty} \leq \frac{\alpha_1}{\sqrt{d_1 d_1}}, \hspace{0.3cm} \left\lVert L_2 \right\rVert_{\infty} \leq \frac{\alpha_2}{\sqrt{d_2 d_2}} \text{ and } \left\lVert L_3 \right\rVert_{\infty} \leq \frac{\alpha_3}{\sqrt{d_3 d_3}}
    \end{equation}
    for some fixed parameters $\alpha_1$, $\alpha_2$ and $\alpha_3$.
\end{assumption}

\vspace{0.2in}

%================= 3rd Assumption==================
\noindent \textbf{Lower bound assumptions on the regularization parameters:}\\
\vspace{0.03in}\\
This assumption places certain lower bounds on the regularization parameters, a standard requirement in the high-dimensional literature \citep{basu2015regularized,negahban2012unified,roy2022regularized}. Note that these bounds involve $\{\mathcal E_t\}_{t=1}^T$, which we currently treat as deterministic or non-random. Later, once we introduce distributional assumptions on $\{\mathcal E_t\}_{t=1}^T$, we will derive appropriate choices of $\lambda_{L_1}$, $\lambda_{L_2}$, $\lambda_{L_3}$, $\lambda_{S_1}$, $\lambda_{S_2}$ and $\lambda_{S_3}$ to ensure that these lower bounds hold with high probability.

\begin{assumption}
\label{ass 3}
 When the errors $\mathcal{E}_{t}$ are deterministic or non-random, the regularization parameters, $\lambda_{L_1}, \lambda_{L_2},  \lambda_{L_3},\lambda_{S_1}, \lambda_{S_2}, \text{ and } \lambda_{S_3}$ satisfy the following constraints:

\begin{align}
    & \lambda_{L_1} \geq 4 \left \lVert {D}_1 \right\rVert_{sp}, \hspace{0.5cm}
    \lambda_{S_1} \geq 4 \left \lVert {D}_1 \right\rVert_{\infty} + \frac{4\gamma \alpha_1}{\sqrt{d_1 d_1}} \nonumber \\ \label{eq2}  
    & \lambda_{L_2} \geq 4 \left \lVert {D}_2 \right\rVert_{sp}, \hspace{0.5cm}
    \lambda_{{S_2}} \geq 4 \left \lVert {D}_2 \right\rVert_{\infty} + \frac{4\gamma \alpha_2}{\sqrt{d_2 d_2}} \nonumber \\
    &\lambda_{L_3} \geq 4 \left \lVert {D}_3 \right\rVert_{sp}, \hspace{0.5cm}
    \lambda_{{S_3}} \geq 4 \left \lVert {D}_3 \right\rVert_{\infty} + \frac{4\gamma \alpha_3}{\sqrt{d_3 d_3}}
\end{align}
where ${D}_1 = \frac{1}{T} \sum_{t=1}^{T}  \mathcal{E}_{t(1)} Y_{\overline{t-1} (1)}^{T} $, ${D}_2 =  \frac{1}{T} \sum_{t=1}^{T}\mathcal{E}_{t(2)} Y_{\overline{t-1} (2)}^{T} $, and ${D}_3 =  \frac{1}{T} \sum_{t=1}^{T}\mathcal{E}_{t(3)} Y_{\overline{t-1} (3)}^{T}$, and $\mathcal{E}_{t(1)}$, $\mathcal{E}_{t(2)}$ and $\mathcal{E}_{t(3)}$ are the row-wise, column-wise and tube-wise matricized versions of $\mathcal{E}_t$.
\end{assumption}

%===============================
\vspace{0.2in}
Under the above assumptions, the following lemma establishes an upper bound to the estimation error $e^2(\hat{L}_1, \hat{L}_2, \hat{L}_3, \hat{S}_1, \hat{S}_2, \hat{S}_3)$ in the case of deterministic or non-random errors $\{\mathcal E_t\}_{t=1}^T$.
\begin{lem}
    \label{lem31}
    Suppose the errors $\{\mathcal{E}_t\}_{t=1}^T$ are deterministic. Then, under assumptions \ref{ass 1},\ref{ass 2} and \ref{ass 3} the estimation error satisfies the following condition:
    \begin{align}
\label{debound}
    &  e^2(\hat{L}_1, \hat{L}_2, \hat{S}_1, \hat{S}_2)\preceq  \lambda_{{L}_1}^2 R_1 + \lambda_{{S}_1}^2 s_1 + \lambda_{{L}_2}^2 R_2+  \lambda_{{S}_2}^2 s_2 + \lambda_{{L}_3}^2 R_3 + \lambda_{{S}_3}^2 s_3
    \end{align}
where the notation '$\preceq$' denotes an upper bound, ignoring all constant factors. 
\end{lem}
Ignoring constants, the above upper bound involves the regularization parameters $\lambda_{L_1}$, $\lambda_{L_2}$, $\lambda_{L_3}$, $\lambda_{S_1}$, $\lambda_{S_2}$ and $\lambda_{S_3}$, the ranks of the low-rank components $L_1$, $L_2$, $L_3$, and the numbers of nonzero elements in the sparse components $S_1$, $S_2$ and $S_3$. In the next step, while we introduce the distributional assumptions on $\{\mathcal E_t\}_{t=1}^T$, we need to find suitable choices for the regularization parameters such that the assumed lower bounds in Assumption \ref{ass 3} hold with high probability. Finally, the estimation error bound can be obtained by substituting these probabilistically valid choices of the regularization parameters into the deterministic bound given in \eqref{debound}. As mentioned earlier, following the standard approach in high-dimensional theory, \cite{r1} also first derived estimation error bounds under deterministic (non-random) errors before extending them to the Gaussian case. Although our derivation follows the same standard strategy used in high-dimensional literature and the expression of the deterministic bound looks similar to the one in \cite{r1}, the two contexts are quite different--the former is in the setting of matrix decomposition, whereas our results are established in the context of high-dimensional tensor autoregression. Thus, the intermediate technical components, such as restricted strong convexity conditions and deviation bounds, are considerably more intricate in our case because of the simultaneous row-wise, column-wise and tube-wise temporal dependencies.   

\vspace{0.2in}
We now extend the above result under a Gaussian distribution assumption on the errors. To that end, define $\mathbb{E}_1$ as a data matrix of order $d_1 \times T d_{2} d_3$, constructed by arranging the time series $\{\mathcal E_{t(1)}\}_{t=1}^T$ side by side. Similarly, let $Y_{-1(1)}$ be a data matrix of order $d_1 \times T d_{2}d_3$, formed by arranging the time series $\{Y_{\overline{t-1} (1)}\}_{t=1}^T$ side by side. Similarly, we construct the data matrix $\mathbb{E}_2$ of order $d_2 \times T d_{1} d_3$ by using $\{\mathcal E_{t(2)}\}_{t=1}^T$ and $Y_{-1(2)}$ of order $d_2 \times T d_{1} d_3$ by using $\{Y_{\overline{t-1} (2)}\}_{t=1}^T$. Finally, the data matrices $\mathbb{E}_3$ and $Y_{-1(3)}$ of order $d_3 \times T d_{1} d_2$ are created in a similar fashion. It is easy to verify that the matrices $D_1$, $D_2$ and $D_3$, defined earlier in assumption \ref{ass 3}, can be expressed as 
$D_1 = \frac{1}{T} \mathbb{E}_1 Y_{-1(1)}^T$,
$D_2 = \frac{1}{T} \mathbb{E}_2 Y_{-1(2)}^T$ and 
$D_3 = \frac{1}{T} \mathbb{E}_3 Y_{-1(3)}^T$.

Now, let $\{p_{1t}\}$ be a process characterized by the columns of $\mathbb{E}_1$, which is a centered, stationary, Gaussian process. Similarly, let $\{p_{2t}\}$ be a process characterized by the columns of $Y_{-1(1)}$. It is assumed that, the process $\{p_{2t}\}$ is also a centered, stationary, Gaussian process, and it is obvious that $Cov(p_{1t},p_{2t})=0\: \forall t$. As in \cite{basu2015regularized}, we first define the spectral density corresponding to the process $\{p_{1t}\}$ as $f_{p_1}(\theta)= \frac{1}{2\pi} \sum_{\ell=-\infty}^{\infty}\Gamma_{p_1}(\ell) e^{-i \ell\theta}, \theta\in[-\pi,\pi]$, where $\Gamma_{p_1}(h)=Cov(p_{1t},p_{1\text{ }\overline{t+h}}), \text{ }t,h \in \mathbb{Z}$. We then assume that the above spectral density exists with its maximum eigenvalue being bounded almost everywhere on $[-\pi,\pi]$. In terms of notation, this implies that $\mathscr{M}(f_{p_1})=\underset{\theta \in [-\pi,\pi]}{\text{ess sup}} \Lambda_{\text{max}}(f_{p_1}(\theta))$ $< \infty$, where $\Lambda_{\text{max}}(f_{p_1}(\theta))$ denotes the maximum eigenvalue of the spectral density $f_{p_1}(\theta)$ and $\underset{\theta \in [-\pi,\pi]}{\text{ess sup}}$ denotes the essential supremum of that maximum eigenvalue over $[-\pi,\pi]$. Similarly, we define $\mathscr{M}(f_{p_2})$ corresponding to the process $\{p_{2t}\}$ and assume that $\mathscr{M}(f_{p_2}) < \infty$. Finally, we define the cross spectral density of the two processes $\{p_{1t}\}$ and $\{p_{2t}\}$ as $f_{p_1,p_2}(\theta)= \frac{1}{2\pi} \sum_{\ell=-\infty}^{\infty}\Gamma_{p_1,p_2}(\ell) e^{-i \ell\theta}, \theta\in[-\pi,\pi]$ where $\Gamma_{p_1,p_2}(h)=Cov(p_{1t},p_{2\text{ }\overline{t+h}}), \text{ }t,h \in \mathbb{Z}$. We assume that the above cross spectral density exists and its maximum eigen value is bounded almost everywhere on $[-\pi,\pi]$. In terms of notation, $\mathscr{M}(f_{p_1,p_2})=\underset{\theta \in [-\pi,\pi]}{\text{ess sup }} \sqrt{\Lambda_{\text{max}}(f^{*}_{p_1,p_2}(\theta) f_{p_1,p_2}(\theta))}$ $< \infty$, where $f^{*}_{p_1,p_2}(\theta)$ is the Hermitian conjugate of the cross-spectral density $f_{p_1,p_2}(\theta)$. We then define $Q_1$ as 

\begin{equation}
    Q_1 =  \mathscr{M}(f_{p_1})+ \mathscr{M}(f_{p_2})+\mathscr{M}(f_{p_1,p_2}).
\end{equation}

Similarly, we define $Q_2$ using $\mathbb{E}_2$ and $Y_{-1(2)}$ and define $Q_3$ using $\mathbb{E}_3$ and $Y_{-1(3)}$. 

\vspace{0.2in}

\begin{thm}
    \label{thm:ebound}
    Suppose that $vec(\mathcal{E}_t)$ are i.i.d with MVN$(0, \Sigma)$, where ${\Sigma} = \big[ {\Sigma_1} \otimes {I}_{d_2} \otimes {I}_{d_3} + {I}_{d_1} \otimes {\Sigma_2} \otimes {I}_{d_3} + {I}_{d_1} \otimes {I}_{d_2} \otimes {\Sigma_3} \big]$ with ${\Sigma_1\in \mathbb{R}^{d_1 \times d_1}}$, ${\Sigma_2\in \mathbb{R}^{d_2 \times d_2}}$ and ${\Sigma_3\in \mathbb{R}^{d_3 \times d_3}}$ being symmetric positive semi-definite matrices, and assume that Assumption \ref{ass 2} holds. It can then be shown that the conditions in Assumptions \ref{ass 1} and \ref{ass 3} are satisfied with high probability and we will have the following:
    \begin{eqnarray*}
 e^2(\hat{L}_1, \hat{L}_2, \hat{L}_3, \hat{S}_1, \hat{S}_2, \hat{S}_3) &\leq& s_1\{c_1Q_1^2 \frac{2 \log d_1}{T}+c_2\frac{\gamma^2\alpha_1^2}{d_1^2}\}+ s_2\{c_3 Q_2^2 \frac{2 \log d_2}{T}+c_4\frac{\gamma^2\alpha_2^2}{d_2^2}\} + \\
 &&s_3\{c_5 Q_3^2 \frac{2 \log d_3}{T}+c_6\frac{\gamma^2\alpha_3^2}{d_3^2}\} +
c_7 R_1 Q_1^2 \frac{2 d_1}{T}  + c_8 R_2 Q_2^2 \frac{2 d_2}{T}  + c_9 R_3 Q_3^2 \frac{2 d_3}{T}  
\end{eqnarray*}
with probability $1- max(e^{-c_{10} \log d_1}, e^{-c_{11} \log d_2}, e^{-c_{12} \log d_3} )$ under suitably chosen constants $c_1$ to  $c_{12}$.
\end{thm}
\vspace{0.1in}

The above bound is interpretable as follows. The terms $s_1Q_1^2\frac{2\log d_1}{T}$, $s_2Q_2^2\frac{2\log d_2}{T}$ and $s_3Q_3^2\frac{2\log d_3}{T}$ are in line with the sparse regularized vector autoregressive case \cite{basu2015regularized}. These terms can be interpreted as follows: the term $s_1Q_1^2\frac{2\log d_1}{T}$ arises as a result of estimating $s_1$ non-zero elements of $d_1 \times d_1$ dimensional matrix $S_1$. Note that, there are ${d_1^2 \choose s_1}$ possible subsets of size $s_1$ and thus the numerator includes the corresponding term with the scaling $\log({d_1^2 \choose s_1})\approx s_1 2\log(d_1)$. A similar interpretation follows for the term $s_2Q_2^2\frac{2\log d_2}{T}$ and $s_3Q_3^2\frac{2\log d_3}{T}$. The term $Q_1^2R_1\frac{2d_1}{T}$ contains $R_1\times 2d_1$ that corresponds to the number of free elements in $L_1$. The terms $Q_2^2R_2\frac{2d_2}{T}$ and $Q_3^2R_3\frac{2d_3}{T}$ can be interpreted in a similar fashion. Finally, the terms $\frac{s_1\gamma^2\alpha_1^2}{d_1^2}$, $\frac{s_2\gamma^2\alpha_2^2}{d_2^2}$ and $\frac{s_3\gamma^2\alpha_3^2}{d_3^2}$ appear due to the non-identifiability of the low-rank and sparse components \citep{roy2022regularized}.

\subsection{Estimation error bound under sub-exponential distribution}

Here we extend the above result to the case when the errors $\mathcal{E}_t$ will have $\alpha$-\textit{sub-exponential tail decay}. As defined in \cite{gotze2019concentration}, a random variable $H$ is said to have $\alpha$-sub-exponential tail decay if the following holds: $Pr\{\lvert H \rvert>t\} \leq c_1 \cdot exp(-\frac{t^\alpha}{c_2})$, for some constants $c_1$ and $c_2$, where the parameter $\alpha \in (0,1] \cup \{2\}$. The above definition covers a variety of distribution depending on the chosen value of $\alpha$. Examples include, \textit{sub-Gaussian} distribution, $\textit{sub-Exponential}$ distribution such as Poisson or Weibull random variables and so on (see \cite{gotze2019concentration}). The following theorem provides the expression of the estimation error bound under sub-exponential tail decay assumption on $\{\mathcal E_t\}_{t=1}^T$. The proof of this result is deferred to the Appendix, which required generalizing the deviation bounds from the Gaussian to the sub-Exponential tail decay setting. We have obtained those generalized deviation bounds in Lemmas \ref{lemma5}–\ref{lemma8} in the Appendix.

\begin{thm}
\label{theo_x_rand}  
Suppose each coordinate of $vec(\mathcal E_t)$ is distributed as $\alpha$-sub-exponential tail decay, and also assume that Assumption \ref{ass 2} holds. Then it can be shown that conditions in Assumptions \ref{ass 1} and \ref{ass 3} are satisfied with high probability and we will have
\begin{eqnarray*}
 e^2(\hat{L}_1, \hat{L}_2, \hat{L}_3, \hat{S}_1, \hat{S}_2, \hat{S}_3) &\leq& s_1\{c_1Q_1^2\frac{\{2\log d_1\}^{\frac{2}{\alpha}}}{T}+c_2\frac{\gamma^2\alpha_1^2}{d_1^2}\}+ s_2\{c_3Q_2^2\frac{\{2\log d_2\}^{\frac{2}{\alpha}}}{T}+c_4\frac{\gamma^2\alpha_2^2}{d_2^2}\}+\\
 &&s_3\{c_5Q_3^2\frac{\{2\log d_3\}^{\frac{2}{\alpha}}}{T}+c_6\frac{\gamma^2\alpha_3^2}{d_3^2}\}+c_7 Q_1^2 R_1 \frac{\{2d_1\}^{\frac{2}{\alpha}}}{T} + c_8 Q_2^2 R_2 \frac{\{2d_2\}^{\frac{2}{\alpha}}}{T}\\
 &&+c_9 Q_3^2 R_3 \frac{\{2d_3\}^{\frac{2}{\alpha}}}{T}.  
\end{eqnarray*}
with probability $1- max(e^{-c_{10} \{log(d_1)\}^{\frac{2}{\alpha}}}, e^{-c_{11} \{log(d_2)\}^{\frac{2}{\alpha}}}, e^{-c_{12} \{log(d_3)\}^{\frac{2}{\alpha}}})$ for some suitably chosen constants $c_1$ to $c_{12}$.
\end{thm}
Note that, the case $\alpha=2$ corresponds to the Gaussian distribution, which is a special case of $\alpha$-sub-exponential tail decay family. Thus, for $\alpha=2$, estimation error bound in Theorem \ref{theo_x_rand} boils down to the bound obtained in Theorem \ref{thm:ebound}.

\section{Simulation studies}
\label{simu}
In this section, we evaluate the performance of our proposed method based on synthetic data under different settings. We first assess estimation quality of our model in Section \ref{Sim:est}. In Section \ref{Sim:pred}, we evaluate predictive performance of our model.           
\noindent
\subsection{Estimation quality} \label{Sim:est}
\noindent\textit{Data generating process:} We begin by describing the procedure for generating the true low-rank components ${L_1}$, ${L_2}$, ${L_3}$ and the true sparse components ${S_1}$, ${S_2}$, ${S_3}$ of our model. To generate ${L_1} \in \mathbb{R}^{d_1 \times d_1}$ with rank $R_1$, we first start with a matrix in $\mathbb{R}^{d_1 \times d_1}$ with entries from Uniform (0,1), and then obtain its singular value decomposition (SVD). We then randomly select $(d_1-R_1)$ diagonal elements of the diagonal matrix $D$ of the above-mentioned SVD, change those elements to zeros while the others remain non-zeros, and name the resulting matrix as ${D_1}$. Finally, the matrix ${L_1}$ with rank $R_1$ can be generated as ${U} {D_1} {V}^T$, where ${U}$ and ${V}$ are the matrices with orthonormal columns from the aforementioned SVD. The matrices ${L_2} \in \mathbb{R}^{d_2 \times d_2}$ with rank $R_2$ and ${L_3} \in \mathbb{R}^{d_3 \times d_3}$ with rank $R_3$ can be generated in a similar fashion.    

To generate the sparse components, we first start with a matrix with all its elements as zeros, then randomly select a small proportion of the elements and replace those zeros with entries from Uniform distribution, whose range is governed by a pre-specified maximum eigenvalue that controls the spectral properties of the matrix. Then the signs of those non-zero elements are decided by tossing a fair coin. The above-mentioned proportion of non-zero elements in the sparse components is referred to as edge-density. Finally, to ensure the stationarity of the generated matrix, we check its maximum absolute eigen value, and if the same is higher than the above-mentioned pre-specified value, we scale down the entries of the matrix in such a way that the condition is satisfied.

Given the true low-rank and sparse transition matrices, we generate the error tensors $\mathcal{E}_t$ in $\mathbb{R}^{d_1 \times d_2 \times d_3}$, where, as mentioned earlier in Sections \ref{model} and \ref{theo}, $vec(\mathcal{E}_t)$'s are drawn independently and identically from a Multivariate Normal distribution with mean zero and covariance matrix ${\Sigma}$, where ${\Sigma} = \big[ {\Sigma_1} \otimes {I}_{d_2} \otimes {I}_{d_3} + {I}_{d_1} \otimes {\Sigma_2} \otimes {I}_{d_3} + {I}_{d_1} \otimes {I}_{d_2} \otimes {\Sigma_3} \big]$ and ${\Sigma_1\in \mathbb{R}^{d_1 \times d_1}}$, ${\Sigma_2\in \mathbb{R}^{d_2 \times d_2}}$ and ${\Sigma_3\in \mathbb{R}^{d_3 \times d_3}}$ are symmetric positive semi-definite matrices. Finally, the data tensors ${\mathcal{Y}_{t} \in \mathbb{R}^{d_1 \times d_2 \times d_3}}$ are generated recursively as $\mathcal{Y}_{t} = \operatorname{fold}_1 [(L_1 + S_1) Y_{\overline{t-1} (1)}] + \operatorname{fold}_2 [ (L_2 + S_2) Y_{\overline{t-1} (2)}] + \operatorname{fold}_3 [ (L_3 + S_3)Y_{\overline{t-1} (3)}] + \mathcal{E}_{t}$. We then employ our proposed algorithm in Section \ref{model} on this simulated data to estimate the model parameters. The regularization parameters $\lambda_{{L_1}}$, $\lambda_{{L_2}}$, $\lambda_{{L_3}}$, $\lambda_{{S_1}}$, $\lambda_{{S_2}}$ and $\lambda_{{S_3}}$ are selected using a grid search method. More specifically, we run the algorithm and obtain estimates of ${L_1}$, ${L_2}$, ${L_3}$, $S_1$, $S_2$ and $S_3$ for different grids of ($\lambda_{{L_1}}$, $\lambda_{{L_2}}$, $\lambda_{{L_3}}$, $\lambda_{{S_1}}$, $\lambda_{{S_2}}$, $\lambda_{{S_3}}$) and select the one for which the ranks of the estimated low-rank components are as close as possible to the ranks of the true ${L_1}$, ${L_2}$ and ${L_3}$, that is ${R_1}$, ${R_2}$ and ${R_3}$ respectively, and also the numbers and positions of the zeros and non-zero elements in the estimated sparse components are as close as possible to the same in the true sparse components. Later in this section, we develop an AIC criteria, which facilitates selection of the optimum values of the regularization parameters when the true ranks and sparsity levels are unknown to us.\\

\noindent \textit{Evaluation criteria}: We primarily use the notion of Relative Error (RE) to evaluate the estimation quality of our proposed method, which is defined as 

$$\frac{\left \lVert {\hat{L}_{1}} - {L_1} \right \rVert_F^{2} + \left \lVert {\hat{L}_{2}} - {L_2} \right \rVert_F^{2}+ \left \lVert {\hat{L}_{3}} - {L_3} \right \rVert_F^{2}+\left \lVert {\hat{S}_{1}} - {S_1} \right \rVert_F^{2} + \left \lVert {\hat{S}_{2}} - {S_2} \right \rVert_F^{2}+ \left \lVert {\hat{S}_{3}} - {S_3} \right \rVert_F^{2}} { \left \lVert{L_1} \right\rVert_F^{2} +  \left \lVert{L_2} \right\rVert_F^{2} + \left \lVert{L_3} \right\rVert_F^{2} +\left \lVert{S_1} \right\rVert_F^{2} +  \left \lVert{S_2} \right\rVert_F^{2}+\left \lVert{S_3} \right\rVert_F^{2}}.$$

The quality of the estimation is indicated by low values of the above relative error. Additionally, we also assess the similarity in rank between estimated transition matrices ${\hat{L}_{1}}$, ${\hat{L}_{2}}$, ${\hat{L}_{3}}$ and the true parameters ${L_1}$, ${L_2}$, ${L_3}$. Alongside that, the measures sensitivity and specificity help to assess the effectiveness of support recovery for the estimation of the sparse components $S_1$, $S_2$ and $S_2$, which are defined as follows
\begin{enumerate}
    \item Specificity for $\hat{S_1}$, denoted by $SP_{S_1}$, is defined as the proportion of true negatives, or alternatively, 1 - False Positive Rate (FPR), where, FPR is defined as
    $$
    \frac{\text{Total number of non-zero elements in} \hspace{0.1 cm} {\hat{S}_1} \hspace{0.1 cm} \text{that are actually zero in} \hspace{0.1 cm} {S_1}   }{\text{Total number of elements in} \hspace{0.1 cm} {S_1} \hspace{0.1 cm} \text{that are actually zero}}
    $$
    \item Sensitivity for $\hat{S_1}$, denoted by $SN_{S_1}$, is defined as the True Positive Rate (TPR) as follows 
    $$
    \frac{ \text{Total number of non-zero elements in} \hspace{0.1cm} {S_1} \hspace{0.1cm} \text{that are correctly classified as non-zero in} \hspace{0.1cm} {\hat{S_1}}  }{\text{Total number of elements in} \hspace{0.1 cm} {S_1} \hspace{0.1 cm} \text{that are actually non-zero}}
    $$
\end{enumerate}
$SP_{S_2}$, $SN_{S_2}$, $SP_{S_3}$ and $SN_{S_3}$ are defined in a similar way. Higher values of specificity and sensitivity, that is, values either close to 1 or exactly 1, are preferable.\\
%============================= 1st simulation table=============
\begin{table}[h]
\caption{Performance Evaluation under setup 1: $d_1 = 15$, $d_2 = 10$, $d_3 = 10$. Relative error, ranks of the estimated low-rank components and sensitivity and specificity of the estimated sparse components are reported for all four sub-cases. As the number of time points increases, estimation quality improves. Also, for any fixed time point, when the true edge densities $e_1$, $e_2$ and $e_3$  increase (keeping the ranks fixed) or the true ranks $R_1$, $R_2$ and $R_3$ increase (keeping the edge densities fixed), the relative error increases, which is in line with our theoretical finding.}\label{tab: eg1}
%===========sub-case1====
%===========sub-case1====
\resizebox{\textwidth}{!}{%
\begin{tabular}{c|cccccccccc}
\multicolumn{1}{c}{} &
\multicolumn{10}{c}{Sub-case 1: $e_1=0.2, e_2=0.2, e_3=0.2, R_1=3, R_2=3, R_3=3$}\\
\cmidrule(lr){2-11}
Time Points
& RE
& $\hat{R}_1$
& $\hat{R}_2$
& $\hat{R}_3$
& $SN_{S_1}$
& $SP_{S_1}$
& $SN_{S_2}$
& $SP_{S_2}$
& $SN_{S_3}$
& $SP_{S_3}$\\
\hline
100 & 0.09 & 3 & 3 & 3 & 0.94 & 0.84 & 1 & 0.9 &0.95& 0.93\\
200 & 0.06 & 3 & 3 & 3 & 0.97 & 0.96 & 0.97 & 0.98 & 0.96&0.95\\
300 & 0.06 & 3 & 3 & 3 & 0.99 & 0.98 & 0.97 & 1 &0.98&0.98\\
\hline
\end{tabular}
}

\bigskip

%==========sub-case 2=============
\resizebox{\textwidth}{!}{%
\begin{tabular}{c|cccccccccc}
\multicolumn{1}{c}{} & \multicolumn{10}{c}{$\mbox{Sub-case 2: } e_1 = 0.4, e_2 = 0.4, e_3=0.4, R_1=3, R_2=3, R_3=3$} \\ 
\cmidrule(lr){2-11}
Time Points 
& RE 
& $\hat{R}_1$
& $\hat{R}_2$
& $\hat{R}_3$
& $SN_{S_1}$
& $SP_{S_1}$
& $SN_{S_2}$
& $SP_{S_2}$
& $SN_{S_3}$
& $SP_{S_3}$\\
\hline
100 & 0.17 & 4 & 3 &4 & 0.85 & 0.81 & 0.91 & 0.85& 0.9 &0.88\\
200 & 0.13 & 3 & 3&3 & 0.89 & 0.93 & 0.95 & 0.98& 0.93&0.94\\
300 & 0.09 & 3 & 3&3 & 0.92 & 0.94 & 0.96 & 1& 0.95&0.98\\
\hline
\end{tabular}
}

\bigskip

%===========Sub-case 3====================
\resizebox{\textwidth}{!}{%
\begin{tabular}{c|cccccccccc}
\multicolumn{1}{c}{} & \multicolumn{10}{c}{$\mbox{Sub-case 3: } e_1 = 0.2, e_2 = 0.2, e_3=0.2, R_1=5, R_2=5, R_3=5$} \\ 
\cmidrule(lr){2-11}
Time Points 
& RE 
& $\hat{R}_1$
& $\hat{R}_2$
& $\hat{R}_3$
& $SN_{S_1}$
& $SP_{S_1}$
& $SN_{S_2}$
& $SP_{S_2}$
& $SN_{S_3}$
& $SP_{S_3}$\\
\hline
100 & 0.21 & 5 & 5 &5 &0.94 & 0.85 & 1 & 0.92 &0.94&0.92 \\
200 & 0.16 & 5 & 5 &5 &0.96 & 0.94 & 0.97 & 0.98&0.95&0.96 \\
300 & 0.11 & 5 & 5 &5 &0.98 & 0.96 & 0.95 & 0.99&0.98&0.97 \\
\hline
\end{tabular}
}

\bigskip

%=========Sub-case 4===================
\resizebox{\textwidth}{!}{%
\begin{tabular}{c|cccccccccc}
\multicolumn{1}{c}{} & \multicolumn{10}{c}{$\mbox{Sub-case 4: } e_1 = 0.4, e_2 = 0.4, e_3 = 0.4, R_1=5, R_2=5, R_3=5$} \\ 
\cmidrule(lr){2-11}
Time Points 
& RE 
& $\hat{R}_1$
& $\hat{R}_2$
& $\hat{R}_3$
& $SN_{S_1}$
& $SP_{S_1}$
& $SN_{S_2}$
& $SP_{S_2}$
& $SN_{S_3}$
& $SP_{S_3}$\\
\hline
100 & 0.23 & 6 & 5 &6 & 0.83 & 0.82 & 0.90 & 0.86 & 0.90&0.89 \\
200 & 0.18 & 5 & 5& 5 & 0.85 & 0.91 & 0.92 & 0.97 &0.91 & 0.9\\
300 & 0.15 & 5 & 5& 5 & 0.92 & 0.93 & 0.95 & 0.99 & 0.95& 0.95\\
\hline
\end{tabular}
}

\end{table}

\noindent \textit{Numerical Results}: We now assess the performance of our model using the above-mentioned metrics under different setup. Each setup here corresponds to a specific combination of the triplet ($d_1$, $d_2$,$d_3$). Additionally, under each setup we have different sub-cases denoting the varying levels of sparsity and different true rank values as discussed next.\\
\begin{itemize}
    \item Setup 1: $d_1$ = 15, $d_2$ = 10, $d_3$ = 10;\\ 
    Setup 2: $d_1$ = 30, $d_2$ = 20, $d_3$ = 20.\\ 
    \vspace{0.1in}
    \item Sub-case 1:  $e_1$ = 0.2, $e_2$ = 0.2, $e_3$ = 0.2, $R_1$ = 3, $R_2$ = 3, $R_3$ = 3;\\
    Sub-case 2:  $e_1$ = 0.4, $e_2$ = 0.4, $e_3$ = 0.4, $R_1$ = 3, $R_2$ = 3, $R_3$ = 3;\\
    Sub-case 3:  $e_1$ = 0.2, $e_2$ = 0.2, $e_3$ = 0.2, $R_1$ = 5, $R_2$ = 5, $R_3$ = 5;\\
    Sub-case 4:  $e_1$ = 0.4, $e_2$ = 0.4, $e_3$ = 0.4, $R_1$ = 5, $R_2$ = 5, $R_3$ = 5, where $e_1$, $e_2$, $e_3$ are the edge densities of $S_1$, $S_2$ and $S_3$ respectively and $R_1$, $R_2$ and $R_3$, as defined earlier, are the ranks of $L_1$, $L_2$ and $L_3$ respectively.  
\end{itemize}
%=============================2nd simulation table=================
\begin{table}[h]
\caption{Performance Evaluation under setup 2: $d_1 = 30$, $d_2 = 20$, $d_3 = 20$. Relative error, ranks of the estimated low-rank components and sensitivity and specificity of the estimated sparse components are reported for all four sub-cases. As the number of time points increases, estimation quality improves. Also, for any fixed time point, when the true edge densities $e_1$, $e_2$ and $e_3$ increase (keeping the ranks fixed) or the true ranks $R_1$, $R_2$ and $R_3$ increase (keeping the edge densities fixed), the relative error increases, which is in line with our theoretical finding.}\label{tab: eg2}
%===========sub-case1====
\resizebox{\textwidth}{!}{%
\begin{tabular}{c|cccccccccc}
\multicolumn{1}{c}{} &
\multicolumn{10}{c}{Sub-case 1: $e_1=0.2, e_2=0.2, e_3=0.2, R_1=3, R_2=3, R_3=3$}\\
\cmidrule(lr){2-11}
Time Points
& RE
& $\hat{R}_1$
& $\hat{R}_2$
& $\hat{R}_3$
& $SN_{S_1}$
& $SP_{S_1}$
& $SN_{S_2}$
& $SP_{S_2}$
& $SN_{S_3}$
& $SP_{S_3}$\\
\hline
100 & 0.11 & 3 & 3 & 3 & 0.93 & 0.82 & 1 & 0.89 &0.93& 0.91\\
200 & 0.09 & 3 & 3 & 3 & 0.96 & 0.95 & 0.95 & 0.98 & 0.95&0.94\\
300 & 0.08 & 3 & 3 & 3 & 0.98 & 0.97 & 0.95 & 1 &0.98&0.97\\
\hline
\end{tabular}
}

\bigskip

%==========sub-case 2=============
\resizebox{\textwidth}{!}{%
\begin{tabular}{c|cccccccccc}
\multicolumn{1}{c}{} & \multicolumn{10}{c}{$\mbox{Sub-case 2: } e_1 = 0.4, e_2 = 0.4, e_3=0.4, R_1=3, R_2=3, R_3=3$} \\ 
\cmidrule(lr){2-11}
Time Points 
& RE 
& $\hat{R}_1$
& $\hat{R}_2$
& $\hat{R}_3$
& $SN_{S_1}$
& $SP_{S_1}$
& $SN_{S_2}$
& $SP_{S_2}$
& $SN_{S_3}$
& $SP_{S_3}$\\
\hline
100 & 0.19 & 4 & 3 &4 & 0.82 & 0.78 & 0.88 & 0.83& 0.87 &0.85\\
200 & 0.14 & 3 & 3&3 & 0.83 & 0.92 & 0.93 & 0.98& 0.9&0.91\\
300 & 0.12 & 3 & 3&3 & 0.90 & 0.90 & 0.93 & 1& 0.93&0.97\\
\hline
\end{tabular}
}

\bigskip

%===========Sub-case 3====================
\resizebox{\textwidth}{!}{%
\begin{tabular}{c|cccccccccc}
\multicolumn{1}{c}{} & \multicolumn{10}{c}{$\mbox{Sub-case 3: } e_1 = 0.2, e_2 = 0.2, e_3=0.2, R_1=5, R_2=5, R_3=5$} \\ 
\cmidrule(lr){2-11}
Time Points 
& RE 
& $\hat{R}_1$
& $\hat{R}_2$
& $\hat{R}_3$
& $SN_{S_1}$
& $SP_{S_1}$
& $SN_{S_2}$
& $SP_{S_2}$
& $SN_{S_3}$
& $SP_{S_3}$\\
\hline
100 & 0.24 & 5 & 5 &5 &0.92 & 0.81 & 1 & 0.89 &0.92&0.90 \\
200 & 0.18 & 5 & 5 &5 &0.95 & 0.93 & 0.95 & 0.98&0.93&0.92 \\
300 & 0.14 & 5 & 5 &5 &0.98 & 0.96 & 0.95 & 0.99&0.98&0.97 \\
\hline
\end{tabular}
}

\bigskip

%=========Sub-case 4===================
\resizebox{\textwidth}{!}{%
\begin{tabular}{c|cccccccccc}
\multicolumn{1}{c}{} & \multicolumn{10}{c}{$\mbox{Sub-case 4: } e_1 = 0.4, e_2 = 0.4, e_3 = 0.4, R_1=5, R_2=5, R_3=5$} \\ 
\cmidrule(lr){2-11}
Time Points 
& RE 
& $\hat{R}_1$
& $\hat{R}_2$
& $\hat{R}_3$
& $SN_{S_1}$
& $SP_{S_1}$
& $SN_{S_2}$
& $SP_{S_2}$
& $SN_{S_3}$
& $SP_{S_3}$\\
\hline
100 & 0.28 & 6 & 5 &6 & 0.81 & 0.78 & 0.88 & 0.83 & 0.87&0.85 \\
200 & 0.24 & 5 & 5& 5 & 0.83 & 0.91 & 0.92 & 0.97 &0.9 & 0.9\\
300 & 0.22 & 5 & 5& 5 & 0.90 & 0.90 & 0.93 & 0.99 & 0.93& 0.95\\
\hline
\end{tabular}
}
\end{table}

Performance evaluation results are summarized in tables \ref{tab: eg1} and \ref{tab: eg2} under aforementioned setups and sub-cases. It is evident from both tables that as the number of time points increases, the relative error decreases. Alongside that, we also see that better support recovery, that is higher sensitivity and specificity, is achieved with higher values of $T$. It is obvious that, relative errors are in general slightly better in Table \ref{tab: eg1} as compared to Table \ref{tab: eg2} as the setup in Table \ref{tab: eg2} has higher burden in terms of parameters. Thus, slightly higher values of $T$ would make the estimation quality in Table \ref{tab: eg2} as good as in Table \ref{tab: eg1}. Finally, it is worth noting that for any fixed setup, say in Table \ref{tab: eg1}, when edge density is increased from 0.2 to 0.4, there is an increase in the relative error. Similarly, for any fixed setup, say in Table \ref{tab: eg1}, when true ranks $R_1$, $R_2$ and $R_3$ are increased, relative error also increases. Similar pattern is observed in Table \ref{tab: eg2} as well. This finding is consistent with the expression of the estimation error bound obtained in Theorems \ref{thm:ebound} and \ref{theo_x_rand}.

\noindent
\subsection{Predictive performance}\label{Sim:pred}
\noindent We now assess the predictive performance of our model and compare it against the Tucker-based TAR model \citep{li2021multi} and the sparse vector autoregressive model \citep{basu2015regularized}. The Tucker-based TAR model, as mentioned earlier in Section \ref{intro}, uses a multiplicative interaction of row-wise, column-wise and tube-wise temporal dependence. On the other hand, to apply the sparse VAR model to our matrix-variate time series, we simply vectorize the tensor data, and apply sparsity regularization on that vector. We first fix a forecast horizon `h'. Then, for each $t^\prime \in \{ T-10, T-9 \dots , T-h\}$, we use all the data up to time point $t^\prime$ to estimate the model parameters, and finally we use that model to predict the value of $\mathcal Y_{t^\prime+h}$, which is denoted by  $\hat{\mathcal Y}_{t^\prime+h}$. Then, for that forecast horizon `h', the Root Mean Squared Error (RMSE) is defined as $ \sqrt{ \frac{1}{10-h+1} \sum_{t'=T-10}^{T-h} \frac{\left \lVert \mathcal Y_{t'+h}- \hat{\mathcal Y}_{t'+h}  \right \rVert_{F}^{2} }{d_1 d_2 d_3} }$, as in \cite{ghosh2018high} and \cite{chakraborty2023bayesian}. To examine the predictive performance of our model, we use a simulated data with $d_1 = 10, d_2 = 15$, $d_3=15$ and $T=80$. The true ranks of ${L_1}$, ${L_2}$ and ${L_3}$ are taken as 3, 3 and 4 respectively, while the true edge densities of ${S_1}$, ${S_2}$ and ${S_3}$ are taken as 0.5, 0.3 and 0.3 respectively. We consider forecast horizon values $h=1,2,3$ and compare the RMSE values of our model with that of the Tucker-based TAR and the sparse vector autoregressive model. As summarized in Table \ref{table-pred perf-simu}, RMSE values for our model are lower than both the Tucker-based TAR and the sparse VAR model, demonstrating better predictive performance of our model. As expected, the sparse VAR model exhibits poor predictive performance due to its naive vectorization of the tensor-variate time series, which disregards the inherent row-column-tube interactions within the data. While the Tucker-based TAR model performs reasonably well in forecasting, the proposed additive TAR consistently outperforms it across all forecasting horizons, highlighting its superior predictive ability alongside other strengths of this model discussed earlier. 

\begin{table}[h]
\caption{Predictive performance using RMSE values. The proposed additive TAR model performs better than the competing Tucker-based TAR model and the sparse VAR model.}\label{table-pred perf-simu}%
\centering
\begin{tabular}{@{}llll@{}}
\toprule
Forecast horizon (h) & Additive TAR & Tucker-based TAR & Sparse VAR\\
\midrule
 1   & 0.521   & 0.538   & 1.020   \\
 2   & 0.525   & 0.536   & 0.767   \\
 3   & 0.528   & 0.534   & 0.767 \\
\bottomrule
\end{tabular}
\end{table}
%\vspace{0.2in}
\noindent\textbf{AIC Criteria}\\

\noindent As mentioned earlier in this section, while working with real data, the true rank and the true sparsity levels are unknown. In such situations, we choose the values of $\lambda_{L_1}$, $\lambda_{S_1}$, $\lambda_{L_2}$, $\lambda_{S_2}$, $\lambda_{L_3}$ and $\lambda_{S_3}$ in such a way that the AIC, as defined below, is minimized.

$$
AIC = T\log\left(\frac{RSS}{T}\right)+ 2 \text{ Rank }(\hat{L}_1)+ 2 \text{ Rank }(\hat{L}_2) + 2 \text{ Rank }(\hat{L}_3)+ 2k_1 + 2k_2 +2k_3
$$
where RSS, the residual sum of square, is defined as $\frac{1}{2T}\sum_{t=1}^{T} \Bigl\lVert  \mathcal{Y}_{t}
- \operatorname{fold}_1\Big[(\hat{L}_{1}+\hat{S}_{1})Y_{\overline{t-1}(1)}\Big]
- \operatorname{fold}_2\Big[(\hat{L}_{2}+\hat{S}_{2})Y_{\overline{t-1}(2)}\Big]
- \operatorname{fold}_3\Big[(\hat{L}_{3}+\hat{S}_{3})Y_{\overline{t-1}(3)}\Big]
\Bigr\rVert_{F}^{2} \nonumber$, and $k_1$, $k_2$ and $k_3$ are the number of non-zero elements in $\hat{S}_1$, $\hat{S}_2$ and $\hat{S}_3$ respectively. This formulation is quite common in the literature, which essentially rewards goodness of fit, and at the same time it 
penalizes overfitting.

\section{Applications in origin-destination demand for NYC Yellow taxi}
\label{real_data}
In this section, we apply our proposed additive TAR model to the New York City Taxi and Limousine Commission (TLC) Yellow Taxi Trip Record dataset, one of the largest publicly available repositories of urban mobility data \citep{jiang2024entropy, xie2021revealing, hochmair2016spatiotemporal}. It contains detailed trip-level information, including pickup and drop-off locations and timestamps, enabling the construction of high-resolution origin-destination (OD) demand series. An OD demand series records the number of trips between each origin and destination pair over successive time intervals. Compared with aggregate regional demand, OD demand captures the complete travel flow between spatial locations, thereby providing richer information for operational decision-making and resource allocation. In particular, accurate modeling and forecasting of OD demand support fleet repositioning, vehicle dispatching, ride matching, congestion management, dynamic pricing, and infrastructure planning \citep{liu2019contextualized,li2021multi}.  

\begin{figure}
    \centering
    \includegraphics[scale=0.7]{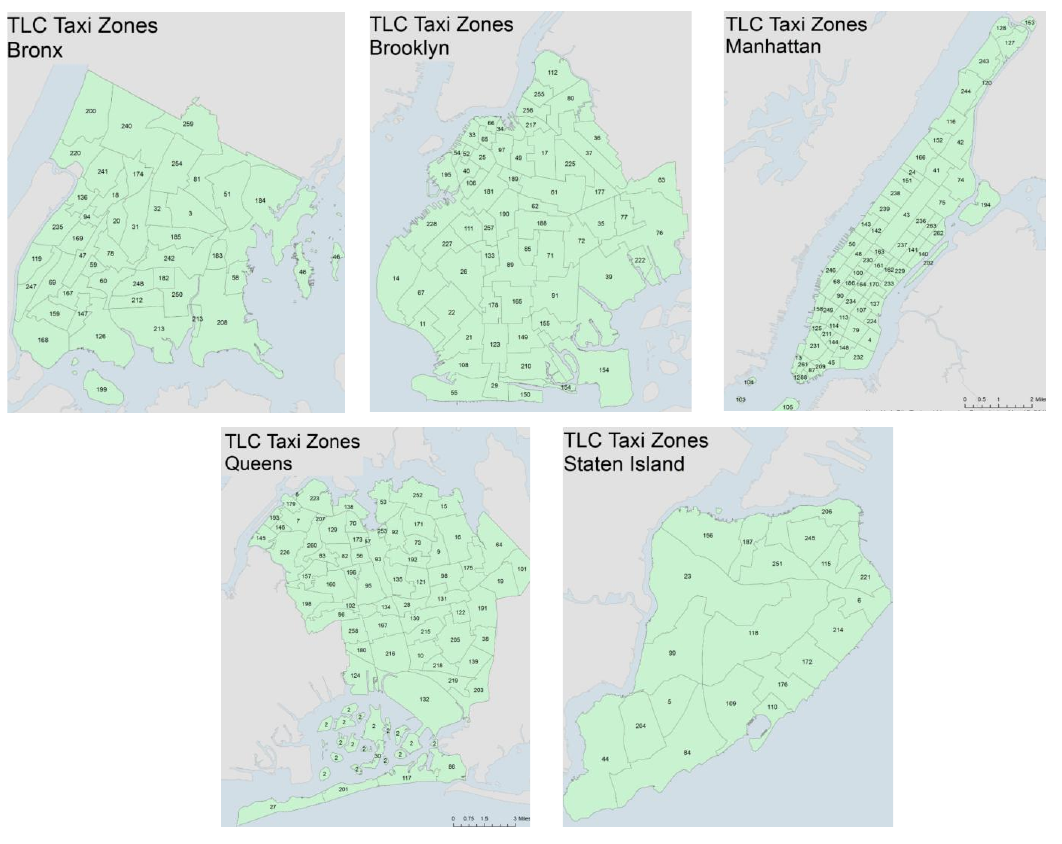}
    \caption{Distribution of trip location IDs across the boroughs Bronx, Brooklyn, Manhattan, Queens and Staten Islands. There is only one trip location ID in EWR borough.}
    \label{fig:zonemap}
\end{figure}

We first aggregate raw trip-level records and construct a sequence of monthly three-dimensional tensors as discussed next. The TLC dataset contains approximately 265 location identifiers distributed over six boroughs of New York City, namely Manhattan, Brooklyn, Queens, Bronx, Staten Island and Newark Airport (EWR) (see Figure \ref{fig:zonemap}). Each trip record includes the pickup and drop-off location identifiers together with the trip start time. To obtain a compact and meaningful representation of taxi demand, the pickup and drop-off location IDs are mapped to their corresponding boroughs. Furthermore, trip start times are classified into five operational time periods reflecting typical traffic conditions: morning peak (6:00 AM--10:59 AM), midday (11:00 AM--4:59 PM), evening peak (5:00 PM--8:59 PM), night (9:00 PM--11:59 PM), and late night (12:00 AM--5:59 AM). Using these aggregated spatial and temporal attributes, we construct a three-dimensional tensor $\mathcal{Y}_t \in \mathbb{R}^{6 \times 6 \times 5}$ for each month $t$, where the first and second dimensions correspond to the origin and destination boroughs, respectively, and the third dimension represents the five trip-start time categories. In particular, the $(i,j,k)^{th}$ entry of $\mathcal{Y}_t$ denotes the total number of taxi trips originated from the $i^{th}$ borough, terminated in the $j^{th}$ borough, and initiated during the $k^{th}$ time category in month $t$. We consider data from 110 months in total from January 2017 to February 2026, constituting the tensor time series $\{\mathcal{Y}_t\}_{t=1}^{T=110}$.

\begin{figure}
    \centering
    \includegraphics[scale=0.9]{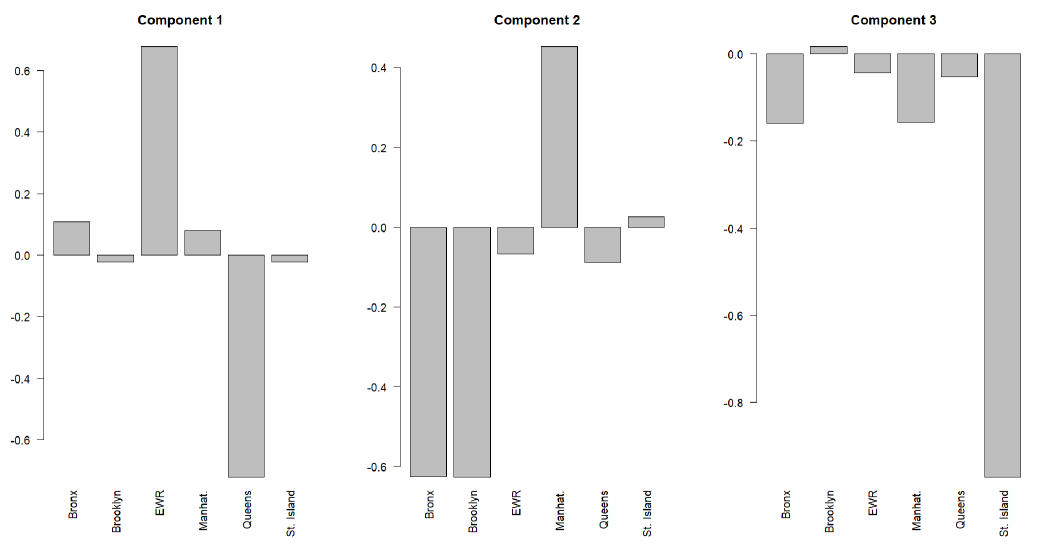}
    \caption{Latent factor structure of $\hat{L}_1$ (rank 3) using a varimax rotation on its leading three singular vectors. The factors appear to be specialized trip generation (airport-related), structural contrasts between
commercial and residential regions, and peripheral isolation (Staten Island).}
    \label{fig:l1}
\end{figure}

We run our proposed algorithm on the above data which estimates three low-rank baseline or `shared' temporal dependence in taxi-demands along the dimension of origin boroughs, destination boroughs, and trip start-time categories, denoted by $\hat{L}_1$, $\hat{L}_2$ and $\hat{L}_3$, respectively. In addition to these baseline temporal connections, additional idiosyncratic temporal connections are captured by the sparse matrices $\hat{S}_1$, $\hat{S}_2$ and $\hat{S}_3$. The estimated matrix $\hat{L}_1 \in \mathbb{R}^{6 \times 6}$ is of rank 3. This low-rank component captures the shared (baseline) temporal effect of past demand across the six origin boroughs on current demand across the same origin boroughs. To interpret this baseline effect, we perform a singular value decomposition of $\hat{L}_1$, extract its three leading singular vectors, and apply a varimax rotation to obtain an interpretable factor structure. As depicted in Figure \ref{fig:l1}, the first factor reflects a contrast between airport-originated trips, centered on EWR, and urban origins such as Queens. This indicates that trip generation from airport locations follows a distinct temporal pattern compared to regular intra-city origins. The second factor captures a core origin structure contrasting Manhattan with outer boroughs such as Bronx and Brooklyn. This can be attributed to the differences in the functional roles of these regions, with Manhattan exhibiting activity-driven trip generation associated with commercial and business centers, while Bronx and Brooklyn primarily contribute residential-origin flows. The third factor isolates Staten Island as trip origin borough, reflecting its geographical separation and limited integration with the broader network. Overall, these findings suggest that the baseline origin-side dynamics may reflect a combination of specialized trip generation (airport-related), structural contrasts between commercial and residential regions, and peripheral isolation.         

\begin{figure}
    \centering
    \includegraphics[scale=0.9]{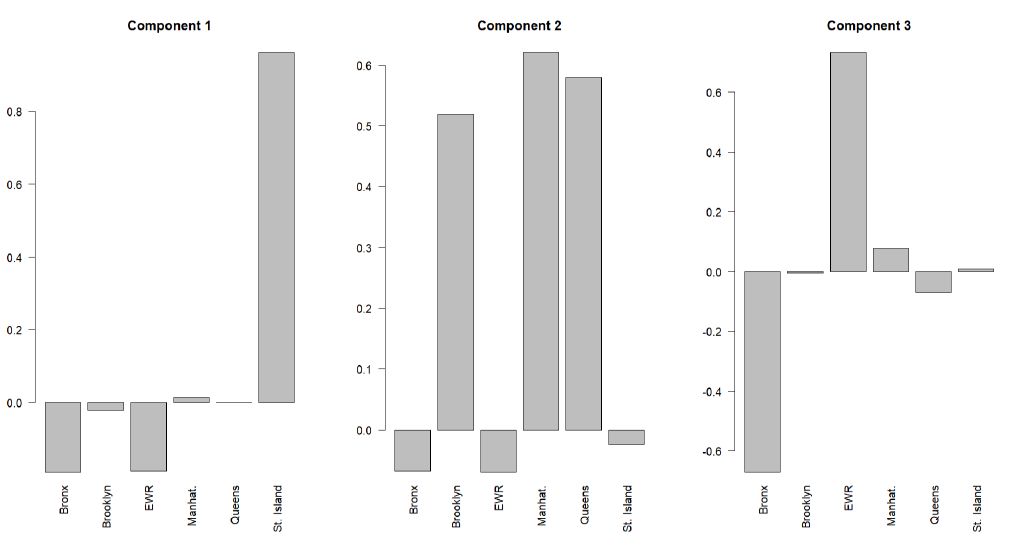}
    \caption{Latent factor structure of $\hat{L}_2$ (rank 3) using a varimax rotation on its leading three singular vectors. The factors appear to be peripheral isolation (Staten Island), core urban demand concentration, and specialized transport flows (airport-related).}
    \label{fig:l2}
\end{figure}

The estimated matrix $\hat{L}_2 \in \mathbb{R}^{6 \times 6}$ is also of rank 3, capturing the shared (baseline) temporal dependence of the number of trips terminating in the six destination boroughs at the current time point on the number of trips terminating in those destination boroughs at the previous time point. As before, Figure \ref{fig:l2} depicts the interpretable factor structure after varimax rotation on the singular values of $\hat{L}_2$. The first factor loads almost exclusively on Staten Island, reflecting its geographical isolation and weak integration with the broader system. This suggests that destination dynamics for Staten Island evolve largely independently of other regions. The second factor captures a core urban destination structure, with strong loadings on Manhattan, Brooklyn, and Queens, reflecting a high-demand regime driven by dense economic activity, commuting flows, and mixed residential–commercial usage. The third factor shows a contrast between airport-related flows, centered on EWR, and more localized destinations such as the Bronx. This highlights the distinct nature of airport-bound travel relative to routine intra-city movements. These findings suggest that the baseline destination-side dynamics may be attributed to combination of peripheral isolation, core urban demand concentration, and specialized transport flows.

\begin{figure}
    \centering
    \includegraphics[scale=0.9]{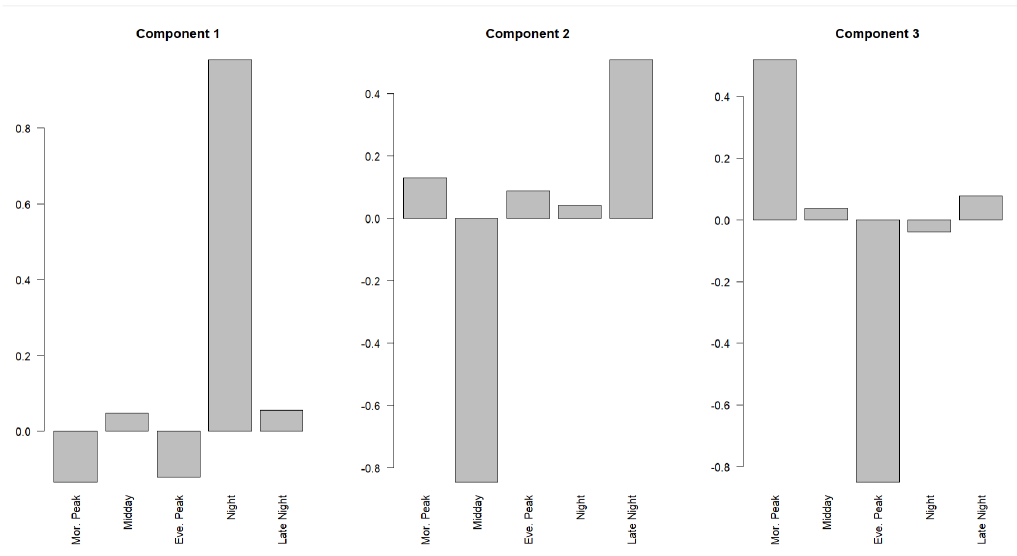}
    \caption{Latent factor structure of $\hat{L}_3$ (rank 3) using a varimax rotation on its leading three singular vectors. The factors appear to be peak-hour asymmetry (morning peak and evening peak), off-peak heterogeneity (midday and late night), and distinct night time behavior.}
    \label{fig:l3}
\end{figure}

\begin{figure}
    \centering
    \includegraphics[scale=0.65]{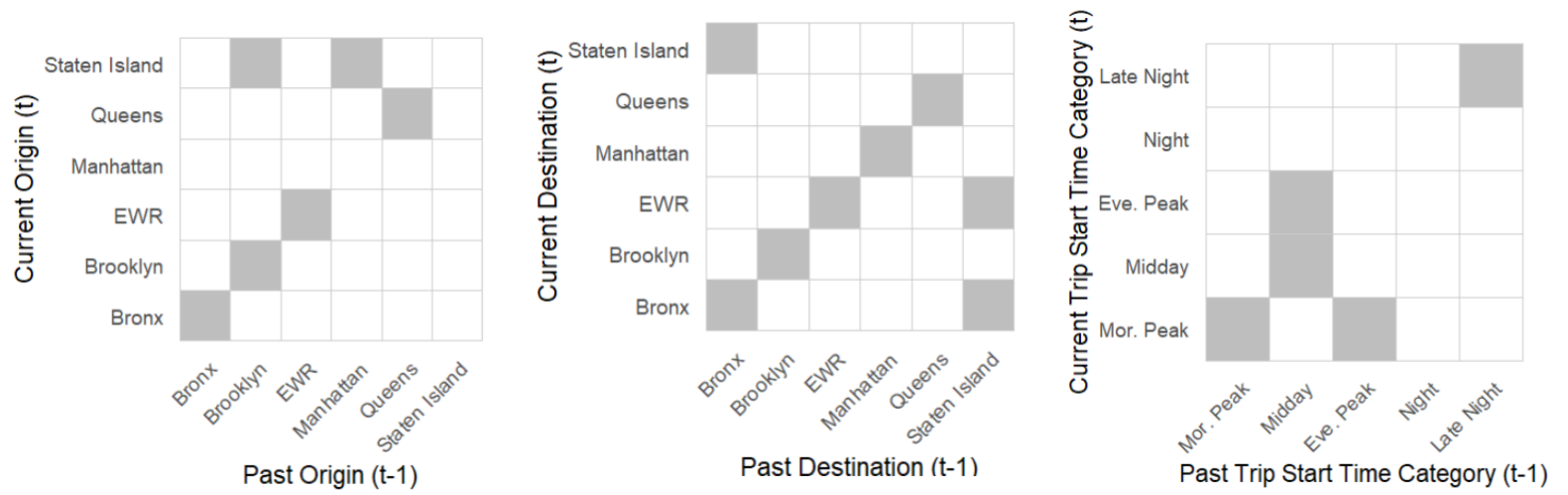}
    \caption{binary heatmaps of the estimated sparse components $\hat{S}_1$, $\hat{S}_2$ and $\hat{S}_3$, which represent additional idiosyncratic temporal dependencies beyond the shared baseline dynamics captured by $\hat{L}_1$, $\hat{L}_2$ and $\hat{L}_3$.}
    \label{fig:sparse}
\end{figure}

Finally, the estimated matrix $\hat{L}_3 \in \mathbb{R}^{5 \times 5}$ is of rank 3, capturing the baseline temporal dependence of the number of trips started at five operational hour categories at the current time point on the number of trips started at those five operational hour categories at the previous time point. As displayed in Figure \ref{fig:l3}, varimax rotation on the singular vectors shows that the first factor loads mostly on the night period. The second factor captures a contrast between midday and late Night, highlighting differences between routine daytime activity and more irregular late-night taxi-demand. The third factor reflects a contrast between Morning Peak and Evening Peak, suggesting asymmetric temporal dynamics between inbound and outbound commuting flows. Thus, baseline operation hour-wise dynamics appear to be governed by a combination of peak-hour asymmetry, off-peak heterogeneity, and distinct nighttime behavior. Figure \ref{fig:sparse} presents the binary heatmaps of the estimated sparse components $\hat{S}_1$, $\hat{S}_2$ and $\hat{S}_3$, which represent additional idiosyncratic temporal dependencies beyond the shared baseline dynamics captured by $\hat{L}_1$, $\hat{L}_2$ and $\hat{L}_3$. As expected, these components are highly sparse, indicating that the dominant temporal dependence is captured by the corresponding low-rank baseline components, with only a few localized idiosyncratic interactions remaining.

To further summarize the joint contribution of the latent origin, destination, and trip-start time structures, we perform a post hoc descriptive analysis by constructing an additive importance index for each of the $3\times 3\times3=27$ possible factor combinations. For each mode, the strength of a latent factor is quantified by the sum of the absolute values of its varimax-rotated loadings and subsequently normalized so that the strengths within each mode sum to one. The importance score of a given origin–destination–time combination is then defined as the sum of the corresponding normalized factor strengths across the three modes. Due to the additive nature of the proposed tensor autoregressive model, this index provides a descriptive measure of the relative prominence of each latent spatial–temporal regime rather than an estimate of a higher-order interaction effect. Ranking these scores facilitates the identification of the most prominent combinations of latent origin, destination, and temporal factors underlying the baseline temporal dynamics. 

\begin{table}[ht]
\centering
\caption{Top 10 latent origin--destination--trip start time factor combinations ranked according to the additive importance index.}
\label{tab:latent_combinations}

\resizebox{\textwidth}{!}{%
\begin{tabular}{clll}
\toprule
Rank & Origin factor & Destination factor & Trip-start time factor \\
\midrule
1 & Commercial vs.\ Residential & Core urban destinations & Midday vs.\ Late Night \\
2 & Commercial vs.\ Residential & Core urban destinations & Morning Peak vs.\ Evening Peak \\
3 & Airport-related origin & Core urban destinations & Midday vs.\ Late Night \\
4 & Commercial vs.\ Residential & Core urban destinations & Night \\
5 & Commercial vs.\ Residential & Airport destination & Midday vs.\ Late Night \\
6 & Airport-related origin & Core urban destinations & Morning Peak vs.\ Evening Peak \\
7 & Commercial vs.\ Residential & Airport destination & Morning Peak vs.\ Evening Peak \\
8 & Staten Island & Core urban destinations & Midday vs.\ Late Night \\
9 & Commercial vs.\ Residential & Peripheral destinations & Midday vs.\ Late Night \\
10 & Airport-related origin & Core urban destinations & Night \\
\bottomrule
\end{tabular}%
}
\end{table}

Finally, we evaluate the predictive performance of our proposed regularized additive TAR model on this dataset and compare it with the competing Tucker-based TAR model and the sparse VAR model. As described earlier in Sections \ref{intro} and \ref{simu}, the Tucker-based TAR model uses a multiplicative interaction of row-wise, column-wise and tube-wise temporal dependence. On the other hand, to apply the sparse VAR model to our tensor-variate time series, we simply vectorize the tensor data, and apply sparsity regularization on that vector. Table \ref{table-pred perf} summarizes the RMSE values, defined in Section \ref{simu}, for all the three models across the forecast horizons 1, 2 and 3. As the table illustrates, RMSE values are consistently lower for our model for all the forecast horizons, indicating improved predictive performance of our method as compared to the Tucker-based TAR model and the sparse VAR model. This aligns with our simulation results presented in Section \ref{simu}, where the sparse VAR model performs notably worse, unsurprisingly, as it vectorizes tensor time series, thereby discarding important structural information intrinsic to the matrix form. While the Tucker-based TAR model performs better than the sparse VAR, it still consistently underperforms compared to our method, further validating the ability of our proposed approach in achieving improved forecasting accuracy.

\begin{table}[h]
\centering
\caption{Predictive performance using RMSE values. The proposed additive TAR model performs better than the competing Tucker-based TAR model and the sparse VAR model.}\label{table-pred perf}%
\begin{tabular}{@{}llll@{}}
\toprule
Forecast horizon (h) & Additive TAR & Tucker-based TAR & Sparse VAR\\
\midrule
 1   & 0.761   & 0.801   & 1.229   \\
 2   & 0.747   & 0.794   & 1.113   \\
 3   & 0.746   & 0.798   & 1.053 \\
\bottomrule
\end{tabular}
\end{table}

\section{Discussion}
\label{disc}
In this work, we propose a high-dimensional regularized additive tensor autoregressive model that captures the temporal dependence among the tensor-valued time series by employing an additive interaction form, wherein the overall temporal connection is represented as the sum of row-wise, column-wise and tube-wise temporal dependence in the data. To accommodate high-dimensionality of the parameters, we then impose low-rank plus sparse regularized structures on row-wise, column-wise and tube-wise transition matrices. As discussed in \cite{zhang2024additive}, this additive interaction form, as opposed to convoluted bilinear representation, offers more comprehensible interpretation of the row-wise and column-wise and tube-wise temporal dependence. Also, with additive form, the penalized transition matrices help in extracting meaningful low-dimensional pattern in the data, whereas, the same with multiplicative bilinear form provides only dimension reduction.

Some future research directions along this line are discussed next. First, this method can be explored in the context of dynamic factor model as well. \cite{wang2019factor} proposed a factor model for matrix-variate time series, where they pre-multiplied and post-multiplied the core factor matrix $F_t$ with the front-loading (or, row-wise loading) $R$ and back-loading (or, column-wise loading) $C$ matrices respectively, yielding the bilinear form $RF_tC^\prime$. In contrast, it would be interesting to explore whether an additive row-wise, column-wise and tube-wise factor-loading representation can be employed by borrowing the idea from this paper. Secondly, we have demonstrated the model in this paper for a tensor with three dimensions that resulted in three types of additive components, namely, row-wise, column-wise and tube-wise temporal dependence. In case of tensor with a higher dimension, say $D>3$, \eqref{eq1} will translate to $\mathcal{Y}_{t} = \sum_{d=1}^D\operatorname{fold}_d [ B_d Y_{\overline{t-1} (d)}]+\mathcal{E}_{t}$, where $Y_{\overline{t-1} (d)}$ is matricized version of $\mathcal{Y}_{t-1}$ along it's $d^{th}$ mode \citep{kolda2009tensor}. Though our proposed algorithm and all the theoretical derivations follow for this case as well, it can be a bit difficult to visualize different mode-wise matricized data in such a case. Future research could explore a new variant of the additive TAR model that provides a more effective visualization of tensor data with more than three dimensions. Finally, to the best of our knowledge, overall Bayesian literature is sparse in the context of matrix and tensor-variate time series models. It would be interesting to see implementation of this additive tensor model and the corresponding theoretical developments under the Bayesian approach.

\bibliographystyle{unsrt}
\bibliography{bibliography}

\appendix

\section{Proof of the theoretical results}

\begin{lem}
    \label{lem-1}
    Let $C_1(L_1,S_1)$, $C_2(L_2,S_2)$ and $C_3(L_3,S_3)$ denote the weighted combinations of the nuclear norm and $l_1$ norm of the regularization parameters
    \begin{flalign}
        &C_1(L_1,S_1) = \left \lVert {L}_{1} \right \rVert _{*} + \frac{\lambda_{{S_1}}}{\lambda_{{L_1}}} \left \lVert {S}_{1} \right \rVert _{1} \nonumber \\
         &C_2(L_2,S_2) = \left \lVert {L}_{2} \right \rVert _{*} + \frac{\lambda_{{S_2}}}{\lambda_{{L_2}}} \left \lVert {S}_{2} \right \rVert _{1} \nonumber \\
          &C_3(L_3,S_3) = \left \lVert {L}_{3} \right \rVert _{*} + \frac{\lambda_{{S_3}}}{\lambda_{{L_3}}} \left \lVert {S}_{3} \right \rVert _{1} 
    \end{flalign}
    Then, for any $R_1 = 1,2\dots d_1$, $R_2 = 1,2 \dots d_2$ and $R_3 = 1,2 \dots d_3$, there exists decomposition of the forms $\hat{\Delta}_{L_{1}} = \hat{\Delta}_{
L_{1}}^{A_1} + \hat{\Delta}_{L_{1}}^{B_1} $,
$\hat{\Delta}_{L_{2}} = \hat{\Delta}_{
L_{2}}^{A_2} + \hat{\Delta}_{L_{2}}^{B_2} $ and $\hat{\Delta}_{L_{3}} = \hat{\Delta}_{
L_{3}}^{A_3} + \hat{\Delta}_{L_{3}}^{B_3} $
with
\textit{rank}$\big(\hat{\Delta}_{L_{1}}^{A_1}$\big) $\leq 2R_1$,
\textit{rank}$\big(\hat{\Delta}_{L_{2}}^{A_2}$\big) $\leq 2R_2$, \textit{rank}$\big(\hat{\Delta}_{L_{3}}^{A_3}$\big) $\leq 2R_3$, $L_1^T\hat{\Delta}_{L_{1}}^{B_1} =0 $, $L_1(\hat{\Delta}_{L_{1}}^{B_1})^T =0$, $L_2^T \hat{\Delta}_{L_{2}}^{B_2} =0 $, $L_2(\hat{\Delta}_{L_{2}}^{B_2})^T =0$, $L_3^T\hat{\Delta}_{L_{3}}^{B_3} =0 $, $L_3(\hat{\Delta}_{L_{3}}^{B_3})^T =0$
and
\begin{flalign}
    & C_1(L_1, S_1) - C_1(L_1 + \hat{\Delta}_{L_{1}}, 
    S_1 + \hat{\Delta}_{S_{1}})
    \leq C_1(\hat{\Delta}_{L_{1}}^{A_1}, \hat{\Delta}_{S_{1}}^{M}) - C_1(\hat{\Delta}_{L_{1}}^{B_1}, \hat{\Delta}_{S_{1}}^{M^{\bot}})
    \label{eq3}
\end{flalign}
\begin{flalign}
    & C_2(L_2,S_2) - C_2(L_2 + \hat{\Delta}_{L_{2}}, 
    S_2 + \hat{\Delta}_{S_{2}})
    \leq
    C_2(\hat{\Delta}_{L_{2}}^{A_2}, \hat{\Delta}_{S_{2}}^{N}) - C_2(\hat{\Delta}_{L_{2}}^{B_2}, \hat{\Delta}_{S_{2}}^{N^{\bot}})
    \label{eq4}
\end{flalign}
\begin{flalign}
    & C_3(L_3,S_3) - C_3(L_3 + \hat{\Delta}_{L_{3}}, 
    S_3 + \hat{\Delta}_{S_{3}})
    \leq
    C_3(\hat{\Delta}_{L_{3}}^{A_3}, \hat{\Delta}_{S_{3}}^{G}) - C_3(\hat{\Delta}_{L_{3}}^{B_3}, \hat{\Delta}_{S_{3}}^{G^{\bot}})
    \label{eq5}
\end{flalign}
\end{lem}

\begin{lem}
    \label{lem-2}
    Suppose that the errors $\mathcal{E}_t$ are deterministic. Let $D_1$, $D_2$ and $D_3$ be the matrices defined as follows:
    \begin{align}
        & {D}_1 = \frac{1}{T} \sum_{t=1}^{T}  \mathcal{E}_{t(1)} Y_{\overline{t-1} (1)}^{T} \\
        &{D}_2 =  \frac{1}{T} \sum_{t=1}^{T}\mathcal{E}_{t(2)} Y_{\overline{t-1} (2)}^{T} \\
        &{D}_3 =  \frac{1}{T} \sum_{t=1}^{T}\mathcal{E}_{t(3)} Y_{\overline{t-1} (3)}^{T}
    \end{align}
    Then, under the conditions $\lambda_{L_1} \geq 4 \left \lVert {D}_1 \right\rVert_{sp}$, $\lambda_{L_2} \geq 4 \left \lVert {D}_2 \right\rVert_{sp}$, $\lambda_{L_3} \geq 4 \left \lVert {D}_3 \right\rVert_{sp}$, $\lambda_{S_1} \geq 4 \left \lVert {D}_1 \right\rVert_{\infty}$, $\lambda_{S_2} \geq 4 \left \lVert {D}_2 \right\rVert_{\infty}$ and $\lambda_{S_3} \geq 4 \left \lVert {D}_3 \right\rVert_{\infty}$, the errors ($\hat{\Delta}_{L_{1}}, \hat{\Delta}_{S_{1}})$, $(\hat{\Delta}_{L_{2}},\hat{\Delta}_{S_{2}}$) and $(\hat{\Delta}_{L_{3}},\hat{\Delta}_{S_{3}}$) will satisfy the following constraints:
\begin{align}
    & C_1(\hat{\Delta}_{L_{1}}^{B_1}, \hat{\Delta}_{S_{1}}^\mathbb{M^{\bot}}) \leq 3 C_1(\hat{\Delta}_{L_{1}}^{A_1}, \hat{\Delta}_{S_{1}}^\mathbb{M})
    \label{rsset}
\end{align} 
\begin{align}
    & C_2(\hat{\Delta}_{L_{2}}^{B_2}, \hat{\Delta}_{S_{2}}^\mathbb{N^{\bot}}) \leq 3 C_2(\hat{\Delta}_{L_{2}}^{A_2}, \hat{\Delta}_{S_{2}}^\mathbb{N})
    \label{rsset1}
    \end{align}
\begin{align}
    & C_3(\hat{\Delta}_{L_{3}}^{B_3}, \hat{\Delta}_{S_{3}}^\mathbb{G^{\bot}}) \leq 3 C_3(\hat{\Delta}_{L_{3}}^{A_3}, \hat{\Delta}_{S_{3}}^\mathbb{G})
    \label{rsset2}
    \end{align}
\end{lem}

\subsection*{\small{\textbf{Basic Inequality}}}

\begin{flalign}
  &\frac{1}{2T} \sum_{t=1}^{T} \left \lVert \operatorname{fold}_1 \Bigg[ (\hat{\Delta}_{L_{1}} + \hat{\Delta}_{S_{1}})  Y_{\overline{t-1} (1)} \Bigg] + \operatorname{fold}_2 \Bigg[ (\hat{\Delta}_{L_{2}} + \hat{\Delta}_{S_{2}})  Y_{\overline{t-1} (2)} \Bigg]+ \operatorname{fold}_3 \Bigg[ (\hat{\Delta}_{L_{3}} + \hat{\Delta}_{S_{3}})  Y_{\overline{t-1} (3)} \Bigg]\right\rVert_{F}^{2} \nonumber
  \\
  & \leq  \frac{1}{T} \sum_{t=1}^{T}
  \biggl \langle \mathcal{E}_{t}, \operatorname{fold}_1 \Bigg[ (\hat{\Delta}_{L_{1}} + \hat{\Delta}_{S_{1}})  Y_{\overline{t-1} (1)} \Bigg] + \operatorname{fold}_2 \Bigg[ (\hat{\Delta}_{L_{2}} + \hat{\Delta}_{S_{2}})  Y_{\overline{t-1} (2)} \Bigg]+ \operatorname{fold}_3 \Bigg[ (\hat{\Delta}_{L_{3}} + \hat{\Delta}_{S_{3}})  Y_{\overline{t-1} (3)} \Bigg] \biggr \rangle  \nonumber \\
  & + \lambda_{{L_{1}}} {C_{1}}({L_{1},S_{1}}) + \lambda_{{L_{2}}} {C_{2}}({L_{2},S_{2}}) + \lambda_{{L_{3}}} {C_{3}}({L_{3},S_{3}}) -  \lambda_{{L_{1}}} {C_{1}}({L_{1} +\hat{\Delta}_{{L_{1}}},S_{1}+\hat{\Delta}_{{S_{1}}}}) \nonumber \\
  & -  \lambda_{{L_{2}}} {C_{2}}({L_{2} +\hat{\Delta}_{{L_{2}}},S_{2}+\hat{\Delta}_{{S_{2}}}})
-  \lambda_{{L_{3}}} {C_{3}}({L_{3} +\hat{\Delta}_{{L_{3}}},S_{3}+\hat{\Delta}_{{S_{3}}}})    
\label{eq6}
\end{flalign}

\begin{proof}

Using our model defined in equation \ref{eq1}, we can write the following.

\begin{flalign}
& \sum_{t=1}^{T} \Bigl\lVert  \mathcal{Y}_{t}
- \operatorname{fold}_1\Big[(\hat{L}_{1}+\hat{S}_{1})Y_{\overline{t-1}(1)}\Big]
- \operatorname{fold}_2\Big[(\hat{L}_{2}+\hat{S}_{2})Y_{\overline{t-1}(2)}\Big]
- \operatorname{fold}_3\Big[(\hat{L}_{3}+\hat{S}_{3})Y_{\overline{t-1}(3)}\Big]
\Bigr\rVert_{F}^{2} \nonumber\\
&= \sum_{t=1}^{T} \Bigl\lVert  \mathcal{E}_{t}
- \Big(\operatorname{fold}_1\big[(\hat{L}_{1}+\hat{S}_{1})Y_{\overline{t-1}(1)}\big]
- \operatorname{fold}_2\big[(\hat{L}_{2}+\hat{S}_{2})Y_{\overline{t-1}(2)}\big]
- \operatorname{fold}_3\big[(\hat{L}_{3}+\hat{S}_{3})Y_{\overline{t-1}(3)}\big]\Big)
\nonumber\\
&\qquad\quad
+ \Big(\operatorname{fold}_1\big[(L_{1}+S_{1})Y_{\overline{t-1}(1)}\big]
- \operatorname{fold}_2\big[(L_{2}+S_{2})Y_{\overline{t-1}(2)}\big]
- \operatorname{fold}_3\big[(L_{3}+S_{3})Y_{\overline{t-1}(3)}\big]\Big)
\Bigr\rVert_{F}^{2} \nonumber\\
&= \sum_{t=1}^{T} \Bigl\lVert  \mathcal{E}_{t}
- \operatorname{fold}_1\Big[\big((\hat{L}_{1}-L_{1})+(\hat{S}_{1}-S_{1})\big)Y_{\overline{t-1}(1)}\Big]
- \operatorname{fold}_2\Big[\big((\hat{L}_{2}-L_{2})+(\hat{S}_{2}-S_{2})\big)Y_{\overline{t-1}(2)}\Big] \nonumber \\
&\qquad\quad
- \operatorname{fold}_3\Big[\big((\hat{L}_{3}-L_{3})+(\hat{S}_{3}-S_{3})\big)Y_{\overline{t-1}(3)}\Big]
\Bigr\rVert_{F}^{2} \nonumber\\
&= \sum_{t=1}^{T} \Bigl\lVert  \mathcal{E}_{t}
- \operatorname{fold}_1\Big[(\hat{\Delta}_{L_{1}}+\hat{\Delta}_{S_{1}})Y_{\overline{t-1}(1)}\Big]
- \operatorname{fold}_2\Big[(\hat{\Delta}_{L_{2}}+\hat{\Delta}_{S_{2}})Y_{\overline{t-1}(2)}\Big]
- \operatorname{fold}_3\Big[(\hat{\Delta}_{L_{3}}+\hat{\Delta}_{S_{3}})Y_{\overline{t-1}(3)}\Big]
\Bigr\rVert_{F}^{2}
\label{eq7}
\end{flalign}

Now, let us define,

$$
    \mathcal{A}_{t} = \operatorname{fold}_1 \Bigg[ (\hat{\Delta}_{L_{1}} + \hat{\Delta}_{S_{1}})  Y_{\overline{t-1} (1)} \Bigg] +\operatorname{fold}_2 \Bigg[ (\hat{\Delta}_{L_{2}} + \hat{\Delta}_{S_{2}})  Y_{\overline{t-1} (2)} \Bigg]+\operatorname{fold}_3 \Bigg[ (\hat{\Delta}_{L_{3}} + \hat{\Delta}_{S_{3}})  Y_{\overline{t-1} (3)} \Bigg]
$$

Then the quantity in equation \ref{eq7} can be rewritten as 
\begin{align}
    & \sum_{t=1}^{T} \left\lVert \mathcal{E}_t- \mathcal{A}_t\right\rVert^{2}_{F} \nonumber \\
    &= \sum_{t=1}^{T} \left\lVert {\mathcal{E}_{t}}\right\rVert _{F}^{2} +  \sum_{t=1}^{T} \left\lVert {\mathcal{A}_{t}}\right\rVert _{F}^{2} - 2 \sum_{t=1}^{T} \biggl \langle {\mathcal{E}_{t}}, {\mathcal{A}_{t}}\biggr \rangle
    \label{eq8}
\end{align}

    Using the optimality of $({\hat{L}_{1}, \hat{L}_{2}, \hat{L}_{3}, \hat{S}_{1}, \hat{S}_{2}}, \hat{S}_{3})$ and the feasibility of  $({{L}_{1}, {L}_{2}, {L}_{3} ,{S}_{1}, {S}_{2}}, {S}_{3})$, we have the following.

    \begin{flalign}
        & \frac{1}{2T} \sum_{t=1}^{T} \left \lVert  \mathcal{Y}_{t} -  \operatorname{fold}_1 \Bigg[ (\hat{L}_{1} + \hat{S}_{1}) Y_{\overline{t-1} (1)}\Bigg] - \operatorname{fold}_2 \Bigg[ (\hat{L}_{2} + \hat{S}_{2}) Y_{\overline{t-1} (2)}\Bigg] - \operatorname{fold}_3 \Bigg[ (\hat{L}_{3} + \hat{S}_{3}) Y_{\overline{t-1} (3)}\Bigg]\right \rVert_{F}^{2} \nonumber \\
        & + \lambda_{{S_{1}}} \left \lVert {\hat{S}_{1}} \right \rVert _{1} + \lambda_{{S_{2}}} \left \lVert {\hat{S}_{2}} \right \rVert _{1} + \lambda_{{S_{3}}} \left \lVert {\hat{S}_{3}} \right \rVert _{1} + \lambda_{{L_{1}}} \left \lVert {\hat{L}_{1}} \right \rVert _{*} + \lambda_{{L_{2}}} \left \lVert {\hat{L}_{2}} \right \rVert _{*} + \lambda_{{L_{3}}} \left \lVert {\hat{L}_{3}} \right \rVert _{*} \nonumber \\
        &\leq
        \frac{1}{2T} \sum_{t=1}^{T} \left \lVert  \mathcal{Y}_{t} -  \operatorname{fold}_1 \Bigg[ ({L}_{1} + {S}_{1}) Y_{\overline{t-1} (1)}\Bigg] - \operatorname{fold}_2 \Bigg[ ({L}_{2} + {S}_{2}) Y_{\overline{t-1} (2)}\Bigg] - \operatorname{fold}_3 \Bigg[ ({L}_{3} + {S}_{3}) Y_{\overline{t-1} (3)}\Bigg]\right \rVert_{F}^{2} \nonumber \\
        & + \lambda_{{S_{1}}} \left \lVert {{S}_{1}} \right \rVert _{1} + \lambda_{{S_{2}}} \left \lVert {{S}_{2}} \right \rVert _{1} + \lambda_{{S_{3}}} \left \lVert {{S}_{3}} \right \rVert _{1} + \lambda_{{L_{1}}} \left \lVert {{L}_{1}} \right \rVert _{*} + \lambda_{{L_{2}}} \left \lVert {{L}_{2}} \right \rVert _{*} + \lambda_{{L_{3}}} \left \lVert {{L}_{3}} \right \rVert _{*} \nonumber \\
        \label{eq9}
    \end{flalign}

    We now combine the decomposition in \ref{eq8} along with the inequality in \ref{eq9} to arrive at the proof of this lemma.

    \begin{flalign}
    & \frac{1}{2T} \sum_{t=1}^{T}  \left\lVert {\mathcal{E}_{t}}\right\rVert_{F}^{2} + \frac{1}{2T} \sum_{t=1}^{T}  \left\lVert {\mathcal{A}_{t}}\right\rVert_{F}^{2} - \frac{1}{T} \sum_{t=1}^{T} \biggl \langle {\mathcal{E}_{t}}, {\mathcal{A}_{t}}\biggr \rangle + \lambda_{{S_{1}}} \left\lVert {\hat{S}_{1}} \right\rVert_{1} + \lambda_{{S_{2}}} \left\lVert {\hat{S}_{2}} \right\rVert_{1} + \lambda_{{S_{3}}}\left\lVert {\hat{S}_{3}} \right\rVert_{1}\nonumber \\ & +\lambda_{{L_{1}}} \left\lVert {\hat{L}_{1}} \right\rVert_{*} 
    + \lambda_{{L_{2}}} \left\lVert {\hat{L}_{2}} \right\rVert_{*} + \lambda_{{L_{3}}} \left\lVert {\hat{L}_{3}} \right\rVert_{*} \nonumber \\
    &\leq
    \frac{1}{2T} \sum_{t=1}^{T}  \left\lVert {\mathcal{E}_{t}}\right\rVert_{F}^{2} + \lambda_{{S_{1}}} \left\lVert {S_{1}} \right\rVert_{1} + \lambda_{{S_{2}}} \left\lVert {S_{2}} \right\rVert_{1} +  \lambda_{{S_{3}}} \left\lVert {S_{3}} \right\rVert_{1}+\lambda_{{L_{1}}} \left\lVert {L}_{1} \right\rVert_{*} + \lambda_{{L_{2}}} \left\lVert {L}_{2} \right\rVert_{*} + \lambda_{{L_{3}}} \left\lVert {L}_{3} \right\rVert_{*}
\end{flalign}
\end{proof}

\subsection*{\small{Proof of Lemma \ref{lem31}}}

\begin{proof}
    Using Assumption \ref{ass 1}, we have the following:  
    \begin{flalign}
    \label{eq10}
    &\frac{1}{2T} \sum_{t=1}^{T} \left\lVert \operatorname{fold}_1 \Bigg[ (\hat{\Delta}_{L_{1}} + \hat{\Delta}_{S_{1}})  Y_{\overline{t-1} (1)} \Bigg] + \operatorname{fold}_1 \Bigg[ (\hat{\Delta}_{L_{1}} + \hat{\Delta}_{S_{1}})  Y_{\overline{t-1} (1)} \Bigg]+ \operatorname{fold}_1 \Bigg[ (\hat{\Delta}_{L_{1}} + \hat{\Delta}_{S_{1}})  Y_{\overline{t-1} (1)} \Bigg] \right\rVert_{F}^{2} \nonumber \\
    & \geq
    \frac{\gamma}{2} \Bigg[ \left \lVert \hat\Delta_{L_{1}}+ \hat\Delta_{{S_{1}}}  \right \rVert ^{2}_{F} + \left \lVert \hat\Delta_{{L_{2}}} + \hat\Delta_{{S_{2}}}  \right \rVert ^{2}_{F} + \left \lVert \hat\Delta_{{L_{3}}} + \hat\Delta_{{S_{3}}}  \right \rVert ^{2}_{F} \Bigg]
\end{flalign}
We now find a lower bound for the right-hand side of the above inequality and an upper bound for the left side. We begin with the derivation of the lower bound first. We see that

\begin{flalign}
    & \frac{\gamma}{2} \Bigg[ \left \lVert \hat\Delta_{L_{1}}+ \hat\Delta_{{S_{1}}}  \right \rVert ^{2}_{F} + \left \lVert \hat\Delta_{{L_{2}}} + \hat\Delta_{{S_{2}}}  \right \rVert ^{2}_{F} + \left \lVert \hat\Delta_{{L_{3}}} + \hat\Delta_{{S_{3}}}  \right \rVert ^{2}_{F} \Bigg] 
    = \frac{\gamma}{2} \Bigg[ \left \lVert \hat\Delta_{L_{1}}\right \rVert ^{2}_{F}+ \left \lVert\hat\Delta_{{S_{1}}}  \right \rVert ^{2}_{F} + 2\biggl \langle {\hat{\Delta}_{{L_{1}}}}, {\hat{\Delta}_{{S_{1}}}} \biggr\rangle \nonumber \\
    &+ \left \lVert \hat\Delta_{L_{2}}\right \rVert ^{2}_{F}+ \left \lVert\hat\Delta_{{S_{2}}}  \right \rVert ^{2}_{F} + 2\biggl \langle {\hat{\Delta}_{{L_{2}}}}, {\hat{\Delta}_{{S_{2}}}} \biggr\rangle+ \left \lVert \hat\Delta_{L_{3}}\right \rVert ^{2}_{F}+ \left \lVert\hat\Delta_{{S_{3}}}  \right \rVert ^{2}_{F} + 2\biggl \langle {\hat{\Delta}_{{L_{3}}}}, {\hat{\Delta}_{{S_{3}}}} \biggr\rangle\Bigg]
\end{flalign}
Using this decomposition, we obtain the following,
\begin{flalign}
\label{eq11}
    &\frac{\gamma}{2} \Bigg[ \left \lVert \hat\Delta_{L_{1}}\right \rVert ^{2}_{F}+ \left \lVert\hat\Delta_{{S_{1}}}  \right \rVert ^{2}_{F} + 
     \left \lVert \hat\Delta_{L_{2}}\right \rVert ^{2}_{F}+ \left \lVert\hat\Delta_{{S_{2}}}  \right \rVert ^{2}_{F} + \left \lVert \hat\Delta_{L_{3}}\right \rVert ^{2}_{F}+ \left \lVert\hat\Delta_{{S_{3}}}  \right \rVert ^{2}_{F} \Bigg]-  \frac{\gamma}{2} \Bigg[ \left \lVert \hat\Delta_{L_{1}}+ \hat\Delta_{{S_{1}}}  \right \rVert ^{2}_{F} \nonumber \\  
     &+ \left \lVert \hat\Delta_{{L_{2}}} + \hat\Delta_{{S_{2}}}  \right \rVert ^{2}_{F} + \left \lVert \hat\Delta_{{L_{3}}} + \hat\Delta_{{S_{3}}}  \right \rVert ^{2}_{F} \Bigg] = -\gamma \biggl\langle \hat\Delta_{L_{1}},\hat\Delta_{{S_{1}}} \biggr\rangle -\gamma \biggl\langle \hat\Delta_{L_{2}},\hat\Delta_{{S_{2}}} \biggr\rangle-\gamma \biggl\langle \hat\Delta_{L_{3}},\hat\Delta_{{S_{3}}} \biggr\rangle
\end{flalign}

From the Dual norm inequality, we may write down the following,
\begin{flalign}
    \gamma \Big| \biggl\langle {\hat{\Delta}_{{L_{1}}}}, {\hat{\Delta}_{{S_{1}}}}   \biggr\rangle\Big| \leq \gamma \left \lVert {\hat{\Delta}_{{L_{1}}}}\right\rVert_{\infty} \left \lVert {\hat{\Delta}_{{S_{1}}}} \right\rVert_{1} \nonumber \\
    \gamma \Big| \biggl\langle {\hat{\Delta}_{{L_{2}}}}, {\hat{\Delta}_{{S_{2}}}}   \biggr\rangle\Big| \leq \gamma \left \lVert {\hat{\Delta}_{{L_{2}}}}\right\rVert_{\infty} \left \lVert {\hat{\Delta}_{{S_{2}}}} \right\rVert_{1} \nonumber \\
    \gamma \Big| \biggl\langle {\hat{\Delta}_{{L_{3}}}}, {\hat{\Delta}_{{S_{3}}}}   \biggr\rangle\Big| \leq \gamma \left \lVert {\hat{\Delta}_{{L_{3}}}}\right\rVert_{\infty} \left \lVert {\hat{\Delta}_{{S_{3}}}} \right\rVert_{1}
\end{flalign}

\begin{flalign}
&\Rightarrow\;
\gamma \Bigl|
\langle \hat{\Delta}_{L_{1}}, \hat{\Delta}_{S_{1}} \rangle
+\langle \hat{\Delta}_{L_{2}}, \hat{\Delta}_{S_{2}} \rangle
+\langle \hat{\Delta}_{L_{3}}, \hat{\Delta}_{S_{3}} \rangle
\Bigr|
\le
\gamma \Bigl[
\|\hat{\Delta}_{L_{1}}\|_{\infty}\,\|\hat{\Delta}_{S_{1}}\|_{1}
+\|\hat{\Delta}_{L_{2}}\|_{\infty}\,\|\hat{\Delta}_{S_{2}}\|_{1}
+\|\hat{\Delta}_{L_{3}}\|_{\infty}\,\|\hat{\Delta}_{S_{3}}\|_{1}
\Bigr] \nonumber\\
&\le\;
\gamma \Bigl[
\bigl(\|\hat{L}_{1}\|_{\infty}+\|L_{1}\|_{\infty}\bigr)\,\|\hat{\Delta}_{S_{1}}\|_{1}
+\bigl(\|\hat{L}_{2}\|_{\infty}+\|L_{2}\|_{\infty}\bigr)\,\|\hat{\Delta}_{S_{2}}\|_{1}
+\bigl(\|\hat{L}_{3}\|_{\infty}+\|L_{3}\|_{\infty}\bigr)\,\|\hat{\Delta}_{S_{3}}\|_{1}
\Bigr] \nonumber\\
&\le\;
\gamma \Bigl[
\frac{2\alpha_{1}}{\sqrt{d_{1}d_{1}}}\,\|\hat{\Delta}_{S_{1}}\|_{1}
+\frac{2\alpha_{2}}{\sqrt{d_{2}d_{2}}}\,\|\hat{\Delta}_{S_{2}}\|_{1}
+\frac{2\alpha_{3}}{\sqrt{d_{3}d_{3}}}\,\|\hat{\Delta}_{S_{3}}\|_{1}
\Bigr] \nonumber
\end{flalign}
Now using equation \ref{eq11}, we may write the following,

\begin{flalign}
    &\frac{\gamma}{2} \Bigg[ \left \lVert \hat\Delta_{L_{1}}+ \hat\Delta_{{S_{1}}}  \right \rVert ^{2}_{F} + \left \lVert \hat\Delta_{{L_{2}}} + \hat\Delta_{{S_{2}}}  \right \rVert ^{2}_{F} + \left \lVert \hat\Delta_{{L_{3}}} + \hat\Delta_{{S_{3}}}  \right \rVert ^{2}_{F} \Bigg] \nonumber \\
    & \geq \frac{\gamma}{2} \Bigg[ \left \lVert \hat\Delta_{L_{1}}\right \rVert ^{2}_{F}+ \left \lVert\hat\Delta_{{S_{1}}}  \right \rVert ^{2}_{F} + 
     \left \lVert \hat\Delta_{L_{2}}\right \rVert ^{2}_{F}+ \left \lVert\hat\Delta_{{S_{2}}}  \right \rVert ^{2}_{F} + \left \lVert \hat\Delta_{L_{3}}\right \rVert ^{2}_{F}+ \left \lVert\hat\Delta_{{S_{3}}}  \right \rVert ^{2}_{F} \Bigg]- \gamma \Bigg[\frac{2 \alpha_1}{\sqrt{d_{1} d_1}} \left \lVert{\hat{\Delta}_{{S_{1}}}}\right\rVert_{1}  \nonumber \\
     &+\frac{2 \alpha_2}{\sqrt{d_{2} d_2}} \left \lVert{\hat{\Delta}_{{S_{2}}}}\right\rVert_{1} + \frac{2 \alpha_3}{\sqrt{d_{3} d_3}} \left \lVert{\hat{\Delta}_{{S_{3}}}}\right\rVert_{1} \Bigg] \nonumber \\
     & \geq \frac{\gamma}{2} \Bigg[ \left \lVert \hat\Delta_{L_{1}}\right \rVert ^{2}_{F}+ \left \lVert\hat\Delta_{{S_{1}}}  \right \rVert ^{2}_{F} + 
     \left \lVert \hat\Delta_{L_{2}}\right \rVert ^{2}_{F}+ \left \lVert\hat\Delta_{{S_{2}}}  \right \rVert ^{2}_{F} + \left \lVert \hat\Delta_{L_{3}}\right \rVert ^{2}_{F}+ \left \lVert\hat\Delta_{{S_{3}}}  \right \rVert ^{2}_{F} \Bigg]- \frac{\lambda_{{S_{1}}}}{2} \left \lVert\hat\Delta_{{S_{1}}}  \right \rVert_{1} \nonumber \\
     &-\frac{\lambda_{{S_{2}}}}{2} \left \lVert\hat\Delta_{{S_{2}}}  \right \rVert_{1}-\frac{\lambda_{{S_{3}}}}{2} \left \lVert\hat\Delta_{{S_{3}}}  \right \rVert_{1} \nonumber \\
     & \geq \frac{\gamma}{2} \Bigg[ \left \lVert \hat\Delta_{L_{1}}\right \rVert ^{2}_{F}+ \left \lVert\hat\Delta_{{S_{1}}}  \right \rVert ^{2}_{F} + 
     \left \lVert \hat\Delta_{L_{2}}\right \rVert ^{2}_{F}+ \left \lVert\hat\Delta_{{S_{2}}}  \right \rVert ^{2}_{F} + \left \lVert \hat\Delta_{L_{3}}\right \rVert ^{2}_{F}+ \left \lVert\hat\Delta_{{S_{3}}}  \right \rVert ^{2}_{F} \Bigg]-\frac{\lambda_{{S_{1}}}}{2} \left \lVert\hat\Delta_{{S_{1}}}  \right \rVert_{1} \nonumber \\
     &-\frac{\lambda_{{S_{2}}}}{2} \left \lVert\hat\Delta_{{S_{2}}}  \right \rVert_{1}-\frac{\lambda_{{S_{3}}}}{2} \left \lVert\hat\Delta_{{S_{3}}}  \right \rVert_{1} - \frac{\lambda_{{L_{1}}}}{2} \left \lVert\hat\Delta_{{L_{1}}}  \right \rVert_{*} - \frac{\lambda_{{L_{2}}}}{2} \left \lVert\hat\Delta_{{L_{2}}}  \right \rVert_{*} -\frac{\lambda_{{L_{3}}}}{2} \left \lVert\hat\Delta_{{L_{3}}}  \right \rVert_{*} \nonumber \\
     & \geq \frac{\gamma}{2} \Bigg[ \left \lVert \hat\Delta_{L_{1}}\right \rVert ^{2}_{F}+ \left \lVert\hat\Delta_{{S_{1}}}  \right \rVert ^{2}_{F} + 
     \left \lVert \hat\Delta_{L_{2}}\right \rVert ^{2}_{F}+ \left \lVert\hat\Delta_{{S_{2}}}  \right \rVert ^{2}_{F} + \left \lVert \hat\Delta_{L_{3}}\right \rVert ^{2}_{F}+ \left \lVert\hat\Delta_{{S_{3}}}  \right \rVert ^{2}_{F} \Bigg]- \frac{\lambda_{{L_{1}}}}{2} C_1(\hat\Delta_{L_{1}},\hat\Delta_{S_{1}}) \nonumber \\
     & - \frac{\lambda_{{L_{2}}}}{2} C_2(\hat\Delta_{L_{2}},\hat\Delta_{S_{2}}) -\frac{\lambda_{{L_{3}}}}{2} C_3(\hat\Delta_{L_{3}},\hat\Delta_{S_{3}}) 
     \label{eqa1}
\end{flalign}
We now obtain an upper bound for the left side of the inequality \ref{eq10}.\\
We have already shown the following in the right-hand side of the Basic Inequality
\begin{flalign}
\label{eq12}
    \frac{1}{2T}\sum_{t=1}^{T} \left \lVert \mathcal{A}_{t} \right\rVert^{2}_{F} &\leq
    \frac{1}{T}\sum_{t=1}^{T} \biggl \langle \mathcal{E}_{t}, \mathcal{A}_{t}\biggr \rangle + \lambda_{L_{1}}\big(C_{1} (L_1,S_1)- C_1(L_1+ \hat\Delta_{L_{1}},S_1+ \hat\Delta_{S_{1}} )\big) \nonumber \\
    &+ \lambda_{L_{2}}\big(C_{2} (L_2,S_2)- C_2(L_2+ \hat\Delta_{L_{2}},S_2+ \hat\Delta_{S_{2}} )\big)+ \lambda_{L_{3}}\big(C_{3} (L_3,S_3)- C_3(L_3+ \hat\Delta_{L_{3}},S_3+ \hat\Delta_{S_{3}} )\big)
\end{flalign}

Now, using the inequalities in \ref{eq3}, \ref{eq4}, \ref{eq5} and \ref{eq12}, we obtain the following
\begin{flalign}
\label{eq13}
    & \frac{1}{2T}\sum_{t=1}^{T} \biggl \lVert \operatorname{fold}_1 \Bigg[ (\hat{\Delta}_{L_{1}} + \hat{\Delta}_{S_{1}})  Y_{\overline{t-1} (1)} \Bigg] +\operatorname{fold}_2 \Bigg[ (\hat{\Delta}_{L_{2}} + \hat{\Delta}_{S_{2}})  Y_{\overline{t-1} (2)} \Bigg]+\operatorname{fold}_3 \Bigg[ (\hat{\Delta}_{L_{3}} + \hat{\Delta}_{S_{3}})  Y_{\overline{t-1} (3)} \Bigg] \biggr \rVert^{2}_{F} \nonumber \\
    & \leq \frac{1}{T} \biggl \langle \mathcal{E}_t, \operatorname{fold}_1 \Bigg[ (\hat{\Delta}_{L_{1}} + \hat{\Delta}_{S_{1}})  Y_{\overline{t-1} (1)} \Bigg] +\operatorname{fold}_2 \Bigg[ (\hat{\Delta}_{L_{2}} + \hat{\Delta}_{S_{2}})  Y_{\overline{t-1} (2)} \Bigg]+\operatorname{fold}_3 \Bigg[ (\hat{\Delta}_{L_{3}} + \hat{\Delta}_{S_{3}})  Y_{\overline{t-1} (3)} \Bigg] \biggr \rangle \nonumber \\
    &+ \lambda_{L_{1}} \bigg( C_1(\hat{\Delta}_{L_{1}}^{A_1}, \hat{\Delta}_{S_{1}}^\mathbb{M}) - C_1(\hat{\Delta}_{L_{1}}^{B_1}, \hat{\Delta}_{S_{1}}^\mathbb{M^{\bot}})\bigg) + \lambda_{L_{2}} \bigg( C_2(\hat{\Delta}_{L_{2}}^{A_2}, \hat{\Delta}_{S_{2}}^\mathbb{N}) - C_2(\hat{\Delta}_{L_{2}}^{B_2}, \hat{\Delta}_{S_{2}}^\mathbb{N^{\bot}})\bigg) \nonumber \\
    &+ \lambda_{L_{3}} \bigg( C_3(\hat{\Delta}_{L_{3}}^{A_3}, \hat{\Delta}_{S_{3}}^\mathbb{G}) - C_3(\hat{\Delta}_{L_{3}}^{B_3}, \hat{\Delta}_{S_{3}}^\mathbb{G^{\bot}})\bigg)
\end{flalign}
Further, we can write
\begin{flalign}
    &\frac{1}{T} \sum_{t=1}^{T} \biggl \langle \mathcal{E}_t, \operatorname{fold}_1 \Bigg[ (\hat{\Delta}_{L_{1}} + \hat{\Delta}_{S_{1}})  Y_{\overline{t-1} (1)} \Bigg] +\operatorname{fold}_2 \Bigg[ (\hat{\Delta}_{L_{2}} + \hat{\Delta}_{S_{2}})  Y_{\overline{t-1} (2)} \Bigg]+\operatorname{fold}_3 \Bigg[ (\hat{\Delta}_{L_{3}} + \hat{\Delta}_{S_{3}})  Y_{\overline{t-1} (3)} \Bigg]\biggr \rangle \nonumber \\
    &= \frac{1}{T} \sum_{t=1}^{T} \biggl \langle \mathcal{E}_t , \operatorname{fold}_1 \Bigg[ (\hat{\Delta}_{L_{1}} + \hat{\Delta}_{S_{1}})  Y_{\overline{t-1} (1)} \Bigg]\biggr \rangle + \frac{1}{T} \sum_{t=1}^{T} \biggl \langle \mathcal{E}_t , \operatorname{fold}_2\Bigg[ (\hat{\Delta}_{L_{2}} + \hat{\Delta}_{S_{2}})  Y_{\overline{t-1} (2)} \Bigg]\biggr \rangle \nonumber \\
    &+
    \frac{1}{T} \sum_{t=1}^{T} \biggl \langle \mathcal{E}_t , \operatorname{fold}_3 \Bigg[ (\hat{\Delta}_{L_{3}} + \hat{\Delta}_{S_{3}})  Y_{\overline{t-1} (3)} \Bigg]\biggr \rangle
\end{flalign}

We may note the following

\begin{flalign}
    &\biggl \langle \mathcal{E}_t, \operatorname{fold}_1 \Bigg[ (\hat{\Delta}_{L_{1}} + \hat{\Delta}_{S_{1}})  Y_{\overline{t-1} (1)} \Bigg]\biggr \rangle \nonumber \\
    &= \biggl \langle \mathcal{E}_{t(1)},  (\hat{\Delta}_{L_{1}} + \hat{\Delta}_{S_{1}})  Y_{\overline{t-1} (1)}  \biggr \rangle \nonumber \\
    &= \operatorname{tr} \bigg(\mathcal{E}_{t(1)}^{T}  (\hat{\Delta}_{L_{1}} + \hat{\Delta}_{S_{1}})  Y_{\overline{t-1} (1)}    \bigg) \nonumber \\
    &= \operatorname{tr} \bigg((\hat{\Delta}_{L_{1}} + \hat{\Delta}_{S_{1}})  Y_{\overline{t-1} (1)} \mathcal{E}_{t(1)}^{T}  \bigg) \nonumber \\
    &= \operatorname{tr} \bigg((\hat{\Delta}_{L_{1}} + \hat{\Delta}_{S_{1}}) (\mathcal{E}_{t(1)}Y_{\overline{t-1} (1)}^{T})^{T} \bigg) \nonumber \\
    &= \biggl \langle \mathcal{E}_{t(1)} Y_{\overline{t-1} (1)}^{T}, (\hat{\Delta}_{L_{1}} + \hat{\Delta}_{S_{1}})  \biggr \rangle
\end{flalign}
Similarly, it is possible to show that

\begin{flalign}
    &\biggl \langle \mathcal{E}_t, \operatorname{fold}_2 \Bigg[ (\hat{\Delta}_{L_{2}} + \hat{\Delta}_{S_{2}})  Y_{\overline{t-1} (2)} \Bigg]\biggr \rangle = \biggl \langle \mathcal{E}_{t(2)} Y_{\overline{t-1} (2)}^{T}, (\hat{\Delta}_{L_{2}} + \hat{\Delta}_{S_{2}})  \biggr \rangle \\
    &\biggl \langle \mathcal{E}_t, \operatorname{fold}_3 \Bigg[ (\hat{\Delta}_{L_{3}} + \hat{\Delta}_{S_{3}})  Y_{\overline{t-1} (3)} \Bigg]\biggr \rangle = \biggl \langle \mathcal{E}_{t(3)} Y_{\overline{t-1} (3)}^{T}, (\hat{\Delta}_{L_{3}} + \hat{\Delta}_{S_{3}})  \biggr \rangle
\end{flalign}

Hence, we have the following

\begin{flalign}
    &\frac{1}{T} \sum_{t=1}^{T} \biggl \langle \mathcal{E}_t , \operatorname{fold}_1 \Bigg[ (\hat{\Delta}_{L_{1}} + \hat{\Delta}_{S_{1}})  Y_{\overline{t-1} (1)} \Bigg]\biggr \rangle \nonumber \\
    &= \frac{1}{T} \sum_{t=1}^{T} \biggl \langle \mathcal{E}_{t(1)} Y_{\overline{t-1} (1)}^{T}, (\hat{\Delta}_{L_{1}} + \hat{\Delta}_{S_{1}})  \biggr \rangle \nonumber \\
    &= \biggl \langle D_{1}, (\hat{\Delta}_{L_{1}} + \hat{\Delta}_{S_{1}}) \biggr \rangle \hspace{0.7 cm} \text{where,} \hspace{0.1cm} D_{1}= \frac{1}{T} \sum_{t=1}^{T} \mathcal{E}_{t(1)} Y_{\overline{t-1} (1)}^{T} \nonumber \\
    & \leq
    \left \lVert {\hat{\Delta}_{{L_{1}}}} \right\rVert_{*}  \left \lVert {D_1} \right\rVert_{sp} + \left \lVert {\hat{\Delta}_{{S_{1}}}} \right\rVert_{1}  \left \lVert {D_1}
    \right\rVert_{\infty} \nonumber \\
    & \leq
     \left \lVert {{D}_1} \right\rVert_{sp} \Bigg[  \left \lVert {\hat{\Delta}_{{L_{1}}}}^{A_1} \right\rVert_{*} + \left \lVert {\hat{\Delta}_{{L_{1}}}}^{B_1} \right\rVert_{*}\Bigg] + 
     \left \lVert {{D}_1} \right\rVert_{\infty} \Bigg[  \left \lVert {\hat{\Delta}_{{S_{1}}}}^{M} \right\rVert_{1} + \left \lVert {\hat{\Delta}_{{S_{1}}}}^{M^{\bot}} \right\rVert_{1}\Bigg]
\end{flalign}

Using the definition of $C_{1}(L_{1}, S_{1})$ in Lemma \ref{lem-1} and the assumptions on the regularizing parameters in Assumption \ref{ass 3}, we get the following

\begin{flalign}
    & \frac{1}{T} \sum_{t=1}^{T} \biggl \langle \mathcal{E}_t , \operatorname{fold}_1 \Bigg[ (\hat{\Delta}_{L_{1}} + \hat{\Delta}_{S_{1}})  Y_{\overline{t-1} (1)} \Bigg]\biggr \rangle \leq
    \frac{\lambda_{{L_1}}}{4}
    \Bigg[ {C_1}\big({\hat{\Delta}_{{L_{1}}}}^{A_1} + {\hat{\Delta}_{{S_{1}}}}^{M} \big) + {C_1}\big({\hat{\Delta}_{{L_{1}}}}^{B_1} + {\hat{\Delta}_{{S_{1}}}}^{M^{\bot}} \big)\Bigg]
\end{flalign}
Using similar steps, one can show that
\begin{flalign}
    & \frac{1}{T} \sum_{t=1}^{T} \biggl \langle \mathcal{E}_t , \operatorname{fold}_2 \Bigg[ (\hat{\Delta}_{L_{2}} + \hat{\Delta}_{S_{2}})  Y_{\overline{t-1} (2)} \Bigg]\biggr \rangle \leq
    \frac{\lambda_{{L_2}}}{4}
    \Bigg[ {C_2}\big({\hat{\Delta}_{{L_{2}}}}^{A_2} + {\hat{\Delta}_{{S_{2}}}}^{N} \big) + {C_2}\big({\hat{\Delta}_{{L_{2}}}}^{B_2} + {\hat{\Delta}_{{S_{2}}}}^{N^{\bot}} \big)\Bigg] \\
    & \frac{1}{T} \sum_{t=1}^{T} \biggl \langle \mathcal{E}_t , \operatorname{fold}_3 \Bigg[ (\hat{\Delta}_{L_{3}} + \hat{\Delta}_{S_{3}})  Y_{\overline{t-1} (3)} \Bigg]\biggr \rangle \leq
    \frac{\lambda_{{L_3}}}{4}
    \Bigg[ {C_3}\big({\hat{\Delta}_{{L_{3}}}}^{A_3} + {\hat{\Delta}_{{S_{3}}}}^{G} \big) + {C_3}\big({\hat{\Delta}_{{L_{3}}}}^{B_3} + {\hat{\Delta}_{{S_{3}}}}^{G^{\bot}} \big)\Bigg]
\end{flalign}
Using these inequalities and the one in \ref{eq13}, we can show the following

\begin{flalign*}
    & \frac{1}{2T} \sum_{t=1}^{T} \left\lVert \operatorname{fold}_1 \Bigg[ (\hat{\Delta}_{L_{1}} + \hat{\Delta}_{S_{1}})  Y_{\overline{t-1} (1)} \Bigg] +\operatorname{fold}_2 \Bigg[ (\hat{\Delta}_{L_{2}} + \hat{\Delta}_{S_{2}})  Y_{\overline{t-1} (2)} \Bigg]+\operatorname{fold}_3 \Bigg[ (\hat{\Delta}_{L_{3}} + \hat{\Delta}_{S_{3}})  Y_{\overline{t-1} (3)} \Bigg]\right\rVert _{F}^{2} \\
    & \leq \frac{\lambda_{{L_1}}}{4}
    \Bigg[ {C_1}\big({\hat{\Delta}_{{L_{1}}}}^{A_1} + {\hat{\Delta}_{{S_{1}}}}^{M} \big) + {C_1}\big({\hat{\Delta}_{{L_{1}}}}^{B_1} + {\hat{\Delta}_{{S_{1}}}}^{M^{\bot}} \big)\Bigg] + \frac{\lambda_{{L_2}}}{4} \Bigg[{C_2}\big({\hat{\Delta}_{{L_{2}}}}^{A_2} + {\hat{\Delta}_{{S_{2}}}}^{N} \big) + {C_2}\big({\hat{\Delta}_{{L_{2}}}}^{B_2} + {\hat{\Delta}_{{S_{2}}}}^{N^{\bot}} \big) \Bigg] \\
    &\qquad\quad
    + \frac{\lambda_{{L_3}}}{4}
    \Bigg[ {C_3}\big({\hat{\Delta}_{{L_{3}}}}^{A_3} + {\hat{\Delta}_{{S_{3}}}}^{G} \big) + {C_3}\big({\hat{\Delta}_{{L_{3}}}}^{B_3} + {\hat{\Delta}_{{S_{3}}}}^{G^{\bot}} \big)\Bigg] + \lambda_{{L_{1}}} \Bigg[{C_1}({\hat{\Delta}_{{L_{1}}}}^{A_1}, {\hat{\Delta}_{{S_{1}}}}^\mathbb{M}) - {C_1}({\hat{\Delta}_{{L_{1}}}}^{B_1}, {\hat{\Delta}_{{S_{1}}}}^\mathbb{M^{\bot}}) \Bigg] \\
    &\qquad\quad
    + \lambda_{{L_{2}}} \Bigg[{C_2}({\hat{\Delta}_{{L_{2}}}}^{A_2}, {\hat{\Delta}_{{S_{2}}}}^\mathbb{N}) - {C_2}({\hat{\Delta}_{{L_{2}}}}^{B_2}, {\hat{\Delta}_{{S_{2}}}}^\mathbb{N^{\bot}}) \Bigg] + \lambda_{{L_{3}}} \Bigg[{C_3}({\hat{\Delta}_{{L_{3}}}}^{A_3}, {\hat{\Delta}_{{S_{3}}}}^\mathbb{G}) - {C_3}({\hat{\Delta}_{{L_{3}}}}^{B_3}, {\hat{\Delta}_{{S_{3}}}}^\mathbb{G^{\bot}}) \Bigg]
\end{flalign*}

This reduces to the following expression,

\begin{flalign}
\label{eq14}
    & \frac{1}{2T} \sum_{t=1}^{T} \left\lVert \operatorname{fold}_1 \Bigg[ (\hat{\Delta}_{L_{1}} + \hat{\Delta}_{S_{1}})  Y_{\overline{t-1} (1)} \Bigg] +\operatorname{fold}_2 \Bigg[ (\hat{\Delta}_{L_{2}} + \hat{\Delta}_{S_{2}})  Y_{\overline{t-1} (2)} \Bigg]+\operatorname{fold}_3 \Bigg[ (\hat{\Delta}_{L_{3}} + \hat{\Delta}_{S_{3}})  Y_{\overline{t-1} (3)} \Bigg]\right\rVert _{F}^{2} \nonumber \\
    & \leq \frac{3}{2} \lambda_{{L_1}} {C_1}({\hat{\Delta}_{{L_{1}}}}^{A_1}, {\hat{\Delta}_{{S_{1}}}}^\mathbb{M}) +
    \frac{3}{2} \lambda_{{L_2}} {C_2}({\hat{\Delta}_{{L_{2}}}}^{A_2}, {\hat{\Delta}_{{S_{2}}}}^\mathbb{N}) + \frac{3}{2} \lambda_{{L_3}} {C_3}({\hat{\Delta}_{{L_{3}}}}^{A_3}, {\hat{\Delta}_{{S_{3}}}}^\mathbb{G})
\end{flalign}

Combining the expressions in \ref{eqa2}, \ref{eqa1} and \ref{eq14}, we get the following,

\begin{flalign}
    & \frac{\gamma}{2} \Bigg[\left \lVert {\hat{\Delta}_{{L_{1}}}}\right\rVert _{F}^{2} +  \left \lVert {\hat{\Delta}_{{S_{1}}}}\right\rVert _{F}^{2} +  \left \lVert {\hat{\Delta}_{{L_{2}}}}\right\rVert _{F}^{2} +  \left \lVert {\hat{\Delta}_{{S_{2}}}}\right\rVert _{F}^{2} +  \left \lVert {\hat{\Delta}_{{L_{3}}}}\right\rVert _{F}^{2} +  \left \lVert {\hat{\Delta}_{{S_{3}}}}\right\rVert _{F}^{2}\Bigg] \nonumber \\
    &\leq \frac{3}{2} \lambda_{{L_1}} {C_1}({\hat{\Delta}_{{L_{1}}}}^{A_1}, {\hat{\Delta}_{{S_{1}}}}^\mathbb{M}) +
    \frac{3}{2} \lambda_{{L_2}} {C_2}({\hat{\Delta}_{{L_{2}}}}^{A_2}, {\hat{\Delta}_{{S_{2}}}}^\mathbb{N}) + \frac{3}{2} \lambda_{{L_3}} {C_3}({\hat{\Delta}_{{L_{3}}}}^{A_3}, {\hat{\Delta}_{{S_{3}}}}^\mathbb{G})\label{eq15} \nonumber \\ 
    & \qquad\quad
    + \frac{\lambda_{{L_1}}}{2} {C_1(\hat{\Delta}_{{L_{1}}}, \hat{\Delta}_{{S_{1}}})} +\frac{\lambda_{{L_2}}}{2} {C_2(\hat{\Delta}_{{L_{2}}}, \hat{\Delta}_{{S_{2}}})} + \frac{\lambda_{{L_3}}}{2} {C_3(\hat{\Delta}_{{L_{3}}}, \hat{\Delta}_{{S_{3}}})}
\end{flalign}

We have the following results,

\begin{align}
    &{C_1}({\hat{\Delta}_{{L_{1}}}}, {\hat{\Delta}_{{S_{1}}}})
    \leq
    {C_1}({\hat{\Delta}_{{L_{1}}}}^{A_1}, {\hat{\Delta}_{{S_{1}}}}^\mathbb{M}) + {C_1}({\hat{\Delta}_{{L_{1}}}}^{B_1}, {\hat{\Delta}_{{S_{1}}}}^\mathbb{M^{\bot}}) \nonumber \\
    & {C_2}({\hat{\Delta}_{{L_{2}}}}, {\hat{\Delta}_{{S_{2}}}})
    \leq
    {C_2}({\hat{\Delta}_{{L_{2}}}}^{A_2}, {\hat{\Delta}_{{S_{2}}}}^\mathbb{N}) + {C_2}({\hat{\Delta}_{{L_{2}}}}^{B_2}, {\hat{\Delta}_{{S_{2}}}}^\mathbb{N^{\bot}}) \nonumber \\
    & {C_3}({\hat{\Delta}_{{L_{3}}}}, {\hat{\Delta}_{{S_{3}}}})
    \leq
    {C_3}({\hat{\Delta}_{{L_{3}}}}^{A_3}, {\hat{\Delta}_{{S_{3}}}}^\mathbb{G}) + {C_3}({\hat{\Delta}_{{L_{3}}}}^{B_3}, {\hat{\Delta}_{{S_{3}}}}^\mathbb{G^{\bot}})
\end{align}

Together with these results and Lemma \ref{lem-2}, we get the following,

\begin{align}
    & {C_1}({\hat{\Delta}_{{L_{1}}}}, {\hat{\Delta}_{{S_{1}}}})
    \leq
    4 {C_1}({\hat{\Delta}_{{L_{1}}}}^{A_1}, {\hat{\Delta}_{{S_{1}}}}^\mathbb{M}) \nonumber \\
    & {C_2}({\hat{\Delta}_{{L_{2}}}}, {\hat{\Delta}_{{S_{2}}}})
    \leq
    4 {C_2}({\hat{\Delta}_{{L_{2}}}}^{A_2}, {\hat{\Delta}_{{S_{2}}}}^\mathbb{N}) \nonumber \\
    & {C_3}({\hat{\Delta}_{{L_{3}}}}, {\hat{\Delta}_{{S_{3}}}})
    \leq
    4 {C_3}({\hat{\Delta}_{{L_{3}}}}^{A_3}, {\hat{\Delta}_{{S_{3}}}}^\mathbb{G})
\end{align}

With the help of these results, we may rewrite \ref{eq15} in the following manner,

\begin{flalign}
    & \frac{\gamma}{2} \Bigg[\left \lVert {\hat{\Delta}_{{L_{1}}}}\right\rVert _{F}^{2} +  \left \lVert {\hat{\Delta}_{{S_{1}}}}\right\rVert _{F}^{2} +  \left \lVert {\hat{\Delta}_{{L_{2}}}}\right\rVert _{F}^{2} +  \left \lVert {\hat{\Delta}_{{S_{2}}}}\right\rVert _{F}^{2} +  \left \lVert {\hat{\Delta}_{{L_{3}}}}\right\rVert _{F}^{2} +  \left \lVert {\hat{\Delta}_{{S_{3}}}}\right\rVert _{F}^{2}\Bigg] \nonumber \\
   &\leq \label{eq16}
   4 \lambda_{{L_1}} {C_1}({\hat{\Delta}_{{L_{1}}}}^{A_1}, {\hat{\Delta}_{{S_{1}}}}^\mathbb{M})
   + 
   4 \lambda_{{L_2}} {C_2}({\hat{\Delta}_{{L_{2}}}}^{A_2}, {\hat{\Delta}_{{S_{2}}}}^\mathbb{N}) + + 
   4 \lambda_{{L_3}} {C_3}({\hat{\Delta}_{{L_{3}}}}^{A_3}, {\hat{\Delta}_{{S_{3}}}}^\mathbb{G})
\end{flalign}

It follows from Lemma \ref{lem-1} that the rank of rank of ${\hat{\Delta}_{{L_{1}}}}^{A_1}$ is at most 2$R_1$, that of ${\hat{\Delta}_{{L_{2}}}}^{A_2}$ is at most 2$R_2$ while that of ${\hat{\Delta}_{{L_{3}}}}^{A_3}$ is at most 2$R_3$. Using these along with the idea of \textit{Compatibility Constant} defined in \cite{agarwal2012noisy}, we get to the following inequalities,
\begin{flalign}
    & \lambda_{{L_1}}{C_1}({\hat{\Delta}_{{L_{1}}}}^{A_1}, {\hat{\Delta}_{{S_{1}}}}^\mathbb{M}) \leq \sqrt{2 R_1} \lambda_{{L_1}} \left \lVert {\hat{\Delta}_{{L_{1}}}}^{A_1} \right\rVert_{F} + \sqrt{s_1} \lambda_{{S_1}}\left \lVert {\hat{\Delta}_{{S_{1}}}}^{M} \right\rVert_{F} \nonumber \\
    &\qquad \quad
    \leq \sqrt{2 R_1} \lambda_{{L_1}}  \left \lVert {\hat{\Delta}_{{L_{1}}}} \right\rVert_{F} + \sqrt{s_1} \lambda_{{S_1}} \left \lVert {\hat{\Delta}_{{S_{1}}}}\right\rVert_{F}
\end{flalign}

\begin{align}
    & \lambda_{{L_2}}{C_2}({\hat{\Delta}_{{L_{2}}}}^{A_2}, {\hat{\Delta}_{{S_{2}}}}^\mathbb{N}) \leq  \sqrt{2 R_2} \lambda_{{L_2}}  \left \lVert {\hat{\Delta}_{{L_{2}}}} \right\rVert_{F} + \sqrt{s_2} \lambda_{{S_2}} \left \lVert {\hat{\Delta}_{{S_{2}}}}\right\rVert_{F}
\end{align}

\begin{align}
    & \lambda_{{L_3}}{C_3}({\hat{\Delta}_{{L_{3}}}}^{A_3}, {\hat{\Delta}_{{S_{3}}}}^\mathbb{G}) \leq  \sqrt{2 R_3} \lambda_{{L_3}}  \left \lVert {\hat{\Delta}_{{L_{3}}}} \right\rVert_{F} + \sqrt{s_3} \lambda_{{S_3}} \left \lVert {\hat{\Delta}_{{S_{3}}}}\right\rVert_{F}
\end{align}

Using these inequalities along with \ref{eq16} and ignoring certain unnecessary constants, we get the following,

\begin{flalign}
    &\Bigg[\left \lVert {\hat{\Delta}_{{L_{1}}}}\right\rVert _{F}^{2} +  \left \lVert {\hat{\Delta}_{{S_{1}}}}\right\rVert _{F}^{2} +  \left \lVert {\hat{\Delta}_{{L_{2}}}}\right\rVert _{F}^{2} +  \left \lVert {\hat{\Delta}_{{S_{2}}}}\right\rVert _{F}^{2} + \left \lVert {\hat{\Delta}_{{L_{3}}}}\right\rVert _{F}^{2} +  \left \lVert {\hat{\Delta}_{{S_{3}}}}\right\rVert _{F}^{2}\Bigg] \nonumber \\
    &\preceq \sqrt{ R_1} \lambda_{{L_1}}  \left \lVert {\hat{\Delta}_{{L_{1}}}} \right\rVert_{F} + \sqrt{ R_2} \lambda_{{L_2}}  \left \lVert {\hat{\Delta}_{{L_{2}}}} \right\rVert_{F} + \sqrt{ R_3} \lambda_{{L_3}}  \left \lVert {\hat{\Delta}_{{L_{3}}}} \right\rVert_{F} + \sqrt{s_1} \lambda_{{S_1}} \left \lVert {\hat{\Delta}_{{S_{1}}}}\right\rVert_{F}\nonumber\\
    &\qquad\quad
    + \sqrt{s_2} \lambda_{{S_2}} \left \lVert {\hat{\Delta}_{{S_{2}}}}\right\rVert_{F} + \sqrt{s_3} \lambda_{{S_3}} \left \lVert {\hat{\Delta}_{{S_{3}}}}\right\rVert_{F}
\end{flalign}

The above equation yields the following,

\begin{flalign}
    & \left \lVert {\hat{\Delta}_{{L_{1}}}}\right\rVert _{F}^{2} +  \left \lVert {\hat{\Delta}_{{S_{1}}}}\right\rVert _{F}^{2} +  \left \lVert {\hat{\Delta}_{{L_{2}}}}\right\rVert _{F}^{2} +  \left \lVert {\hat{\Delta}_{{S_{2}}}}\right\rVert _{F}^{2}+ \left \lVert {\hat{\Delta}_{{L_{3}}}}\right\rVert _{F}^{2} +  \left \lVert {\hat{\Delta}_{{S_{3}}}}\right\rVert _{F}^{2} \nonumber \\
    &\preceq
     R_1 \lambda_{{L_1}}^2 + R_2 \lambda_{{L_2}}^2 +  R_3 \lambda_{{L_3}}^2 
     + s_1 \lambda_{{S_1}}^2 +  s_2 \lambda_{{S_2}}^2 + s_3 \lambda_{{S_3}}^2
\end{flalign}
This concludes the proof of the lemma.
\end{proof}

\subsection{Proof of error bound under Gaussian distribution}

\begin{proof}
    We start by establishing that the following hold with high probability.

    \begin{flalign*}
        \lambda_{{S_1}} \geq  4 \left \lVert {\mathcal{D}_1} \right\rVert_{\infty} + \frac{4\gamma \alpha_1}{\sqrt{d_1 d_1}} \\
        \lambda_{{S_2}} \geq  4 \left \lVert {\mathcal{D}_2} \right\rVert_{\infty} + \frac{4\gamma \alpha_2}{\sqrt{d_2 d_2}}\\
        \lambda_{{S_3}} \geq  4 \left \lVert {\mathcal{D}_3} \right\rVert_{\infty} + \frac{4\gamma \alpha_3}{\sqrt{d_3 d_3}}
    \end{flalign*}
    Using the definition from Section \ref{theo}, and applying Proposition 2.4(b) in \cite{basu2015regularized} to the matrices $E_1$ and $Y_{-1(1)}$, we can say that $\exists$ a $c>0$ $\ni$ for any $u$, $v$ $\in \mathbb{R}^{d_1}$ with $ \left \lVert u \right \rVert \leq 1$,$ \left \lVert v \right \rVert \leq 1$ and for any $\eta >0$, we get
    \begin{align}
    & P \Bigg[ \left \lvert u^T \Big( \frac{{E_1} {Y_{-1(1)}{^{T}}}}{T}  \Big) v \right \rvert > 2\pi Q_1 \eta \Bigg]
    \leq
    6 \hspace{0.1cm} \text{exp}[-cT min \{\eta, \eta^2\}]
\end{align}
Using the similar approach as the proof of Proposition 4.3 in \cite{basu2015regularized}, we take the union bound over the $d_1^2$ possible choices of $u \in \{e_1, e_2, \dots e_{d_{1}} \}$ and $v \in \{e_1, e_2, \dots e_{d_{1}} \}$ to get the following:
\begin{align}
    & P \Bigg[ \frac{\left \lVert {E_1}{Y_{-1(1)}^{{T}}}  \right \rVert _{\infty}}{T} > 2\pi Q_1 \eta \Bigg]
    \leq
    6 \hspace{0.1cm} \text{exp}[-cT min \{\eta, \eta^2\} + 2 \text{log} (d_1)]
\end{align}
Now, we take $\eta = \sqrt{\frac{2\text{log}(d_1)}{T}}$, and get,
\begin{align}
    &  P \Bigg[ \frac{\left \lVert {E_1}{Y_{-1(1)}^{^{T}}}  \right \rVert _{\infty}}{T}> 2\pi Q_1 \sqrt{\frac{2\text{log}(d_1)}{T}} \Bigg]
    \leq
    6 \hspace{0.1cm} \text{exp}[-c_1 \text{log}(d_1)]
\end{align}
for a suitably chosen constant $c_1$. Thus, we choose $\lambda_{{S_{1}}} = k_1 Q_1 \sqrt{\frac{2\text{log}(d_1)}{T}} + \frac{4 \gamma \alpha_{1}}{\sqrt{d_{1}d_{1}}}$, for some suitably chosen constant $k_1$. Along similar lines, one can show that
\begin{align}
    &  P \Bigg[ \frac{\left \lVert {E_2}{Y_{-1(2)}^{^{T}}}   \right \rVert _{\infty}}{T}> 2\pi Q_2 \sqrt{\frac{2\text{log}(d_2)}{T}} \Bigg]
    \leq
    6 \hspace{0.1cm} \text{exp}[-c_2 \text{log}(d_2)]
\end{align}
for a suitably chosen constant $c_2$. So we choose $\lambda_{{S_{2}}}$ as $k_2 Q_2 \sqrt{\frac{2\text{log}(d_2)}{T}} + \frac{4 \gamma \alpha_{2}}{\sqrt{d_{2}d_{2}}} $ for some suitable chosen constant $k_2$. 
\begin{align}
    &  P \Bigg[ \frac{\left \lVert {E_3}{Y_{-1(3)}^{^{T}}}   \right \rVert _{\infty}}{T}> 2\pi Q_3 \sqrt{\frac{2\text{log}(d_3)}{T}} \Bigg]
    \leq
    6 \hspace{0.1cm} \text{exp}[-c_3 \text{log}(d_3)]
\end{align}
for a suitably chosen constant $c_3$. Therefore we choose $\lambda_{{S_{3}}}$ as $k_3 Q_3 \sqrt{\frac{2\text{log}(d_3)}{T}} + \frac{4 \gamma \alpha_{2}}{\sqrt{d_{3}d_{3}}} $ for some suitable chosen constant $k_3$.

\vspace{0.1in}
We now show that $\lambda_{L_1} \geq 4 \left \lVert \mathcal{D}_1 \right\rVert_{sp}$, $\lambda_{L_2} \geq 4 \left \lVert \mathcal{D}_2 \right\rVert_{sp}$ and $\lambda_{L_3} \geq 4 \left \lVert \mathcal{D}_3 \right\rVert_{sp}$ are satisfied with high probability. For that, let $\mathcal{S}^{d_1-1}$ denote the unit ball for $R^{d_1}$. We discretize this unit ball using $\epsilon$-net $\mathcal{N}$ with cardinality at most $(1+\frac{2}{\epsilon})^{d_1}$. Now following the same argument as in Lemma F.2 of \cite{basu2015regularized}
\begin{equation}
    \underset{u \in \mathcal{S}^{d_1-1}, v \in \mathcal{S}^{d_1-1}}{sup} \lvert u^\prime \frac{(\mathbb{E}_1Y_{-1(1)}^{T})}{T}v\lvert \leq k \underset{u \in \mathcal{N}, v \in \mathcal{N}}{sup} \lvert u^\prime \frac{(\mathbb{E}_1Y_{-1(1)}^{T})}{T}v\lvert  
\end{equation}
for some suitable chosen constant $k$. Now, following the previous approach and taking union bound over $(1+\frac{2}{\epsilon})^{2d_1}$ choices of $u$ and $v$ we get,
\begin{equation}
    Pr\{\frac{\norm{\mathbb{E}_1Y_{-1}^{(1)^T}}_{sp}}{T} > 2\pi \text{ } k\eta Q_1\} \leq 6\text{ }exp[-cT \text{ }min\{\eta^2,\eta\}+ 2d_1\log(1+\frac{2}{\epsilon})] 
\end{equation}
Hence we choose $\eta=\sqrt{\frac{c_12d_1\log(1+\frac{2}{\epsilon})}{cT}}$ and the above equation reduces to the following
\begin{equation}
    Pr\{\frac{\norm{\mathbb{E}_1Y_{-1(1)}^{(T}}_{sp}}{T} > 2\pi \text{ } k\sqrt{\frac{c_12d_1\log(1+\frac{2}{\epsilon})}{cT}} Q_1\} \leq 6\text{ }exp[-c_4d_1] 
\end{equation}

for a suitable chosen constant $c_4$. Hence we choose $\lambda_{L_1}=k_1^*Q_1\sqrt{\frac{2d_1}{T}}$, for a suitable chosen constant $k_1^*$. Using a similar reasoning, it can be shown that 
\begin{equation}
    Pr\{\frac{\norm{\mathbb{E}_2Y_{-1(2)}^{T}}_{sp}}{T} > 2\pi \text{ } k\sqrt{\frac{c_12d_2\log(1+\frac{2}{\epsilon})}{cT}} Q_2\} \leq 6\text{ }exp[-c_5d_2] 
\end{equation}

for a suitable chosen constant $c_5$. So we choose $\lambda_{L_2}=k_2^*Q_2\sqrt{\frac{2d_2}{T}}$, for a suitable chosen constant $k_2^*$. 
\begin{equation}
    Pr\{\frac{\norm{\mathbb{E}_3Y_{-1(3)}^{T}}_{sp}}{T} > 2\pi \text{ } k\sqrt{\frac{c_12d_3\log(1+\frac{2}{\epsilon})}{cT}} Q_3\} \leq 6\text{ }exp[-c_6d_3] 
\end{equation}

for a suitable chosen constant $c_6$. So we choose $\lambda_{L_3}=k_3^*Q_3\sqrt{\frac{2d_3}{T}}$, for a suitable chosen constant $k_3^*$.
Now the proof follows by substituting these choices of the regularizer parameters in the bound obtained in Lemma \ref{lem31}. 
\end{proof}

\subsection{Proof of error bound under sub-exponential tail decay}
\paragraph{}Before presenting the proof of Theorem \ref{theo_x_rand}, we first state and prove the following lemmas, which will be useful for the proof. To that end, we first define the linear process of the following form:
\begin{equation}\label{generic}
    X_t = \sum_{l=0}^{\infty}B_lw_{t-l}
\end{equation}
In the case where the process is Gaussian, the $w_t$'s correspond to Gaussian white noise process. However, we assume that $w_t$ is a white noise process whose coordinates gave the following $\alpha$-sub-exponential tail decay, that is, there exist two constants $a,b$ such that the following holds:
\begin{equation}\label{genericdef}
    Pr\{\lvert w_{tj}\rvert\geq \xi\}\leq a \text{ }exp(-\frac{\xi^\alpha}{b}), \forall \xi >0
\end{equation}
The following lemma generalizes a Hanson-Wright type concentration inequality to the samples from a linear process $X_t$ as defined in equation (\ref{generic}).
\medskip
\begin{lem}
\label{lemma5}
Consider some generic $p$-dimensional linear processes given in the form of $X_t=\sum_{l=0}^{\infty}\Phi_lu_{t-l}$, where $u_t$'s are i.i.d. and their coordinates follow $\alpha$-sub-exponential tail decay, as characterized by equation (\ref{genericdef}). Denote its realization by $X \in \mathbb{R}^{n \times p}$ with $n$ consecutive observations stacked in its rows. Then, for a deterministic $np \times np$ matrix $A$, there exists some constant $C$ such that the following holds:
\begin{equation}
    \small{Pr\{\lvert {Vec(X^\prime)}^\prime\ A Vec(X^\prime)- Exp[{Vec(X^\prime)}^\prime\ A Vec(X^\prime)]\rvert >2\pi\eta\mathscr{M}(f_X)\}\leq \tau(\eta,\alpha,a)},
\end{equation}
where, $\mathscr{M}(f_X)$ is as defined in the main paper and 
\begin{equation}
    \tau(\eta,\alpha,A)=2 exp[-C \text{min} \{\frac{\eta^2}{rank(A)\norm{A}^2_{op}},{(\frac{\eta}{\norm{A}_{op}})}^{\frac{\alpha}{2}}\}]
\end{equation}
\end{lem}
\begin{proof}
Let $Vec(X^\prime)\stackrel{d}{=}\Omega^{\frac{1}{2}}Z$, where $\Omega$ is the covariance matrix of the $np$-dimensional random vector $Vec(X^\prime)$ and $Z$ satisfies $Exp(Z)=0$ and $Exp(ZZ^\prime)$ is $I_{np}$. Now applying Proposition 1.1 in \cite{gotze2019concentration} gives
\begin{align*}
  &Pr\{\lvert {Vec(X^\prime)}^\prime\ A Vec(X^\prime)- Exp[{Vec(X^\prime)}^\prime\ A Vec(X^\prime)]\rvert >2\pi\eta\mathscr{M}(f_X)\}\\
  =&Pr\{\lvert Z^\prime\Omega^{\frac{1}{2}}A\Omega^{\frac{1}{2}}Z- Exp[Z^\prime\Omega^{\frac{1}{2}}A\Omega^{\frac{1}{2}}Z]\rvert >2\pi\eta\mathscr{M}(f_X)\}\\
  \leq&2 exp(-c_0 \cdot \nu(\Omega^{\frac{1}{2}}A\Omega^{\frac{1}{2}},\alpha,2\pi\eta\mathscr{M}(f_X)))
\end{align*}
where
\begin{equation}
    \nu(A,\alpha,t)=min\{\frac{t^2}{M^4\norm{A}^2_F},{(\frac{t}{M^2\norm{A}_{op}})}^{\frac{\alpha}{2}}\}
\end{equation}
Here both $c_0$ and $M$ are constants that depend on $a$ and $b$. Next, we consider the bounds for various norms on $\Omega^{\frac{1}{2}}A\Omega^{\frac{1}{2}}$ as follows:
\begin{itemize}
    \item $\norm{\Omega^{\frac{1}{2}}A\Omega^{\frac{1}{2}}}_{op}\leq \norm{\Omega}_{op}\norm{A}_{op}\leq 2 \pi  \mathscr{M}(f_X) \norm{A}_{op}$, where the last inequality follows from Proposition 2.3 in \cite{basu2015regularized}, which applies to general linear process.\\
    \item
    $\norm{\Omega^{\frac{1}{2}}A\Omega^{\frac{1}{2}}}_{F}\leq \sqrt{rank(\Omega^{\frac{1}{2}}A\Omega^{\frac{1}{2}})}\norm{\Omega^{\frac{1}{2}}A\Omega^{\frac{1}{2}}}_{op}\leq 2\pi\sqrt{rank(A)}\norm{A}_{op}\mathscr{M}(f_X)$
\end{itemize}
The proof follows by putting these bounds in $\nu(\Omega^{\frac{1}{2}}A\Omega^{\frac{1}{2}},\alpha,2\pi\eta\mathscr{M}(f_X))$.  
\end{proof}
\medskip
\paragraph{} Our next lemma is a generalization of Proposition 2.4 in \cite{basu2015regularized}, to the case where the underlying process is characterized by the equations (\ref{generic}) and (\ref{genericdef}).
\begin{lem}
\label{lemma6}
Consider a generic $p$-dimensional linear process in the form of $X_t=\sum_{l=0}^{\infty}\Phi_lu_{t-l}$, where  the coordinates of $u_t$ have $\alpha$-sub-exponential tail decay as characterizd by equation (\ref{genericdef}). Let $\Sigma_X(0)=Cov(X_t, X_t)$. Denote the realization of $X_t$ by $X\in \mathbb{R}^{n\times p}$ and the sample covariance by $S=\frac{1}{n}X^\prime X$. Then
\begin{enumerate}
    \item  For unit vectors $v_1$ and $v_2$ satisfying $\norm{v_1}\leq 1$, $\norm{v_2}\leq 1$, the following bound holds:
    \begin{align*}
        &Pr\{\lvert v_1^\prime (S-\Sigma_X(0))v_1\rvert > 2 \pi  \eta \mathscr{M}(f_X)\} \leq \tau^\prime (\eta, \alpha, n)\\
        \text{and }&Pr\{\lvert v_1^\prime (S-\Sigma_X(0))v_2\rvert > 6 \pi  \eta \mathscr{M}(f_X)\} \leq 2\tau^\prime (\eta, \alpha, n)
    \end{align*}
    \item Consider a $q$-dimensional linear process $Z_t=\sum_{l=0}^{\infty}\Psi_lw_{t-l}$, where the coordinates of $w_t$ have $\alpha$-sub-exponential tail decay, as characterized by equation (\ref{genericdef}). Also $Cov(X_t, Z_t)=0 \forall t$ and the data matrix $Z \in \mathbb{R}^{n \times q}$ is similarly defined. Then the following bound holds:
    \begin{align*}
        &Pr\{\lvert v_1^\prime (X^\prime Z)v_2\rvert\ > 2\pi \eta (\mathscr{M}(f_X)+\mathscr{M}(f_Z)+\mathscr{M}(f_{X,Z}))\}\leq 3 \tau^\prime(\eta,\alpha,n),\\
        &\text{ where }\mathscr{M}(f_{X,Z}) \text{ is defined the same way as in the main paper }.  
    \end{align*}
\end{enumerate}
Here $\tau^\prime$ is defined as $\tau ^\prime(\eta,\alpha,n)=c_1 exp[-c_2 min\{n\eta^2,{(n\eta)}^{\frac{\alpha}{2}}\}]$, for some constants $c_1$ and $c_2$.
\end{lem}

\begin{proof}
First we note that with $A=I_n$ and the definition of $\tau(\eta,\alpha,A)$ the following holds for some constant $C>0$
\begin{equation}
    \tau(n\eta,\alpha,A)=2 exp[-C min\{n\eta^2,{(n\eta)}^{\frac{\alpha}{2}}\}]
\end{equation}
Let $y_t=v_1^\prime X_t$ and $Y=Xv_1 \in \mathbb{R}^n$ be $n$ consecutive observations of the scalar process $\{y_t\}$. Then, we will have $v_1^\prime S v_1 \stackrel{d}{=} \frac{1}{n} Y^\prime Y$ and $v_1^\prime\Sigma_X(0)v_1=Exp[\frac{Y^\prime Y}{n}]$. Applying Lemma \ref{lemma5} to the process $\{y_t\}$ with $A=I_n$ (since moment properties are preserved under linear transformation), we obtain the following:
\begin{equation}
\begin{split}
    Pr\{\lvert v_1^\prime (S- \Sigma_X(0))v_1\rvert> 2\pi \eta \mathscr{M}(f_Y)\}&=Pr\{\lvert Y^\prime Y -Exp(Y^\prime Y)\rvert > 2 \pi n \eta \mathscr{M}(f_Y)\}\\
    &\leq \tau^\prime(\eta, \alpha,n)
\end{split}
\end{equation}
Further by Lemma C.6 of Sun et al. (2018), it follows that $\mathscr{M}(f_Y)\leq \norm{v_1}^2\mathscr{M}(f_X)=\mathscr{M}(f_X)$. Hence the following bound holds:
\begin{equation}
    Pr\{\lvert v_1^\prime (S- \Sigma_X(0))v_1\rvert> 2\pi \eta \mathscr{M}(f_X)\}\leq \tau^\prime(\eta, \alpha,n)
\end{equation}
This proves the first part in (i). The rest of the proof follows along the similar lines to the derivation of Proposition 2.4 in \cite{basu2015regularized} and an outline is as follows:

For $\lvert v_1^\prime (S- \Sigma_X(0))v_2\rvert$, one considers the following decomposition:
\begin{equation}
\begin{split}
2\lvert v_1^\prime (S- \Sigma_X(0))v_2\rvert &\leq \lvert v_1^\prime (S- \Sigma_X(0))v_1\rvert + \lvert v_2^\prime (S- \Sigma_X(0))v_2\rvert+ \lvert {(v_1+v_2)}^\prime\\
&(S- \Sigma_X(0)){(v_1+v_2)}\rvert
\end{split}
\end{equation}
with $\norm{(v_1+v_2)}\leq 2$. Now repeating the steps as in (i) for each of the three components above yields the desired result.

For $\lvert v_1^\prime(X^\prime Z)v_2\rvert$, let $\tilde{y_t}=v_2^\prime Z_t$ and thus $v_1^\prime (X^\prime Z)v_2=\frac{1}{n}\sum_{t=1}^{n}y_t\tilde{y_t}$ and it satisfies the following decomposition:
\begin{align*}
    \frac{2}{n}\sum_{t=1}^{n}y_t\tilde{y_t} &= [\frac{1}{n}\sum_{t=1}^n{(y_t+\tilde{y_t})}^2-Var(y_t+\tilde{y_t})]-[\frac{1}{n}\sum_{t=1}^n{y_t}^2-Var(y_t)]\\
    &-[\frac{1}{n}\sum_{t=1}^n{\tilde{y_t}}^2-Var(\tilde{y_t})]\\
    &=\frac{1}{n}[G^\prime G -Exp(G^\prime G)]-\frac{1}{n}[Y^\prime Y -Exp(Y^\prime Y)]\\
    &-\frac{1}{n}[{\tilde{Y}}^\prime \tilde{Y} -Exp({\tilde{Y}}^\prime \tilde{Y})],
\end{align*}
where $g_t=y_t+\tilde{y_t}$ is the summation process and $G$ and $\tilde{Y}$ are defined analogously to the definition of $Y$. Now the proof of (ii) follows by repeating the same steps as in the proof of the second part of (i) and noting the fact that $\mathscr{M}(f_G) \leq \mathscr{M}(f_Z) + \mathscr{M}(f_X)+\mathscr{M}(f_{X,Z})$. 
\end{proof}
\medskip
Our next lemma can be considered as a generalization of the deviation bound derived in \cite{basu2015regularized}.
\begin{lem}
\label{lemma7}
\label{ext1}
There exist positive constants $C$, $c_1$ and $c_2$ such that the following deviation bound holds:
\begin{equation}
    \norm{X^\prime E}_{\infty} \leq C \mathbb{Q} \frac{{(\log p+ \log q)}^{\frac{1}{\alpha}}}{\sqrt{n}}
\end{equation}
with probability at least $1-c_1 \exp\{-c_2{(\log(pq))}^{\frac{2}{\alpha}}\}$, for any random realizations $X \in \mathbb{R}^{n \times p}$ and $E \in \mathbb{R}^{n \times q}$, drawn from the $p$-dimensional linear processes $\{X_t\}$ and $q$-dimensional linear processes $\{\ep_t\}$ respectively, where the coordinates of $X_t$ and $\ep_t$ have $\alpha$-sub-exponential tail decay and $\mathbb{Q}=\mathscr{M}(f_X)+\mathscr{M}(f_{\ep})+\mathscr{M}(f_{X,\ep})$  
\end{lem}
\begin{proof}
The proof follows by applying part (ii) of Lemma \ref{lemma6} with $v_1=e_i$ and $v_2=e_j$, then taking union bound over all $pq$ elements and finally choosing $\eta = c_0 \mathbb{Q} \frac{{(\log p+ \log q)}^{\frac{1}{\alpha}}}{\sqrt{n}}$ for some suitably chosen constant $c_0$. 
\end{proof}
\medskip
The following lemma verifies the Restricted Strong Convexity condition and thus can be considered as a generalization of Proposition 4.2 of \cite{basu2015regularized}.
\begin{lem}\label{lemma8}
Consider a random realization $X\in\mathbb{R}^{n \times p}$ drawn from the $p$-dimensional linear process $X_t=\sum_{l=0}^{\infty}\Phi_lu_{t-l}$, where each coordinates of $u_t$ has $\alpha$-sub-exponential tail decay. Then RSC holds for $X$ with parameter $\alpha_{RSC}=\pi m(f_X)$ and tolerance $\tau=c_0 \alpha_{RSC}\frac{\log p}{n^{\frac{\alpha}{2}}}$ with probability at least $1-c_1exp(-c_2n^{\frac{\alpha}{2}})$, where the definition of RSC and $m(f_X)$ are the same as defined in \cite{basu2015regularized}.
\end{lem}
\begin{proof}
Let $S=\frac{1}{n}X^\prime X$. First suppose that we have the following:
\begin{equation}\label{toshow}
    \frac{1}{2}v^\prime S v = \frac{1}{2}v^\prime (\frac{X^\prime X}{n}) v \geq \frac{\alpha_{RSC}}{2}\norm{v}_2^2-\tau\norm{v}_1^2, \forall v \in \mathbb{R}^p
\end{equation}
Then, for all $\Delta \in \mathbb{R}^{p \times p}$ and letting $\Delta_j$ denote the $j^{th}$ column, the RSC condition automatically holds since
\begin{align*}
    \frac{1}{2T}\norm{X\Delta}_{F}^2 &=\frac{1}{2} \sum_{j=1}^p \Delta_j^\prime(\frac{X^\prime X}{n})\Delta_j\\
    &\geq \frac{\alpha_{RSC}}{2}\sum_{j=1}^p \norm{\Delta_j}_2^2-\tau \sum_{j=1}^p \norm{\Delta_j}_1^2\\
    &\geq \frac{\alpha_{RSC}}{2} \norm{\Delta}_{F}^2 - \tau \norm{\Delta}_1^2
\end{align*}
Therefore it suffices to verify that (\ref{toshow}) holds. Now applying the discretization argument as in Lemma F.2 and Lemma F.3 in \cite{basu2015regularized}, define $\mathbb{K}(2s)=\{v \in \mathbb{R}^p,\norm{v} \leq 1,\norm{v}_0 \leq 2s\}$ and taking the union bound in this $2s$-sparse cone gives the following inequality:
\begin{align*}
    &Pr\{\underset{v \in \mathbb{K}(2s)}{sup} \lvert v^\prime (S-\Sigma_X(0))v\rvert > 2\pi \mathscr{M}(f_X)\eta\}\\
    \leq&2 \cdot min\{p^s,{(21e \cdot \frac{p}{s})}^s\} \tau^\prime(\eta, \alpha,n)\\
    =&2c_1 exp[-c_2 \text{ }min\{n\eta^2,{(n\eta)}^{\frac{\alpha}{2}}\}+ s\text{ }min \{\log p, \log(21e \frac{p}{s})\}]
\end{align*}
Let $\eta = \frac{m(f_X)}{54\mathscr{M}(f_X)}$. Then applying the results from Lemma 12 in \cite{loh2011high} with $\Gamma = S- \Sigma_X(0)$ and $\delta = \pi \frac{m(f_X)}{27}$ the following holds:
\begin{equation}
    \frac{1}{2}v^\prime S v \geq \frac{\alpha_{RSC}}{2}\norm{v}^2- \frac{\alpha_{RSC}}{2s}\norm{v}_1^2, \text{where } \alpha_{RSC}= \pi m(f_X)
\end{equation}
with probability at least $1-2 min\{p^s,{(21e \cdot \frac{p}{s})}^s\} \tau^\prime(\eta, \alpha,n)$. By letting $s=c_0 \frac{n^{\frac{\alpha}{2}}}{\log p}$ for some small constant $c_0$, $\tau$ can be expressed as $\tau=c_0 \alpha_{RSC}\frac{\log p}{n^{\frac{\alpha}{2}}}$ and thus the bound holds with the probability as given in the statement.
\end{proof} 
\subsection*{\small{Proof of Theorem \ref{theo_x_rand}}}

\begin{proof}
At first, our job is to find suitable choices of $\lambda_{S_1}$, $\lambda_{S_2}$ and $\lambda_{S_3}$ so that $\lambda_{{S_1}} \geq  4 \left \lVert {\mathcal{D}_1} \right\rVert_{\infty} + \frac{4\gamma \alpha_1}{\sqrt{d_1 d_1}}$, $\lambda_{{S_2}} \geq  4 \left \lVert {\mathcal{D}_2} \right\rVert_{\infty} + \frac{4\gamma \alpha_2}{\sqrt{d_2 d_2}}$ and $\lambda_{{S_3}} \geq  4 \left \lVert {\mathcal{D}_3} \right\rVert_{\infty} + \frac{4\gamma \alpha_3}{\sqrt{d_3 d_3}}$ are satisfied with high probability. To that end, applying Lemma \ref{ext1} on  $\left \lVert {E_1}{Y_{-1(1)}^{^{T}}}  \right \rVert _{\infty}$ we get, 

\begin{align}
    &  P \Bigg[ \frac{\left \lVert {E_1}{Y_{-1(1)}^{^{T}}}  \right \rVert _{\infty}}{T}> c_1 Q_1 \frac{\{2\log d_1\}^{1/\alpha}}{\sqrt{T}} \Bigg]
    \leq
    c_2 \hspace{0.1cm} \text{exp}[-c_3 (\log d_1)^{\frac{2}{\alpha}}]
\end{align}

for some suitably chosen constants $c_1, c_2, c_3$.  Thus, we choose $\lambda_{{S_{1}}} = k_1 Q_1 \frac{\{2\text{log}(d_1)\}^{1/\alpha}}{\sqrt{T}} + \frac{4 \gamma \alpha_{1}}{\sqrt{d_{1}d_{1}}}$, for some suitably chosen constant $k_1$. Following a similar reasoning, it can be shown that 

\begin{align}
    &  P \Bigg[ \frac{\left \lVert {E_2}{Y_{-1(2)}^{T}}  \right \rVert _{\infty}}{T}> c_4 Q_2 \frac{\{2\log d_2\}^{1/\alpha}}{\sqrt{T}} \Bigg]
    \leq
    c_5 \hspace{0.1cm} \text{exp}[-c_6 (\log d_2)^{\frac{2}{\alpha}}]
\end{align}
for some suitably chosen constants $c_4, c_5, c_6$.  Thus, we choose $\lambda_{{S_{2}}} = k_2 Q_2 \frac{\{2\text{log}(d_2)\}^{1/\alpha}}{\sqrt{T}} + \frac{4 \gamma \alpha_{2}}{\sqrt{d_{2}d_{2}}}$, for some suitably chosen constant $k_2$. Following a similar approach, we choose $\lambda_{{S_{3}}} = k_3 Q_3 \frac{\{2\text{log}(d_3)\}^{1/\alpha}}{\sqrt{T}} + \frac{4 \gamma \alpha_{3}}{\sqrt{d_{3}d_{3}}}$.\\

Next, we need to choose $\lambda_{L_1}$, $\lambda_{L_2}$ and $\lambda_{L_3}$ in such a way that $\lambda_{L_1} \geq 4 \left \lVert \mathcal{D}_1 \right\rVert_{sp}$, $\lambda_{L_2} \geq 4 \left \lVert \mathcal{D}_2 \right\rVert_{sp}$ and $\lambda_{L_3} \geq 4 \left \lVert \mathcal{D}_3 \right\rVert_{sp}$ are satisfied with high probability. To that end, let $\mathcal{S}^{d_1-1}$ denote the unit ball for $R^{d_1}$. We discretize this unit ball using $\epsilon$-net $\mathcal{N}$ with cardinality at most $(1+\frac{2}{\epsilon})^{d_1}$. Now following the same argument as in Lemma F.2 of \cite{basu2015regularized}, for small enough $\epsilon > 0$,
\begin{equation}
    \underset{u \in \mathcal{S}^{d_1-1}, v \in \mathcal{S}^{d_1-1}}{sup} \lvert u^\prime \frac{(\mathbb{E}_1Y_{-1(1)}^{^T})}{T}v\lvert \leq k \underset{u \in \mathcal{N}, v \in \mathcal{N}}{sup} \lvert u^\prime \frac{(\mathbb{E}_1Y_{-1(1)}^{^T})}{T}v\lvert  
\end{equation}
for some suitable chosen constant $k$. Now, as before, taking union bound over $(1+\frac{2}{\epsilon})^{2d_1}$ choices of $u$ and $v$, Lemma \ref{lemma6} gives the following:
\begin{equation}
    Pr\{\frac{\norm{\mathbb{E}_1Y_{-1(1)}^{^T}}_{sp}}{T} > 2\pi \text{ } k\eta Q_1\} \leq c_1\text{ }exp[-c_2 \text{ }min\{T\eta^2,(T\eta)^{\alpha/2}\}+ 2d_1\log(1+\frac{2}{\epsilon})] 
\end{equation}
Hence we choose $\eta=(2d_1)^{1/\alpha}\sqrt{\frac{c_1\log(1+\frac{2}{\epsilon})}{T}}$ and the above equation boils down to 
\begin{equation}
    Pr\{\frac{\norm{\mathbb{E}_1Y_{-1(1)}^{^T}}_{sp}}{T} > 2\pi \text{ } k(2d_1)^{1/\alpha}\sqrt{\frac{c_1\log(1+\frac{2}{\epsilon})}{T}} Q_1\} \leq 6\text{ }exp[-c_3(d_1)^{2/\alpha}] 
\end{equation}

for a suitable chosen constant $c_3$. So we choose $\lambda_{L_1}=k_1^*Q_1\frac{(2d_1)^{1/\alpha}}{\sqrt{T}}$, for a suitable chosen constant $k_1^*$. Using a similar argument, one can choose $\lambda_{L_2}=k_2^*Q_2\frac{(2d_2)^{1/\alpha}}{\sqrt{T}}$ and $\lambda_{L_3}=k_3^*Q_3\frac{(2d_3)^{1/\alpha}}{\sqrt{T}}$, for suitable chosen constants $k_2^*$ and $k_3^*$. Now the proof of the theorem follows by using these choices of the regularizer parameters and putting the same in the bound obtained in Lemma \ref{lem31}.\\
\end{proof}

\end{document}